\documentclass[twocolumn,twocolappendix]{aastex701}

\usepackage{tabularx} %
\usepackage{amsmath,amssymb}  %
\usepackage[dvipsnames]{xcolor}
\usepackage{graphicx} %

\usepackage{textcomp, gensymb, bm}
\usepackage{caption}
\usepackage{subcaption}
\usepackage{breqn}
\usepackage{ulem}

\newcommand{\revision}[1]{\textcolor{black}{#1}}

\begin{document}

\title{Modern tidal interaction models for rapid binary population synthesis: I. Methods}

\author[orcid=0000-0001-7344-5260,gname=Veome,sname=Kapil]{Veome Kapil}
\affiliation{William H. Miller III Department of Physics and Astronomy, Johns Hopkins University, \\3400 N. Charles Street, Baltimore, Maryland, 21218, USA}
\email{vkapil1@jh.edu}  

\author[orcid=0000-0002-6134-8946,gname=Ilya,sname=Mandel]{Ilya Mandel}
\affiliation{Monash Centre for Astrophysics, School of Physics and Astronomy, Monash University, Clayton, Victoria 3800, Australia}
\affiliation{OzGrav: Australian Research Council Centre of Excellence for Gravitational Wave Discovery, Clayton, VIC 3800, Australia}
\email{ilya.mandel@monash.edu}

\author[orcid=0000-0001-7113-723X,gname=Evgeni,sname=Grishin]{Evgeni Grishin}
\affiliation{Monash Centre for Astrophysics, School of Physics and Astronomy, Monash University, Clayton, Victoria 3800, Australia}
\affiliation{OzGrav: Australian Research Council Centre of Excellence for Gravitational Wave Discovery, Clayton, VIC 3800, Australia}
\email{}

\author[orcid=0000-0002-4544-0750,gname=Jim,sname=Fuller]{Jim Fuller}
\affiliation{TAPIR, Mailcode 350-17, California Institute of Technology, Pasadena, CA 91125, USA}
\email{}

\author[orcid=0000-0003-1530-2557,gname=Jeff,sname=Riley]{Jeff Riley}
\affiliation{Monash Centre for Astrophysics, School of Physics and Astronomy, Monash University, Clayton, Victoria 3800, Australia}
\email{}

\author[orcid=0000-0003-0751-5130,gname=Emanuele,sname=Berti]{Emanuele Berti}
\affiliation{William H. Miller III Department of Physics and Astronomy, Johns Hopkins University, \\3400 N. Charles Street, Baltimore, Maryland, 21218, USA}
\email{berti@jhu.edu}

\begin{abstract}
In this work, we present an updated prescription of contemporary tidal dissipation theory adapted for rapid binary population synthesis. Our simplified expressions encode the dependence of tidal dissipation on stellar structure, stratification, and tidal forcing frequency, while remaining computationally efficient. We implement these prescriptions in the rapid population synthesis code COMPAS, and demonstrate the self-consistent coupling of tides with stellar evolution and binary properties such as orbital periods, spins, and eccentricities for several representative binary systems. When compared with commonly used tidal prescriptions, our equilibrium tidal dissipation efficiencies can be stronger by 1-2 orders of magnitude for low mass main sequence and giant type stars, and dynamical tides can be stronger by 1-7 orders of magnitude due to the explicit dependence on internal stellar structure and the presence of inertial wave dissipation. Despite our simplistic approach, our models agree with detailed stellar simulations to within an order of magnitude across tidal dissipation mechanisms.
\end{abstract}

\keywords{\uat{Stellar physics}{1621} --- \uat{Binary stars}{154} --- \uat{Tides}{1702} --- \uat{Tidal interaction}{1699}}

\section{Introduction}

Tidal interactions play a central role in shaping the orbital and spin evolution of close binary stars and planetary systems. Tidal dissipation in stars is commonly divided into equilibrium tides and dynamical tides. Equilibrium tides describe the large-scale, quasi-hydrostatic deformations of a star and are primarily dissipated by viscous processes in convective envelopes, making them most relevant for low-mass stars and evolved stars with extended convection zones \citep{zahn_tidal_1977, hut_tidal_1981, Ogilvie:2014dwa}. Dynamical tides arise from the excitation of internal gravity waves in radiative regions and inertial waves in convective regions for rapidly rotating stars. These may dominate tidal dissipation in massive stars, compact stars, or rapidly rotating systems \citep{Zahn:1975A&A....41..329Z, Ogilvie:2012gm, mathis2015variation}.

Observations of stellar binaries in open clusters \citep{Meibom_2005,Hansen:2010nq, Raghavan2010survey, geller2021stellar}, heartbeat stars \citep{Fuller:2017bzf, hambleton2018kic}, and tidally interacting hot Jupiters \citep{Penev_2018, dawson2018origins, am2025} offer valuable insights into the efficiency of tidal energy dissipation across stellar and planetary interiors. New binary catalogs, such as Gaia’s successive data releases, have also yielded unprecedented samples of binary and multiple systems with period and eccentricity observations, enabling empirical inference of tidal dissipation strength \citep{Moe:2017icj,Gaia:2018ydn, vallenari2023gaia,tokovinin2020orbits,El-Badry:2024vjt}. Furthermore, gravitational-wave observations by LIGO-Virgo-KAGRA \citep{LIGOScientific:2018mvr, LIGOScientific:2020ibl, KAGRA:2021vkt} have opened an entirely new window into the fates of massive stellar binaries, the properties of which may carry imprints of binary stellar evolutionary processes such as mass transfer and tidal synchronization. In stellar origin compact object binaries, tidal interactions during earlier evolutionary phases can influence stellar spins, orbital separations, eccentricities, and mass loss, thereby shaping the spin magnitudes and mass distributions inferred from gravitational-wave observations \citep[e.g.,][]{Gerosa:2013laa, Gerosa:2018wbw, Qin:2018vaa, Bavera:2020inc, Steinle:2022rhj}. Tidal dissipation may also play an important role in the orbital evolution of wind-fed neutron star high-mass X-ray binaries \citep[e.g.,][]{Levine:1999cy}.

A viable method for studying populations of stellar and compact binaries is rapid binary population synthesis. For population synthesis predictions to be effective, several binary evolutionary processes, including tides, must be modeled accurately and efficiently so that we can evolve millions of binary stellar systems and compare them against observational data. In practice, most population synthesis implementations adopt highly idealized treatments of equilibrium and dynamical tides that reduce the stellar response to a small number of global parameters, such as envelope mass or a fixed convective turnover timescale \citep[e.g.,][]{Hurley:2002rf, Belczynski:2005mr, COMPASTeam:2021tbl, Fragos:2022jik, Mirouh2023:10.1093/mnras/stad2048}. These tidal dissipation models are based on well-established formalisms \citep{Zahn:1975A&A....41..329Z, hut_tidal_1981, Goodman:1998yg}, but may not self-consistently incorporate up-to-date information about stellar interiors or tidal frequencies.

There has been significant progress on the theoretical modeling of tidal dissipation, and the dependence on stellar structure, stratification, and rotation rates is now understood to a better degree \citep{Barker:2010ru, Ogilvie:2012gm,  Esseldeurs2024A&A...690A.266E}. \citet{Duguid:10.1093/mnras/staa2216} provided an update to equilibrium tides with new frequency-dependent scalings derived from detailed hydrodynamical simulations, which suggest that equilibrium tidal dissipation can vary by orders of magnitude over a star’s lifetime as the convective envelope structure and rotation rate evolves. Parallel developments in dynamical tides have shown how the excitation and damping of gravity and inertial waves depend sensitively on the internal stratification. Notably, works such as \citet{Kushnir:2017}, \citet{Ahuir:2021} and \citet{Esseldeurs2024A&A...690A.266E} have provided updated formalisms for characterizing tidal dissipation in the radiative zones of stars. These models offer the promise of more realistic tidal dissipation across a wider range of stellar masses and evolutionary phases. However, model accuracy comes at the cost of detailed stellar structure profiles, which are typically obtained from stellar evolution codes such as MESA~\citep{Paxton:2010ji, Paxton:2013pj} or STAREVOL ~\citep{Siess:2000hq, amard2019first}, making them computationally prohibitive for direct use in large-scale population synthesis.

In this work, we attempt to bridge this gap between detailed tidal dissipation theory and rapid binary population synthesis. We present a unified tidal prescription, implemented in COMPAS~\citep{COMPASTeam:2021tbl, COMPAS:2025}, which self-consistently simulates both equilibrium and dynamical tides alongside stellar evolution, stellar rotation, mass transfer, supernovae, and other binary evolutionary processes. Our approach preserves the dominant dependencies on stellar interior structure and companion properties while remaining computationally efficient. Our framework can enable population-level predictions for orbital evolution, synchronization, and compact object spins that incorporate state-of-the-art tidal physics, which can be directly compared against observations ranging from open clusters to gravitational-wave sources. We present an application to binary compact object populations in Paper II of this series.

This paper is organized as follows. In Sec.~\ref{sec:theory} we review the theory of tides in binary systems, and introduce our tidal prescription. We focus on equilibrium tides in Sec.~\ref{sec:equilibrium_tides}, dynamical tides from internal gravity waves in Sec~\ref{sec:dynamical_tides_igw}, and dynamical tides from inertial waves in Sec.~\ref{sec:dynamical_tides_iw}. In Sec.~\ref{sec:applications}, we apply our model to four representative binary configurations, examining their relevant tidal Love numbers, tidal timescales, and binary evolution while comparing against classical tidal dissipation models.  
Finally, we summarize our main findings in Sec.~\ref{sec:conclusion}. 
In Appendix~\ref{app:comparison} we compare our model for equilibrium and dynamical tides with previous literature, and in Appendix~\ref{app:z77_model} we review for completeness our implementation of the 
`Z77' tidal dissipation model from \citet{zahn_tidal_1977} that we use as a reference.

\section{Theoretical framework}
\label{sec:theory}
Consider a spherical star, star~1, with mass $M_*$ and radius $R_*$, orbiting a companion, star~2, of mass $M_2$, at a semi-major axis $a$. In the reference frame centered on, and co-rotating with star~1, the tidal potential of star~2 can be expressed as a multipole expansion in solid spherical harmonics of degree $\ell\geq2$, azimuthal wave number $m$, and temporal harmonics of the orbital motion $n$, as \citep{Press:1977, Ogilvie:2014dwa, Idini_2021, Dewberry_2022}
\begin{multline}
    \Psi_{\rm tidal} = \text{Re} \sum_{\ell=2}^{\infty}\sum_{m = 0}^{\ell} \sum_{n = -\infty}^{\infty} [\mathcal{A}_{\ell mn} \left( \frac{r}{R_*} \right)^\ell \\ \times Y_\ell^m (\theta, \phi) \exp{(-i \omega_t t)}].
\end{multline}
Here $r$ is the radial coordinate measured from the center of star~1, and
\begin{equation}
    \mathcal{A}_{\ell mn} = A_{\ell mn}(e) \frac{G M_*}{R_*} \left(\frac{M_2}{M_*}\right) \left(\frac{R_*}{a}\right)^{\ell+1}
    \label{eq:tidal_potential_amplitude}
\end{equation}
is a pre-factor that depends on a dimensionless complex coefficient $A_{\ell mn} (e)$ \citep{Ogilvie:2014dwa}. In principle $A_{\ell mn}$ can depend on both eccentricity and obliquity, but we only consider co-planar orbits in this work such that $A_{\ell mn} = A_{\ell mn}(e)$. In general, the exact formula for $A_{\ell mn}$ will depend on the tidal mode being considered (see Table 1 of \citealp{Ogilvie:2014dwa}).
For instance, to leading order for a circular, coplanar orbit ($\ell=2, m=2, n=2$), the amplitude of the tidal potential becomes
\begin{align}
    \mathcal{A}_{222} &= \sqrt{\frac{6 \pi}{5}}\frac{M_2}{(M_* + M_2)} \omega_{\rm orb}^2 \\
    &= \sqrt{\frac{6 \pi}{5}} G \frac{M_2}{a^3}. 
    \label{eq:tidal_potential_222}
\end{align}

The periodic variation of the separation vector due to Keplerian motion is encoded in the $\exp{(-i \omega_t t)}$ term. We take the real component of the entire expression to obtain the real-valued tidal potential. The argument of this exponential is called the tidal frequency,
\begin{equation}
    \omega_t = n \omega_{\rm orb} - m \Omega_{\rm spin},
\end{equation}
and it is the orbital frequency of the tidal potential due to star~2 in the reference frame that is co-rotating with star~1, where $\omega_{\rm orb} = \sqrt{G (M_*+ M_2)/a^3}$ is the mean orbital angular velocity. In the simple case of circular orbits, only the $m=n$ terms are relevant.

Note that we have adopted a convention in which $m~\geq~0$ and $\omega_t$ can be negative or positive. This is equivalent to allowing $-\ell\leq m\leq \ell$ but restricting the tidal frequency to be positive, i.e., $\omega_t =|n\omega_{\rm orb}-m\Omega_{\rm spin}|$.  In our convention, positive $\omega_t$ frequencies represent prograde tides, and negative $\omega_t$ frequencies represent retrograde tides. 

The resulting gravitational perturbation generated by the tidal deformation of star~1 due to the tidal potential of star~2 can be expressed as 
\begin{multline}
    \Phi_{\rm perturb} = \text{Re}\sum_{\ell=2}^{\infty}\sum_{m = 0}^{\ell}\sum_{n = -\infty}^{\infty} [\mathcal{B}_{\ell mn} \left( \frac{r}{R_*} \right)^{-(\ell+1)} \\
    \times Y_\ell^m (\theta, \phi) \exp{(-i \omega_t t)}].
\end{multline}
As before, $r$ represents the distance from the center of star~1 in its co-rotating reference frame.
A convenient way to express the tidal response of star~1 to an external tidal potential from star~2 is the dimensionless and complex potential Love number \citep{Love1909, Idini_2021}
\begin{equation}
    k_{\ell,n}^m (\omega_t) = \frac{\Phi_{{\rm perturb}, \ell mn}}{\Psi_{{\rm tidal}, \ell mn}},
\end{equation}
which captures the frequency-dependent response of the gravitational potential of star~1 to the tidal potential of star~2.
The imaginary part of the tidal Love number $\text{Im}[k_{\ell,n}^m (\omega_t)]$ captures dissipative behavior, and thus, the transfer of energy and angular momentum due to tides. Naturally, we can use the imaginary component of the tidal Love number to express the power $\mathcal{P}$ and torque $\mathcal{T}$ associated with the tidal force of star~2 on star~1 \citep{Ogilvie:2014dwa} as
\begin{align}
    \mathcal{P} &= \frac{(2\ell+1) \omega_t R_*|\mathcal{A}_{\ell mn}|^2}{8 \pi G} \text{Im}[k_{\ell,n}^m], \\
    |\mathcal{T}| &= \frac{(2\ell+1) m R_*|\mathcal{A}_{\ell mn}|^2}{8 \pi G} \text{Im}[k_{\ell,n}^m].
\end{align}
The torque acts to change the rotation velocity of the body, and the resulting binary separation and eccentricity can be computed from the conservation of angular momentum.
Following \citet{Ogilvie:2014dwa}, we can express an equivalent parameterization of the tidal response as
\begin{equation}
    \text{Im}[k_{\ell,n}^m (\omega_t)] = \text{sgn}(\omega_t) \frac{3}{2 Q'}.
\end{equation}
Here, $Q'$ is the modified or effective tidal quality factor. 

In the rest of this paper, we will work with dimensionless, frequency-dependent, imaginary tidal Love numbers $\text{Im}[k_{\ell,n}^m]$. In particular, we will limit our expressions to only the $\ell=2$ terms, which are dominant for tidal dissipation. 

The orbital evolution equations in the presence of tidal dissipation for the semi-major axis $a$, eccentricity $e$ and rotational angular frequency $\Omega_{\rm spin}$, to leading order in eccentricity, can now be written following \citet{zahn_tidal_1977} as
\begin{multline}
    \frac{da}{dt} = -\frac{3}{\omega_{\rm orb}} \biggl(\frac{M_* + M_2}{M_*}\biggr) \frac{G M_2}{R_*^2} \biggl(\frac{R_*}{a}\biggr)^7 \\
    \cdot \biggl[\text{Im}[k_{2,2}^2] 
    + e^2 \biggl( \frac{3}{4} \text{Im}[k_{2,1}^0] + \frac{1}{8} \text{Im}[k_{2,1}^2] \\ 
    - 5 \text{Im}[k_{2,2}^2] + \frac{147}{8} \text{Im}[k_{2,3}^2] \biggr) + \mathcal{O}(e^4) \biggr],
    \label{eq:tidal_dadt}
\end{multline}

\begin{multline}
    \frac{de}{dt} = - \frac{3}{4} \frac{e}{\omega_{\rm orb}} \biggl(\frac{M_* + M_2}{M_*}\biggr) \frac{G M_2}{R_*^3} \biggl( \frac{R_*}{a}\biggr)^8 \\
    \cdot \biggl[\frac{3}{2} \text{Im}[k_{2,1}^0] - \frac{1}{4} \text{Im}[k_{2,1}^2] - \text{Im}[k_{2,2}^2] + \frac{49}{4} \text{Im}[k_{2,3}^2]  \\
     + \mathcal{O}(e^2) \biggr],
     \label{eq:tidal_dedt}
\end{multline}

\begin{multline}
    I\frac{d\Omega_{\rm spin}}{dt} = \frac{3}{2} \frac{G M_2^2}{R_*} \biggl(\frac{R_*}{a}\biggr)^6 \\
    \cdot \biggl[\text{Im}[k_{2,2}^2]
     + e^2 \biggl(\frac{1}{4} \text{Im}[k_{2,1}^2]  - 5 \text{Im}[k_{2,2}^2] + \frac{49}{4} \text{Im}[k_{2,3}^2]\biggr) \\
     + \mathcal{O}(e^4) \biggr],
     \label{eq:tidal_dOmegadt}
\end{multline}
where $I$ is the moment of inertia of star~1. These equations can be shown to be identical to Eqs.~(6) - (8) from \citet{Ogilvie:2014dwa} in the circular limit. 
The above equations are meant to be self-consistent, such that they conserve angular momentum as long as they are applied simultaneously. However, this may not always be the case within a discrete simulation framework such as COMPAS, particularly for high eccentricities, where the truncation of higher order terms may become relevant. The original secular evolution equations are derived by first computing the effect of tidal torque on $a$ and $e$, and then calculating the spin evolution by conservation of angular momentum. In our implementation, we respect the same order of operations. In particular, we first evolve $a$, $e$, and $\Omega$ for each star using Eqs.~\eqref{eq:tidal_dadt}, \eqref{eq:tidal_dedt}, and \eqref{eq:tidal_dOmegadt}, respectively, and then apply an overall constant scaling factor to $\Omega_1$ and $\Omega_2$ to ensure that total angular momentum is conserved.

Evidently, tidal dissipation has a strong dependence on the inverse binary separation. A star in an eccentric binary may experience stronger tidal torques at periastron passage than at apastron, and may be spun up past $\Omega_{\rm spin} = \omega_{\rm orb}$ due to eccentricity. We allow stellar rotation rates in eccentric binaries to attain the `pseudo-synchronization' limit \citep{hut_tidal_1981}, whereby the steady state of stellar rotation is higher relative to the average orbital velocity due to eccentricity. The limiting rotational frequency in the weak friction limit is given by
\begin{equation}
    \Omega_{\rm ps} = \frac{1 + \frac{15}{2}e^2 + \frac{45}{8}e^4 + \frac{5}{16}e^6}{(1+3e^2 + \frac{3}{8}e^4)(1-e^2)^{3/2}} \omega_{\rm orb}.
    \label{eq:pseudosync_hut1981}
\end{equation}
In practice, this limit may be reached before the  $\frac{d \Omega_{\rm spin} }{dt}~=~0$ equilibrium point from Eq.~\eqref{eq:tidal_dOmegadt}. We nevertheless enforce the spin limit from Eq.~\eqref{eq:pseudosync_hut1981} to remain consistent with other tidal models in the literature. Observe how $\Omega_{\rm ps} \rightarrow \omega_{\rm orb}$ as $e \rightarrow 0$, while $\Omega_{\rm ps} > \omega_{\rm orb}$ if $e>0$. The pseudo-synchronization condition reduces to the usual synchronization state once the binary becomes circular.

\begin{figure*}
    \centering
    \includegraphics[width=0.8\linewidth, trim={0 0 0 0}]{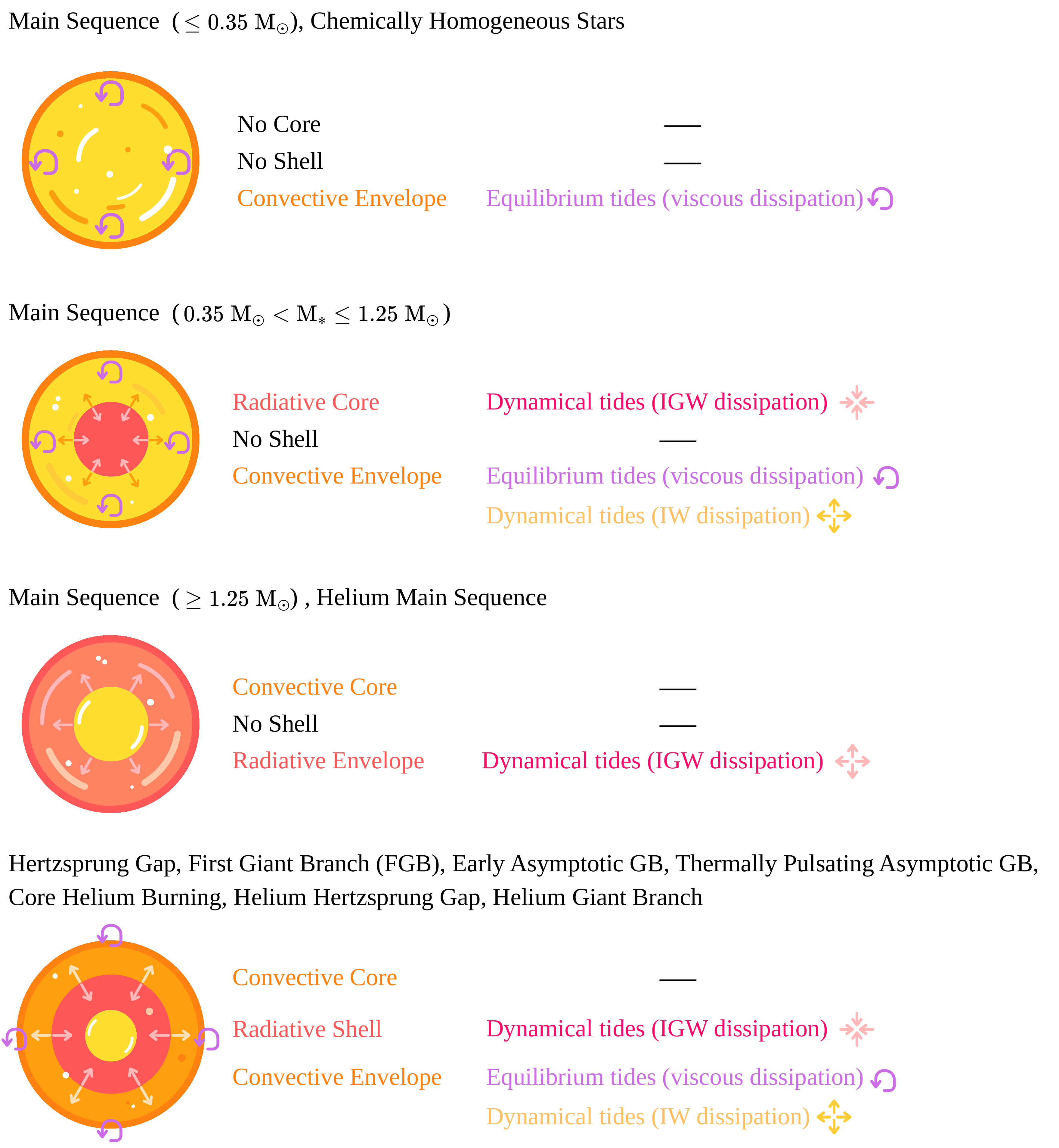}
    \caption{Overview of tidal effects considered in this work, for stars with various internal structures. Convective envelopes may experience a combination of equilibrium tides from viscous dissipation,  dynamical tides from internal gravity wave (IGW) dissipation, and dynamical tides from inertial wave (IW) dissipation. Boundaries between radiative and convective zones may excite dynamical tides from IGWs  and IWs, which are dissipated in the radiative and convective zones of the star, respectively. In our models, convective cores do not experience tidal dissipation.}
    \label{fig:tides_plan}
\end{figure*}

The tidal response $\text{Im}[k_{\ell,n}^m]$ is a property of a given star, and depends on the structure and tidal dissipation mechanisms within the star. The rest of this section is dedicated to summarizing the Love numbers for various dissipation mechanisms and the types of star they are relevant to. A visual summary of the various stellar types and tidal dissipation mechanisms considered in this work is provided in Fig.~\ref{fig:tides_plan}.

\subsection{Equilibrium tides (viscous dissipation)}
\label{sec:equilibrium_tides}
When a star is deformed due to the tidal potential, it experiences an equilibrium tide. This tide is dissipated via turbulent friction in the convective regions of a star. In our model, we only apply viscous dissipation to stars with convective envelopes. In other words, we ignore the effects of viscous dissipation inside a convective core, assuming the core would be too small to meaningfully contribute to tidal effects. This may not be a good simplifying assumption for stars with extended convective cores, but we leave addressing this nuance to future work.
Under our assumptions, equilibrium tides are only relevant for main sequence stars below $1.25 M_\odot$ and evolved stars with convective envelopes. Following Eq.~(23) of \citet{Barker:2020MNRAS.498.2270B}, we can express the tidal response in the presence of equilibrium tides as
\begin{equation}
    \text{Im}[k_{\ell,n}^m] = \text{sgn}(\omega_t) \times \frac{3}{2} \frac{16 \pi G}{3 (2\ell+1) R_*^{2\ell+1} |\mathcal{A}_{\ell mn}|^2} \frac{D_\nu}{|\omega_t|}.
\end{equation}
Here, $\mathcal{A}_{\ell mn}$ is the same tidal potential amplitude as in Eq.~\eqref{eq:tidal_potential_amplitude}.
The viscous dissipation of the tide is given by
\begin{equation} \label{eq:dnu_integral}
    D_\nu = \frac{1}{2} \omega_t^2 \int r^2 \rho_{\rm conv}(r) \nu(r) D_{\ell mn}(r) \text{d}r,
\end{equation}
where the integral is over the convective zone of the star, and $\rho_{\rm conv}(r)$ is the density of the convective region.
 $D_{\ell mn}$ is a dimensionless term that depends on the indices $(\ell,m,n)$ through the tidal displacement field $\xi$, and is given by
\begin{dmath}
    D_{\ell mn}(r) = 3 \left| \frac{d \xi_{r}}{dr}  - \frac{1}{3 r^2} \frac{d}{dr} (r^2 \xi_{r}) + \frac{\ell (\ell+1)}{3} \frac{\xi_{h}}{r}\right|^2 + \ell (\ell+1) \left| \frac{\xi_{r}}{r} + r \frac{d}{dr} \left( \frac{\xi_{h}}{r} \right)\right|^2  + (\ell-1) \ell (\ell+1) (\ell+2) \left| \frac{\xi_{h}}{r} \right|^2,
\end{dmath}
where
\begin{align}
    \xi_{r} &= -\frac{\Phi_{{\rm perturb},\ell mn} + \Psi_{{\rm tidal}, \ell mn}}{g}, \label{eq:xi_rl}\\
    \xi_{h} &= \frac{1}{\ell (\ell+1)} \left(2 \xi_{r} + r \frac{d \xi_{r}}{dr} \right),
\end{align}
represent the radial and horizontal components of the displacement field, respectively, which have units of length. The quantity $g(r)$ is the gravitational acceleration inside the star. 

We can evaluate the integral from Eq.~\eqref{eq:dnu_integral} assuming that the convective region goes from radial coordinate $R_{\rm conv}$ to $R_{*}$, and making the simplifying assumption of constant density $\rho_{\rm conv}$ in the convective zone. We set $\Phi~+~\Psi~\approx~\Psi$ under the Cowling approximation, and assume $g \approx G M_* / r^2$ near the outer boundary for a thin convective shell. Note that this assumption is not valid in the deep interior of the star, or for deep convective envelopes. 

\revision{In the simplified picture where the density of the convective region is constant, we treat the effective turbulent viscosity $\nu$ as independent of $r$. The viscosity may still be frequency dependent, and we rely on the fits from \citet{Duguid:10.1093/mnras/staa2216} to capture the suppression at high frequencies as}
\begin{equation}
\label{eq:viscosity_Duguid}
    \nu(\omega_t) = u_{\rm c} l_{\rm c} \times \begin{cases} 
      5 & \frac{|\omega_t|}{\omega_{\rm c}} < 10^{-2}, \\
      \frac{1}{2} \left( \frac{|\omega_t|}{\omega_{\rm c}} \right)^{-1/2} & \frac{|\omega_t|}{\omega_{\rm c}} \in [10^{-2}, 5] \\
      \frac{25}{\sqrt{20}} \left( \frac{|\omega_t|}{\omega_{\rm c}} \right)^{-2} & \frac{|\omega_t|}{\omega_{\rm c}} > 5,
   \end{cases},
\end{equation} 
where $u_{\rm c}$ is the convective velocity, $l_{\rm c}$ is the mixing-length, and $\omega_{\rm c} = u_{\rm c} / l_{\rm c}$ is the convective turnover frequency.

Putting all the pieces together, the imaginary component of the tidal Love number becomes
\begin{equation}
    \text{Im}[k_{2,n}^m] = \frac{224 \pi }{15} \frac{(R_*^9 - R_{\rm conv}^9) }{G M_*^2 R_*^{5}}  \omega_t  \rho_{\rm conv} \nu(\omega_t).
    \label{eq:imknm_equilibrium}
\end{equation}
 Note that the $(m,n)$ indices appear only in the tidal frequency $\omega_t$, and do not otherwise affect the pre-factor of the imaginary Love number. This is thanks to the dependence of $\text{Im}[k_{2,n}^m]$ on ${\Psi_{\ell mn}}/{|\mathcal{A}_{\ell mn}|}$, whereby all $\ell~=~2$ terms have the same behavior.

Evidently, the tidal dissipation strength requires some knowledge of the radial coordinate of the convective envelope $R_{\rm conv}$ and its density $\rho_{\rm conv}$. COMPAS uses prescriptions from \citet{Hurley:2000pk} and \citet{Picker:2024vqb} to estimate the mass of the convective envelope of a star across its lifetime, and the \citet{Hurley:2002rf} model for the envelope height. Given these parameters, we assume a constant convective envelope density
\begin{equation}
    \rho_{\rm conv} \approx \frac{M_{\rm conv}}{\frac{4}{3} \pi (R_*^3 - R_{\rm conv}^3)}.
\end{equation}

To estimate the viscosity $\nu(\omega_t)$, we first use Eq.~(31) from \citet{Hurley:2002rf} to calculate the convective timescale $t_c$, such that
\begin{equation}
    t_c = \left( \frac{M_{\rm conv} R_{\rm conv, ext} (R_* - \frac{1}{2} R_{\rm conv, ext})}{3 L_*} \right)^{1/3},
\end{equation}
where $L_*$ is the luminosity of the star, and $R_{\rm conv,ext}=R_*-R_{\rm conv}$ is the radial extent of the convective envelope, such that the convective turnover timescale is estimated at the middle of the convective envelope. \revision{Note that by not assuming a specific choice of units in the above equation, we avoid the need for the extra constant ahead of Eq.~(31) in \citet{Hurley:2002rf}.} We approximate the convective mixing length $l_c$ by half the radial extent of the convective region, i.e.,
\begin{equation}
    l_c \approx R_{\rm conv,ext}/2.
    \label{eq:lc_convective}
\end{equation}
The above approximation may be good for stars with shallow convective envelopes, where the scale height and the depth of the convective envelope are similar. For a deep convective envelope (like those developed in evolved stars), the local dissipation within any convective shell should be weighted by both the local convective viscosity and the local tidal distortion. The former goes to zero at the surface, and the latter goes to zero at the center, so the dissipation would occur at some intermediate layer. In practice, this means our estimate for the mixing length $l_c$ is likely overestimated for massive, evolved stars with extended convective envelopes. Since the tidal dissipation in Eq.~\eqref{eq:imknm_equilibrium} scales as $\nu~\propto~l_c^2 / t_c~\propto~l_c^{5/3}$, overestimating the length scale by a factor of 10 could result in an overestimation of viscous dissipation strength by a factor of 50.

From here, the convective velocity can be estimated as 
\begin{equation}
    u_c = l_c/t_c,
\end{equation}
and $\nu$ can then be obtained as per Eq.~\eqref{eq:viscosity_Duguid}.
We compare the frequency dependent tidal Love number for a solar-type star to other results from the literature in Appendix~\ref{app:equilibrium_comparison}. To summarize those results, the strength of equilibrium tides with our simplified prescription is within an order of magnitude of detailed simulations, whereas it may be up to 2 orders of magnitude higher than empirical results. Improved estimation of frequency-dependent viscosity of stars at high tidal frequencies \citep{Duguid:10.1093/mnras/staa2216,vidal2020efficiency, Bavera:2020inc}, as well as more precise estimates of the convective scale heights for extended convective envelopes, may bring our simulations closer to observational estimates.

\subsection{Dynamical tides (internal gravity waves)}
\label{sec:dynamical_tides_igw}
Dynamical tides can arise from the dissipation of internal gravity waves (IGWs), given the presence of at least one convective-radiative boundary in a star. \revision{As we denote using the pink text in Fig.~\ref{fig:tides_plan}, dynamical tides from IGW dissipation can possibly affect almost all stars with $M_*> 0.35 M_\odot$ in our models.}

\subsubsection{Convective core -- radiative shell boundary} \label{subsec:gw_conv_core}
\revision{MS stars with $M_* \geq 1.25 M_\odot$ can develop an inner convective core interfacing with an outer radiative envelope. For the IGWs that are excited from this interface and dissipated in the radiative shell (shown as the outward pointing pink arrows in Fig.~\ref{fig:tides_plan}),} we rely on the prescriptions from \citet{zahn_tidal_1977} and  \citet{hut_tidal_1981}, using the modern re-parameterization suggested by \citet{Kushnir:2017}:
\begin{equation}
    \text{Im}[k_{2,n}^m] = E_{2, {\rm core}} \; s_{n,m} ^{8/3} \times \text{sgn}(n\omega_{\rm orb} - m\Omega_{\rm spin}),
\end{equation}
where 
\begin{equation}
    E_{2, {\rm core}} = \frac{2}{3} \left( \frac{R_c}{R_*} \right)^{9} \left( \frac{M_*}{M_c} \right)^{4/3} \beta_{2, {\rm core}} \frac{\rho_c}{\bar{\rho}_c} \left( 1 - \frac{\rho_c}{\bar{\rho}_c} \right)^2 \label{eq:E2_kushnir}
\end{equation}
and
\begin{equation}
    s_{n,m} = |n\omega_{\rm orb} - m\Omega_{\rm spin}| \left(\frac{R_*^3}{G M_*}\right)^{1/2}.    
\end{equation} 
Here, the subscript `$c$' denotes a quantity relating to the (boundary of the) convective core of the primary star, given that it interfaces with a radiative envelope. For instance, $\rho_c$ is the local density at radius $R_c$, and $\bar{\rho}_c$ is the mean density of the star inside radius $R_c$. Although $E_{2, {\rm core}}$ is not explicitly defined as above in \citet{Kushnir:2017}, it follows by comparing expressions (1) and (8) of their paper. Note that $E_{2, {\rm core}}$ and $s_{n,m}$ are dimensionless by construction. To a reasonable approximation, we can simplify the imaginary tidal Love number to write
\begin{dmath}
    \text{Im}[k_{2,n}^m] \approx 0.1 \frac{2}{3} \left( \frac{R_c}{R_*} \right)^{5} \\ 
    \left((n\omega_{\rm orb} - m\Omega_{\rm spin}) \sqrt{\frac{R_c^3}{G M_c}}\right)^{8/3},
\end{dmath}
where we have set the dimensionless coefficient $\beta_{2, {\rm core}}$ to 1 based on \citet{Kushnir:2017}. We also approximate $\frac{\rho_c}{\bar{\rho}_c} \left( 1 - \frac{\rho_c}{\bar{\rho}_c} \right)^2~\approx~0.1$ as an order of magnitude estimate, which follows from the assumption that $\frac{\rho_c}{\bar{\rho}_c} \approx \frac{1}{2}$. 

For the convective core radius evolution of a star on the main sequence (MS) in COMPAS, we use the following equation derived by fitting a model to 1D stellar simulations performed by \cite{Shikauchi:2024yqd}. First, the core radius at zero-age main sequence (ZAMS) is given by
\begin{equation}
    \frac{R_{c, \rm ZAMS}}{R_\odot} = \frac{M_{*, \rm ZAMS}}{M_\odot} \times (0.06 + 0.05 \times e^{-M_{*, \rm ZAMS} / 61.57 M_\odot}).
    \label{eq:rc_fit_shikauchi}
\end{equation}
We then linearly interpolate the core radius to its final value at terminal age main sequence (TAMS) as
\begin{equation}
    R_{c}(\tau) = R_{c,\rm ZAMS} - \tau (R_{c,\rm ZAMS}-R_{c,\rm TAMS}),
    \label{eq:rc_tams_evolution}
\end{equation}
where $\tau$ is the dimensionless fractional age along the MS, ranging from 0 at ZAMS to 1 at TAMS. 

\subsubsection{Radiative zone -- convective envelope boundary}
\revision{Next, we consider the IGW dissipation contribution from the boundary of a radiative inner zone and an outer convective envelope, which affects MS stars with $0.35 M_\odot < M_* \leq 1.25 M_\odot$ and evolved stars with a prominent convective envelope surrounding a radiative inner shell (see inward pointing pink arrows in Fig.~\ref{fig:tides_plan}).} We parameterize the imaginary tidal Love number from this dissipation as
\begin{equation}
    \text{Im}[k_{2,n}^m] = E_{2, {\rm env}} \; s_{n,m} ^{8/3} \times \text{sgn}(n\omega_{\rm orb} - m\Omega_{\rm spin}).\label{eq:igw_rad_core_conv_envelope}
\end{equation}
To estimate the $E_{2, {\rm env}}$ factor, we begin with the power dissipated by the dynamical tide from a radiative--convective envelope boundary located at radial coordinate $R_{\rm conv}$ \citep{Goodman:1998yg, Terquem:1998ya}
\begin{dmath}
    \dot{E}_0 = \frac{3^{2/3}}{8 \pi} \Gamma^2\left(\frac{1}{3}\right) [\ell (\ell + 1)]^{-4/3} \omega_t^{11/3} \\ \times  \left( \rho r^5 \left| \frac{\mathrm{d} N^2}{\mathrm{d}\ln r} \right|^{-1/3} \left|\frac{\partial \xi_r^{\rm dyn}}{\partial r} \right|^2 \right)_{r= R_{\rm conv}}.
    \label{eq:energy_flux_dynamical_env}
\end{dmath}
The torque associated with this dissipation is given by $T = \frac{m}{\omega_t} \dot{E}_0$, and the imaginary component of the $\ell=2$ tidal Love number is given by \citep{Peale:1999ay, Barker:2010ru}
\begin{equation}
    \text{Im}[k_{2,n}^m] = \frac{2}{3} \frac{\dot{E}_0}{\omega_t} \left( \frac{M_* + M_2}{M_2}\right)^2 \frac{\omega_{\rm dyn}^2}{M_* R_*^2} \frac{1}{\omega_{\rm orb}^4},
\end{equation}
where $\omega_{\rm dyn}^2 = \frac{G M_*}{R_*^3}$ is the squared dynamical frequency of the star. By plugging Eq.~\eqref{eq:energy_flux_dynamical_env} into the above equation, we can obtain the full form of the tidal Love number. However, several terms in Eq.~\eqref{eq:energy_flux_dynamical_env} are difficult to compute with approximate stellar models. As such, we rely on Eq.~(131) of \citet{Ahuir:2021} to provide a tractable form of the energy dissipation. With the appropriate substitutions, we obtain the following expression for $E_{2, {\rm env}}$:
\begin{dmath}
    E_{2, {\rm env}} = \frac{3^{11/3} \Gamma^2(1/3)}{40 \pi} [\ell (\ell+1)]^{-4/3} \frac{R_*^3}{G M_*} \\ \times \left| \frac{\text{d} N^2}{\text{d ln } r} \right|^{-1/3}_{R_{\rm conv}} \mathcal{E}(\alpha, \beta).
\end{dmath}
In the above equation, $N$ is the Brunt-V\"ais\"al\"a frequency, and
\begin{equation}
    \mathcal{E}(\alpha, \beta) = \frac{\alpha^{11} (\frac{M_{\rm conv}}{M_*}) (1-\gamma)^2}{\beta^2 (1-\alpha^3) (1-\alpha)^2} \left( \frac{2 \alpha}{3} -1 \right)^2
\end{equation}
is a function of $\alpha \equiv R_{\rm conv}/R$, $\beta \equiv M_{\rm rad}/M_*$, and $\gamma \equiv \rho_{\rm conv} / \rho_{\rm rad}$. Here, $M_{\rm rad}$ refers to the mass of the radiative zone contained within radius $R_{\rm conv}$, and $M_{\rm conv}$ is the mass in the outer convective envelope that stretches from $R_{\rm conv}$ to $R_*$.

Since we do not typically have access to the density profiles of stars in population synthesis, we must make several assumptions to estimate $\left| \frac{{\rm d} N^2}{{\rm d} \ln r} \right|$ at the boundary. Formally, $N^2~=~0$ in the convective envelope, so we work in the regime of a thin convective shell, such that $M_{\rm conv}  \ll M_{\rm rad}$, and $\rho_{\rm conv} \ll \rho_{\rm rad}$.
The Brunt-V\"ais\"al\"a frequency is given by 
\begin{equation}
    N^2 = - g_0 \left( \frac{1}{\rho_0} \frac{\partial \rho}{\partial r}  + \frac{g_0}{c_s^2}\right), 
    \label{eq:brunt_vaisala_def}
\end{equation}
where $g_0$ is the local gravitational acceleration, $\rho_0$ is the local density of the star, and $c_s$ is the sound speed. For the sake of simplicity, we will use the average values of these quantities at the radiative-convective boundary, such that 
\begin{align}
    g_0|_{R_{\rm conv}} &= \frac{G M_{\rm rad}}{R_{\rm conv}^2}, \\
    \rho_0|_{R_{\rm conv}} &= \frac{3 M_{\rm rad}}{4 \pi R_{\rm conv}^3}, \\
    c_s^2|_{R_{\rm conv}} &= \frac{G M_{\rm rad}}{R_{\rm conv}}.
\end{align}
The density near the boundary can then be expanded as
\begin{equation}
    \rho(r) \approx \rho(R_{\rm conv}) + \left[
 \frac{\text{d} \rho(r)}{\text{d} r}\right]_{R_{\rm conv}} (r - R_{\rm conv}).
\end{equation}
In the limit that $\rho_{\rm conv} \ll \rho_{\rm rad}$, or equivalently, $\rho(r) \ll \rho(R_{\rm conv})~=~\rho_0|_{R_{\rm conv}}$, we get
\begin{equation}
    \left[\frac{\text{d} \rho(r)}{\text{d} r}\right]_{R_{\rm conv}} \approx - \frac{3 M_{\rm rad}}{4 \pi R_{\rm conv}^3} \left(\frac{1}{r - R_{\rm conv}}\right).
\end{equation}
\\
Examining all these values, we find it sufficient to express the Brunt-V\"ais\"al\"a frequency in the form
\begin{equation}
    N^2 \approx \frac{G M_{\rm rad}}{R_{\rm conv}^3}.
\end{equation}
 As $N^2$ varies on the scale of the local scale height $H$, the derivative with respect to $\text{ln}\, r$ may be approximated by
\begin{equation}
    \left| \frac{\text{d} N^2}{\text{d ln } r} \right|_{R_{\rm conv}} \approx \frac{G M_{\rm rad}}{R_{\rm conv}^3} \frac{R_{\rm conv}}{H}.
\end{equation}

Near the surface of the star, we approximate the scale height by $H\approx R_*-R_{\rm conv}$, which is the radial extent of the convective envelope. Therefore,
\begin{align}
    \left| \frac{\text{d} N^2}{\text{d ln } r} \right|_{R_{\rm conv}} \approx \frac{G M_{\rm rad}}{R_{\rm conv}^2 (R_*-R_{\rm conv})}.
\end{align}
For high enough tidal amplitudes (see \cite{Barker:2010ru}), the excited gravity waves may overturn stratification and break near the center of the star, leading to efficient IGW dissipation. The corresponding critical orbital period for solar-type stars is approximately 3 days on MS, and it decreases as the star evolves \citep{Esseldeurs2024A&A...690A.266E}. This period is likely even lower for more massive stars (see Fig.~4 of \citet{Esseldeurs2024A&A...690A.266E}), and the critical period should always be lower than even the shortest period binaries we consider in this work. As such, we omit modeling the wave breaking condition for dynamical tides in our simulations, assuming instead that IGW dissipation is always efficient.

\subsection{Dynamical tides (inertial waves)}
\label{sec:dynamical_tides_iw}
Dynamical tides, when dissipated via inertial waves (IWs), act on the convective zones of a star (see yellow arrows in Fig.~\ref{fig:tides_plan}). This dissipation only contributes when $\omega_{\rm orb} \leq 2 \Omega_{\rm spin}$~\citep{Ogilvie:2012gm, Barker:2020MNRAS.498.2270B} for the $n=2$, $m=2$ component of the tidal Love number. We will limit the discussion of dynamical tides from IW dissipation to the $(2,2)$ component to remain consistent with our references, but our model can easily be extended to include additional modes. We also neglect IW dissipation in convective cores under the assumption that the core boundary satisfies $R_c/R_* \approx 0$. The only contribution from inertial waves in our model comes from systems with a radiative core (or inter-shell) interfacing with an outer convective envelope. In these stars, the $(\ell=2, n=2, m=2)$ component of tidal dissipation, averaged over orbital frequency, can be written using Eq.~(B3) of \citet{Ogilvie:2012gm} as

\begin{widetext}
\begin{dmath}
    \int_{-\infty}^{\infty} \text{Im}[k_{2,2}^2] \frac{{\rm d} \omega_{\rm orb}}{\omega_{\rm orb}} = \frac{100 \pi}{63} \epsilon^2 \left( \frac{\alpha^5}{1 - \alpha^5} \right) (1-\gamma)^2 (1-\alpha)^4 \\
    \times \left(1 + 2\alpha + 3\alpha^2 + \frac{3}{2}\alpha^3 \right)^2 \left[1 + \left( \frac{1-\gamma}{\gamma} \right) \alpha^3 \right] \\
    \times \left[ 1 + \frac{3}{2}\gamma + \frac{5}{2\gamma} \left( 1 + \frac{1}{2}\gamma - \frac{3}{2}\gamma^2 \right) \alpha^3 - \frac{9}{4}(1-\gamma)\alpha^5 \right]^{-2},
\end{dmath}
\end{widetext}
with $\epsilon = \Omega_{\rm spin} / \left( GM_* / R_*^3 \right)^{1/2}$, $\alpha = R_{\rm conv}/R_*$, and $\gamma=\rho_{\rm conv} / \rho_{\rm rad}$ as before. Note that in this calculation, the star is assumed to be a piecewise-homogeneous fluid, such that the density is
\begin{equation}
    \rho(r) = 
    \begin{cases}
        \rho_{\rm rad}, & 0 < r < \alpha R_* \\
        \rho_{\rm conv}, & \alpha R_* < r < R_*
    \end{cases}
\end{equation}

To test the validity of our assumptions, we compare the strength of dynamical tides from inertial and gravity wave dissipation for a low-mass system shown in Fig.~7 of \citet{Ahuir:2021}. The comparison is provided in Appendix~\ref{app:dynamical_comparison}. Our estimates agree at the order of magnitude level with detailed simulations.

\subsection{Practical Caveats}
 A major assumption in our models is that of uniform stellar rotation. In practice, differential rotation can complicate the predicted tidal dissipation of spinning stars~\citep[e.g.,][]{Kumar:1996dw}. We ignore this effect in our work, but recognize that a more careful tidal model including differential rotation may be required to robustly predict binary outcomes. Furthermore, the tidal dissipation Eqs.~\eqref{eq:tidal_dadt} - \eqref{eq:tidal_dOmegadt} are formally valid in the low eccentricity regime only, although we do not impose any specific eccentricity limits in our implementation. 
 In some cases (high eccentricity, high spin), it is technically possible for the $e^2$ terms to spin up the binary beyond pseudo-synchronization. Since we do not carefully model tidal dissipation for very eccentric systems, we defer to the pseudo-synchronization limit from Eq.~\eqref{eq:pseudosync_hut1981} as an upper limit on the spins of stars in our simulations. If the behavior of our tidal dissipation Eqs.~\eqref{eq:tidal_dadt} - \eqref{eq:tidal_dOmegadt} is to push the binary beyond this point, we ignore the $\mathcal{O}(e^2)$ terms as a numerical stop-gap to allow the binary to spin back down to pseudo-synchronization.

 In principle, every convective-radiative boundary in a star may contribute to IGW dissipation as long as the gravity waves can be fully damped inside the radiation zone. However the underlying stellar models in population synthesis are very simplified, and we do not have a reliable way of modeling the locations of all the convective-radiative boundaries that may exist in a star with three or more layers. As such, we choose to ignore the contribution of IGW dissipation in stars with more than two layers and instead assume that the IGWs are not able to propagate close to either the core or the envelope, where they may efficiently be dissipated. 
 
For the sake of our population synthesis implementation, we assume that the wave-breaking condition~\citep{Barker:2010ru} is always satisfied, and that gravity waves are efficiently dissipated in the radiation zone regardless of the companion mass or tidal frequency. On a practical note, IGW dissipation is typically sub-dominant to equilibrium or even IW tides \revision{for stars with convective envelopes, such that this assumption does not significantly impact simulation outcomes. For stars with radiative cores or extended radiative envelopes where IGW dissipation is most relevant, the companion mass required to meet the wave breaking depends on several factors, such as stellar age and tidal period \citep{Barker:2010ru, Barker:2020MNRAS.498.2270B}.} \revision{Furthermore, in regimes where the wave-breaking condition is not met, such as stars with convective cores where IGWs may otherwise form weakly dissipative standing modes, strong internal magnetic fields can still lead to efficient tidal dissipation via the conversion of IGWs into magnetic waves \citep{fuller2015asteroseismology, lecoanet2017conversion,duguid2024efficient}. These mechanisms provide further justification for our simplifying assumption of efficient IGW dissipation across the relevant stellar populations.}

\section{Tidal evolution with individual binary systems}
\label{sec:applications}

With our tidal dissipation model (henceforth labeled the `fiducial' model) implemented in COMPAS, we can now evolve binary systems rapidly and self-consistently in the presence of tidal interactions. We compare our model to the tidal dissipation behavior from \citet{zahn_tidal_1977}, or (nearly) equivalently, \citet{Hurley:2002rf}, which are both widely used as bases for tidal dissipation in population synthesis. We will henceforth call this reference model `Z77'. We provide details of its implementation in Appendix~\ref{app:z77_model}, \revision{with a short summary of the key differences between our fiducial model and the Z77 model in Appendix~\ref{app:fiducial_vs_z77}.} Just like the fiducial tidal model, we implement the Z77 model self-consistently in COMPAS and show the resulting binary evolution where relevant.

In this section, we simulate four sets of binary systems starting from ZAMS at solar metallicity with zero natal spin, chosen to encompass the various stellar types and structures that our model can be applied to. For each stellar type, we first simulate a grid of binaries in the plane of initial orbital period $P_{\rm orb, ZAMS}$ and initial eccentricity $e_{\rm ZAMS}$. We examine the evolution of binaries over this grid, identifying which configurations are likely to become circularized within their lifetimes and should be studied further. Although not quantitative, we also determine an approximate circularization period of our simulated binaries via qualitative inspection of the grid.

We then simulate a single binary with both, the fiducial and Z77 tidal models in COMPAS, and plot several relevant quantities over its evolution such as the stellar structure and the strength of equilibrium and dynamical tides over time. These simulations are performed with COMPAS v3.29.00.
As necessary, we use the following abbreviations for various stellar types in COMPAS: Main Sequence (MS), Hertzsprung Gap (HG), First Giant Branch (FGB), Core Helium Burning (CHeB), Early Asymptotic Giant Branch (EAGB), Thermally Pulsing Asymptotic Giant Branch (TPAGB), Helium Main Sequence (HeMS), and Helium Hertzsprung Gap (HeHG).

\subsection{Convective main sequence stars}
\label{sec:low_mass}
Low-mass stars ($ M_*\leq 0.35 M_\odot$) are expected to be entirely convective~\citep{Chabrier:1997vx}, and binaries composed of such stars have been observed to tidally circularize at orbital periods between 2 and 8 days \citep{Hansen:2010nq, Maxted2023Univ....9..498M}. These observed circularization periods provide an empirical benchmark for equilibrium tidal dissipation.

\begin{figure}
    \centering
    \includegraphics[width=\columnwidth]{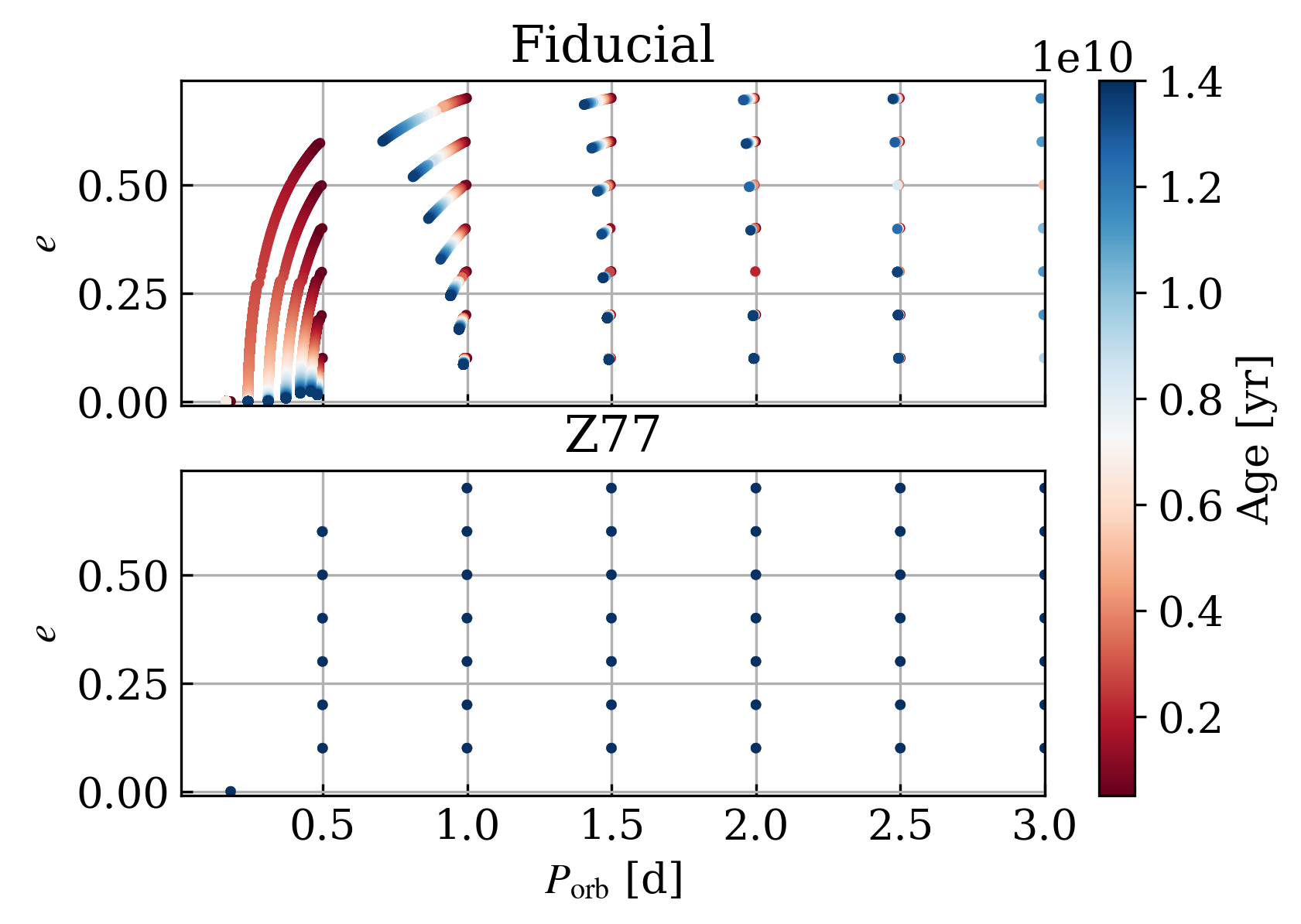}
    \caption{Evolutionary tracks of $e$ and $P_{\rm orb}$ under equilibrium tides for $0.3M_\odot+0.3M_\odot$ binaries, over a grid of initial orbital periods and initial eccentricities. Each binary is initialized on a grid in the $P_{\rm orb}-e$ plane, and we plot the orbital period and eccentricity of each binary for every time step in COMPAS. The colors depict the age of each evolutionary snapshot in years. The binaries are only evolved up to 14 Gyr, approximately the age of the Universe. The top panel shows the period-eccentricity evolution under our fiducial tidal model, and the bottom panel shows evolution under the Z77 model. None of the binaries evolve meaningfully in this plane over their simulated lifetimes with the Z77 model.}
    \label{fig:low_mass_grid_ecc_porb}
\end{figure}

To understand the implications of our tidal models for the circularization periods of low-mass binaries, we simulate a grid of $0.3M_\odot+0.3M_\odot$ binaries at solar metallicity with varying initial orbital periods and eccentricities. In Fig.~\ref{fig:low_mass_grid_ecc_porb} we show the resulting period-eccentricity evolution with our fiducial model as well as the Z77 model. With the fiducial model, binaries with initial orbital periods below $\sim 0.5$ days can tidally circularize via equilibrium tides, while wider binaries do not. It is worth noting that, even with the updated viscous dissipation models in our work, equilibrium tides are not as strong as empirical period-eccentricity distributions of convective binaries would suggest. Our findings are consistent with \citet{Barker:2020MNRAS.498.2270B, barker2022tidal}, who find that frequency-dependent equilibrium tides alone are not sufficient to explain the observed period-eccentricity distributions of low-mass and solar-type binaries. \revision{\citet{terquem2021new} and \citet{terquem2021circularization} argue that viscous dissipation may be underestimated for convective stars in the established theory, although  \citet{barker2021interaction} call this new result into question.} We compare our results to detailed simulations in Appendix~\ref{app:equilibrium_comparison}.

The eccentricity dissipation is even weaker with the Z77 model, where the simulated binaries appear not to lose any appreciable eccentricity over their lifetimes for the simulated grid. This also agrees with other studies based on the equilibrium tidal dissipation formalism from \citet{zahn_tidal_1977} \citep[e.g., ][]{claret1997circularization, Meibom_2005}. An important caveat here is that COMPAS initializes stars on MS, and thus, does not include tidal effects on the pre-main sequence (PMS) phase where stars are larger and may experience stronger tidal dissipation. Additionally, we initialize all our stars to be non-spinning at ZAMS. Conventional tidal predictions involve a dominant phase of PMS synchronization and circularization, followed by very little eccentricity evolution on the MS~\citep{zahn1989tidal}. The missing PMS phase from our simulations could bridge the gap between prediction and observation, enabling binaries to become circularized at larger orbital periods.

\begin{figure*}
    \centering
    
    \begin{subfigure}[t]{0.47\textwidth}
        \centering
        \includegraphics[trim={10 0 -30 0}, width=\textwidth]{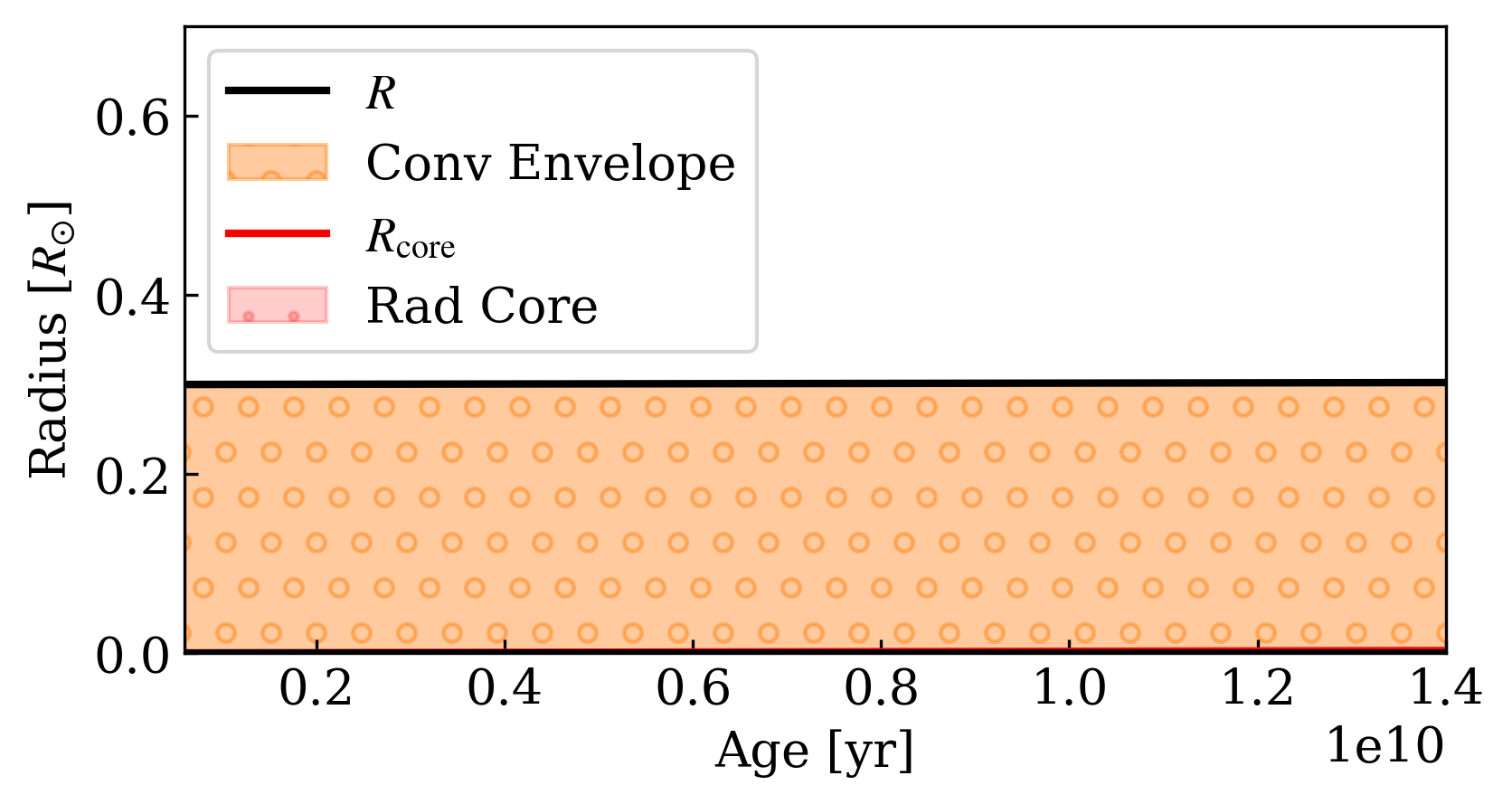}
        \caption{Kippenhahn diagram of the primary star.\\$\;$}
        \label{fig:low_mass_kippenhahn}
    \end{subfigure}
    \hfill
    \begin{subfigure}[t]{0.47\textwidth}
        \centering
        \includegraphics[trim={0 0 0 0}, width=\textwidth, height=0.19\textheight]{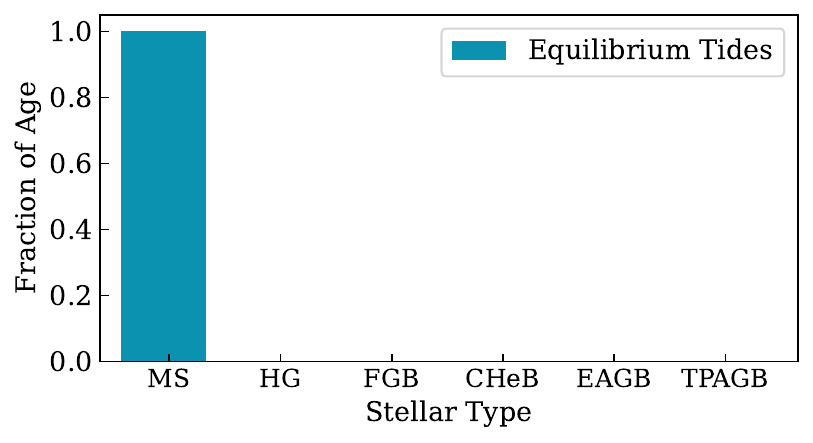}
        \caption{Fraction of the star's age that each tidal mechanism is applicable, per stellar type encountered over the evolution.}
        \label{fig:low_mass_tides}
    \end{subfigure}
        
    \begin{subfigure}[t]{0.47\textwidth}
        \centering
        \includegraphics[trim={27 -10 20 0}, width=\textwidth]{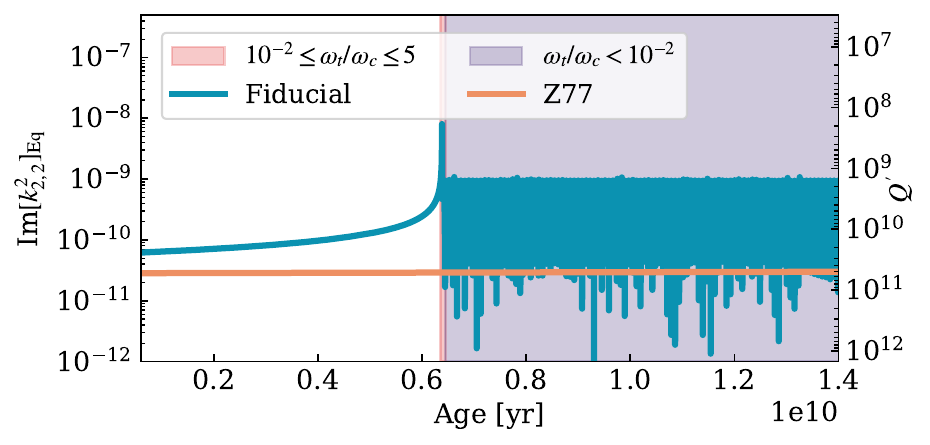}
        \caption{Absolute strength of the $\ell=2$, $m=2$, $n=2$ equilibrium tidal Love number (left y-axis) and the tidal quality factor (right y-axis) of the primary star.}
        \label{fig:low_mass_imk22}
    \end{subfigure}
    \hfill
    \begin{subfigure}[t]{0.47\textwidth}
        \centering
        \includegraphics[trim={14 0 0 0}, width=\textwidth]{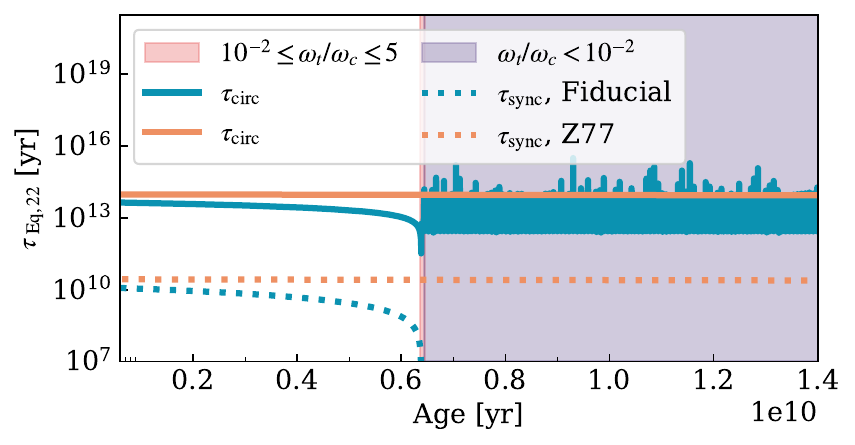}
        \caption{Equilibrium tidal timescales for circularization and synchronization for the primary star.}
        \label{fig:low_mass_tau_sync_circ}
    \end{subfigure}

    \begin{subfigure}[t]{0.47\textwidth}
        \centering
        \includegraphics[trim={10 0 -30 0}, width=\textwidth]{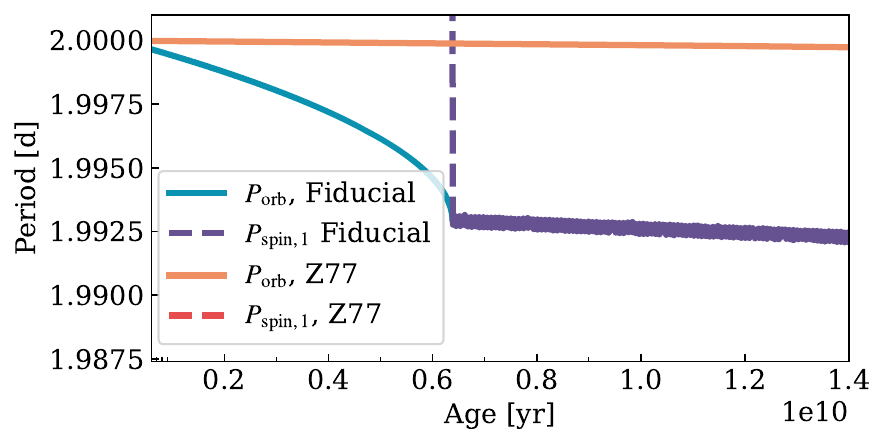}
        \caption{Orbital and primary rotation periods under the fiducial and Z77 models.}
        \label{fig:low_mass_period}
    \end{subfigure}
    \hfill
    \begin{subfigure}[t]{0.47\textwidth}
        \centering
        \includegraphics[trim={14 0 0 0}, width=\textwidth]{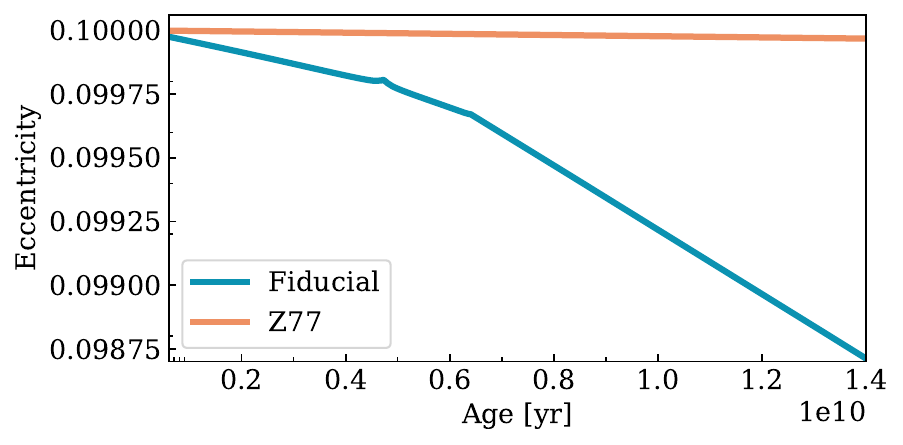}
        \caption{Eccentricity evolution of the binary under the fiducial and Z77 models.}
        \label{fig:low_mass_ecc}
    \end{subfigure}
    \caption{Stellar and tidal evolution for a 0.3 $M_\odot$ + 0.3 $M_\odot$ binary with $P_{\rm orb, ZAMS}~=~2$ days and $e_{\rm ZAMS}=0.1$. The binary is simulated for 14 Gyr, the approximate age of the Universe.}
    
    \label{fig:low_mass}
\end{figure*}

Next, we simulate a binary in COMPAS which highlights the relevant timescales of equilibrium tidal dissipation under our model. We choose the component masses to be $M_{\rm 1, ZAMS} = M_{\rm 2, ZAMS} = 0.3 M_\odot$, while the initial orbital period and eccentricity are $P_{\rm orb, ZAMS}~=~2$~days and $e_{\rm ZAMS}=0.1$. We use the default COMPAS behavior and initiate the stars with zero rotational velocities. Higher initial spins will modify $\omega_t$ and impact viscous dissipation.

We show the time evolution of various stellar and tidal quantities for the primary star in Fig.~\ref{fig:low_mass}. Since the binary components are equal in mass, the evolution for the companion star will be identical. The binary is simulated for $13.8$ Gyr, the approximate age of the Universe. 

From the Kippenhahn diagram in Fig.~\ref{fig:low_mass_kippenhahn}, we can see that the structure of each star in COMPAS is dominated by a convective envelope, as expected.  We show the fraction of the primary star's simulated age spent as a given stellar type and under a given tidal dissipation mechanism in  Fig.~\ref{fig:low_mass_tides}. The low-mass star remains on the main sequence throughout the entirety of the simulation, and only experiences equilibrium tides due to viscous dissipation.

We show the $(\ell=2,n=2,m=2)$ component of the imaginary tidal Love number from viscous dissipation in Fig.~\ref{fig:low_mass_imk22}. As shown in Fig.~\ref{fig:low_mass_tau_sync_circ}, the synchronization timescale is $\mathcal{O}(10^{10})$ yr at the beginning of our fiducial simulation, but gradually decreases as the primary spins up. On the other hand, the Z77 binary experiences a constant $\mathcal{O}(10^{10})$ yr synchronization timescale over the entire simulation. The circularization timescale is roughly $\mathcal{O}(10^{14})$ yr in both models, which exceeds the length of the simulation. 

Equilibrium tides in our fiducial model are roughly 2 times stronger than the Z77 model at early times, and several orders of magnitude stronger once the primary is rotating sufficiently rapidly. The Z77 model relies on a constant term $f_{\rm conv} \sim \mathcal{O}(1)$ to set the strength of equilibrium tides, whereas the newer prescriptions depend on the stellar structure as well as the tidal displacement field. Another key difference in our model is the explicit frequency dependence of the viscous dissipation efficiency. As tides synchronize the binary, there is a consistent decrease in the $(2,2)$ component of the tidal frequency, $\omega_{t, 22} = 2(\omega_{\rm orb} - \Omega_{\rm spin})$. The imaginary Love number scales as $|\omega_{t, 22}|^{-1}$ (as can be seen from Eq.~\eqref{eq:viscosity_Duguid} and Eq.~\eqref{eq:imknm_equilibrium}), leading to increasingly strong tides with rapidly approaching synchronization (decreasing $\omega_{t, 22}$). The cycle continues until around $0.6~\times~10^{10}$~yr, when the binary nearly synchronizes and the tidal frequency drops below $|\omega_{t, 22}|/\omega_c = 5$. The imaginary Love number is now $\propto \omega_{t,22}$, and decreases sharply over the short-lived $10^{-2} \leq|\omega_t|/\omega_c \leq 5$ phase, finally settling into a stable (but noisy) phase once $|\omega_{t, 22}|/\omega_c \leq 10^{-2}$. Beyond this point, the effective viscosity no longer depends on $\omega_{t, 22}$, and the Love number scales as $\omega_{t, 22}$. Due to finite time-stepping near synchronization, we see numerical artifacts in Fig.~\ref{fig:low_mass_imk22} and Fig.~\ref{fig:low_mass_tau_sync_circ} after $0.6~\times~10^{10}$ yr.

The synchronization of the primary star is visible in Fig.~\ref{fig:low_mass_period} as well, where the primary stellar rotation period becomes the same as the orbital period at $0.6~\times~10^{10}$ yr and remains synchronized until the end of the simulation. This behavior does not occur for the Z77 simulation, where equilibrium tides remain too weak to spin up the stars sufficiently.

Equilibrium tides under our model also reduce the binary eccentricity from 0.1 to 0.0987, as can be seen in Fig.~\ref{fig:low_mass_ecc}. Z77 tides are weaker in comparison, and are unable to dissipate any meaningful eccentricity.

\subsection{Radiative core main sequence stars}
\label{sec:int_mass}
\begin{figure}
    \centering
    \includegraphics[width=\columnwidth]{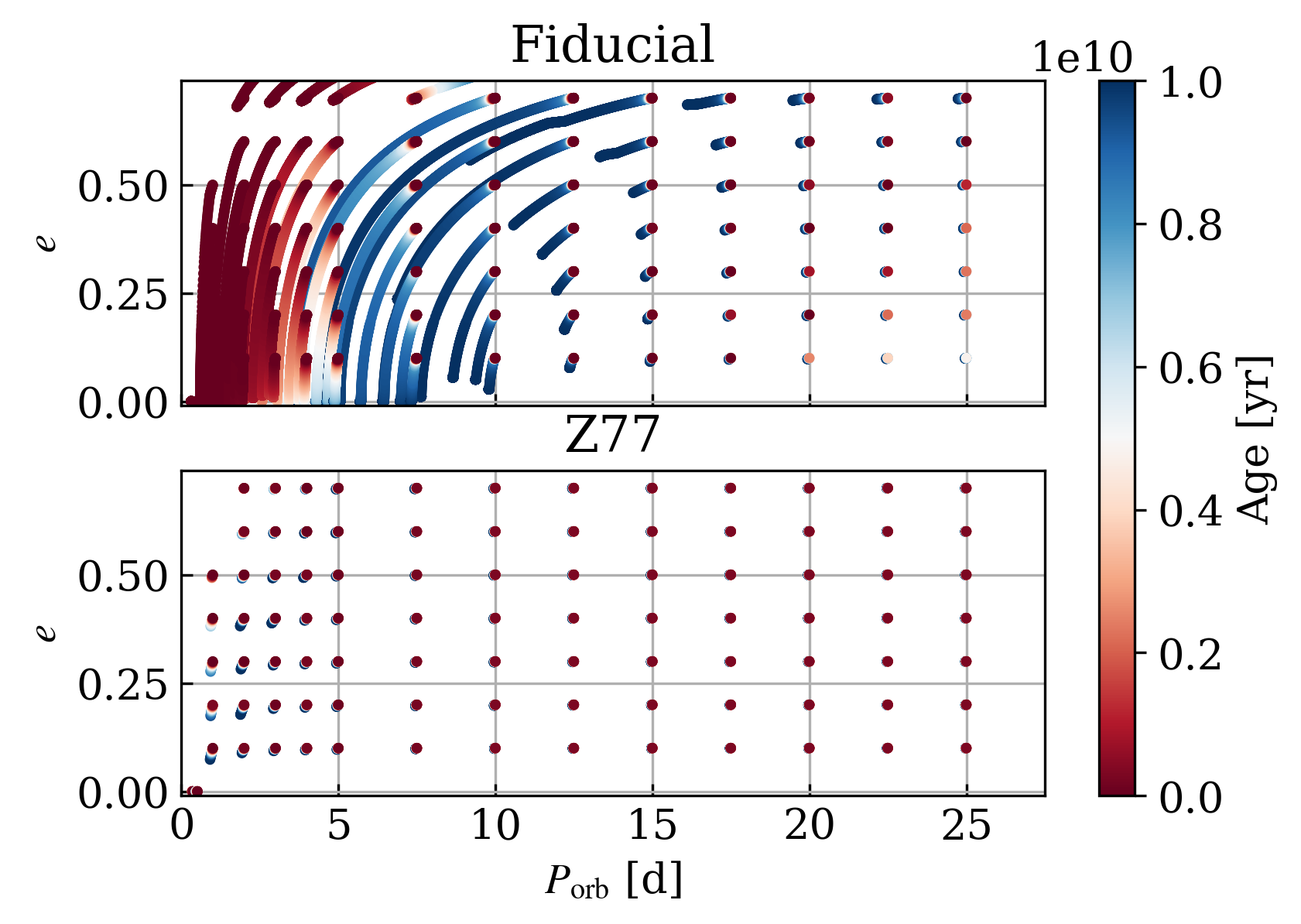}
    \caption{Time evolution of $e$ and $P_{\rm orb}$ for $1M_\odot+1M_\odot$ binaries over a grid of initial orbital periods and initial eccentricities, with our fiducial tides model (top panel) and the Z77 tides model (bottom panel). Each binary is initiated at a grid point, and the colors depict the age of each simulation snapshot in yr. The binaries are only evolved up to their MS lifetimes (approximately 10 Gyr) in this section.}
    \label{fig:int_mass_grid_ecc_porb}
\end{figure}

Solar-type stars have radiative cores and convective envelopes, and are thus expected to experience equilibrium as well as dynamical tides. 
As before, we investigate the general period-eccentricity evolution of solar-type binaries by simulating a grid in COMPAS with our fiducial model as well as the Z77 model for comparison. Following the default COMPAS behavior, all stars are initialized to be non-rotating at ZAMS. To limit our discussion to the MS, we only evolve the binaries for their MS lifetime of 10 Gyr. We show the evolutionary tracks in Fig.~\ref{fig:int_mass_grid_ecc_porb} for both sets of tidal models considered in this work. 

Binaries with initial orbital periods of $\leq 10$~days manage to circularize within their MS lifetimes under the fiducial model, as long as they start off with eccentricities below 0.6. At high eccentricities, the assumptions underlying our secular tidal evolution equations become less reliable, and the resulting behavior should be interpreted qualitatively. When a binary has non-zero eccentricity, its component spins are allowed to exceed synchronization as per Eq.~\eqref{eq:pseudosync_hut1981}. For a high enough eccentricity and appropriately high stellar rotational frequencies, Eq.~\eqref{eq:tidal_dadt} and Eq.~\eqref{eq:tidal_dedt} show that tidal dissipation may contribute positively to the eccentricity and orbital separation. This is the case for high eccentricity binaries with our fiducial tidal model (top panel of Fig.~\ref{fig:int_mass_grid_ecc_porb}), which become wider and more eccentric as the stars spin up.

The Z77 binaries, on the other hand, display no appreciable eccentricity dissipation for the simulated grid. Both equilibrium and dynamical tidal dissipation for these binaries are several orders of magnitude weaker than our fiducial tides. We have seen a comparison of equilibrium tides in the previous section. Dynamical tides from IGW dissipation in the Z77 model are expected to be weak for low-mass stars with relatively small radiative zones~\citep{Zahn:1975A&A....41..329Z, zahn_tidal_1977}, although their formalism only applied to stars with radiative envelopes and convective cores (as opposed to solar-type stars with radiative cores). Later work \citep[e.g., ][]{Terquem:1998ya, Goodman:1998yg, Ogilvie:2007tja} extended the original formalism to stars with central radiative cores, and found dynamical tides that were stronger than those predicted by Z77, but still at least one order of magnitude below equilibrium tides and IW dissipation from the convective zone. Therefore, we trust the overall qualitative behavior observed in our Z77 simulations, where even equilibrium tides are too weak to circularize any simulated binary over our grid.

\begin{figure*}
    \centering
    \begin{subfigure}[t]{0.47\textwidth}
        \centering
        \includegraphics[trim={-7 0 0 0}, width=0.9\columnwidth]{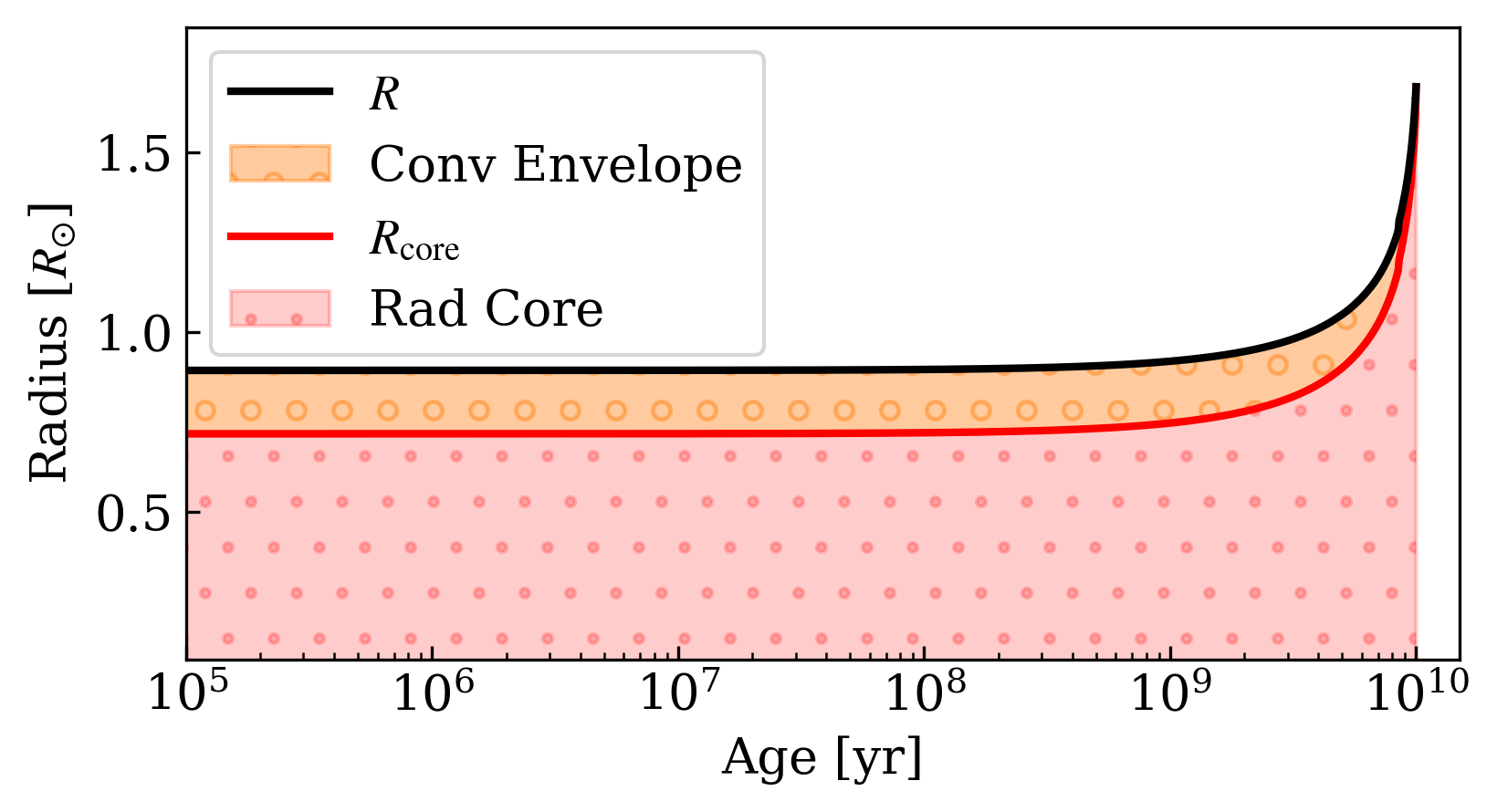}
        \caption{Kippenhahn diagram of the primary star while on MS.}
        \label{fig:int_mass_kipp}
    \end{subfigure}
    \hfill
    \begin{subfigure}[t]{0.47\textwidth}
        \centering
        \includegraphics[trim={0 0 10 0}, width=0.85\columnwidth]{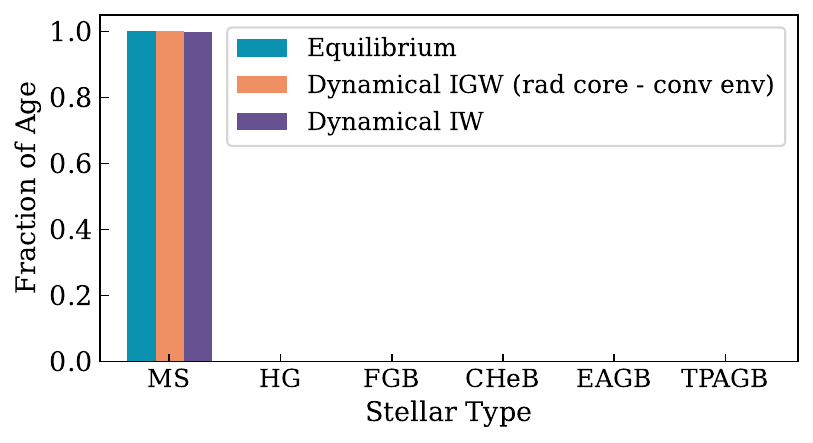}
        \caption{Fraction of time that each tidal mechanism is applicable, up to MS.}
        \label{fig:int_mass_tides}
    \end{subfigure}
    \hfill
    \begin{subfigure}[t]{0.47\textwidth}
        \centering
        \includegraphics[trim={-10 0 30 0}, width=\columnwidth]{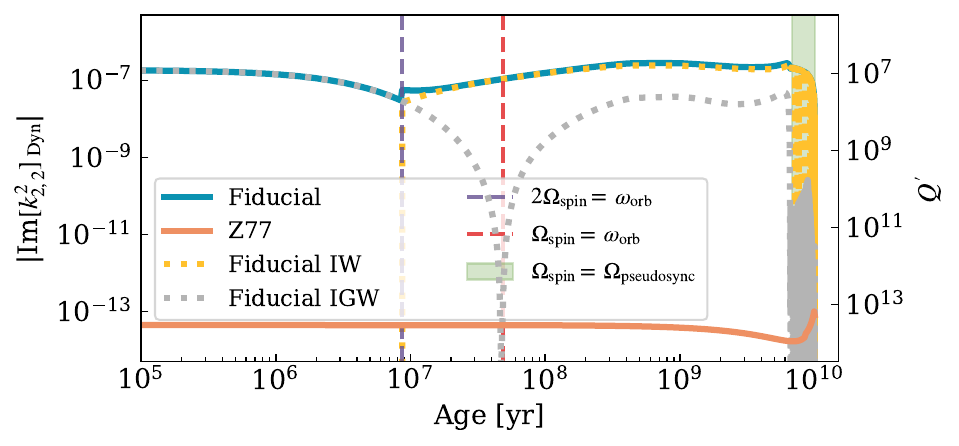}
        \caption{Absolute strength of the $\ell=2$, $m=2$, $n=2$ dynamical tidal Love number (left y-axis) and the tidal quality factor (right y-axis) of the primary star. The dotted yellow and gray curves show the contribution from IW and IGW dissipation, respectively.}
        \label{fig:int_mass_dyn}
    \end{subfigure}
    \hfill
    \begin{subfigure}[t]{0.47\textwidth}
        \centering
        \includegraphics[width=0.9\columnwidth]{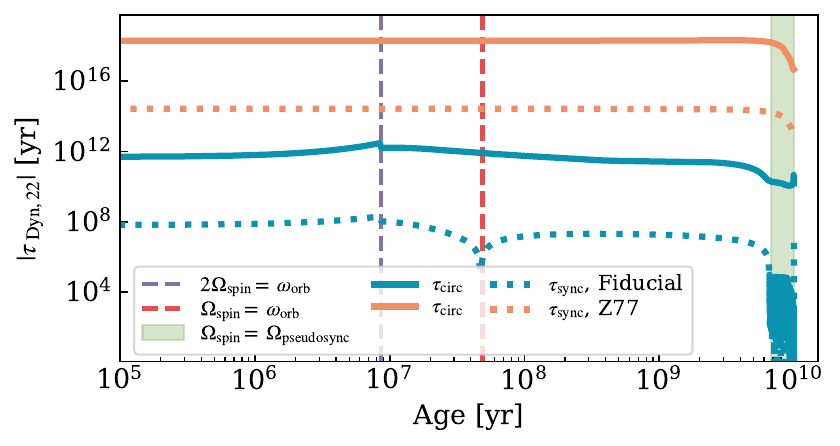}
        \caption{Circularization and synchronization timescales from dynamical tides on the primary star.}
        \label{fig:int_mass_dyn_tau}
    \end{subfigure}
    \begin{subfigure}[t]{0.47\textwidth}
        \centering
        \includegraphics[trim={-10 0 30 0}, width=\columnwidth]{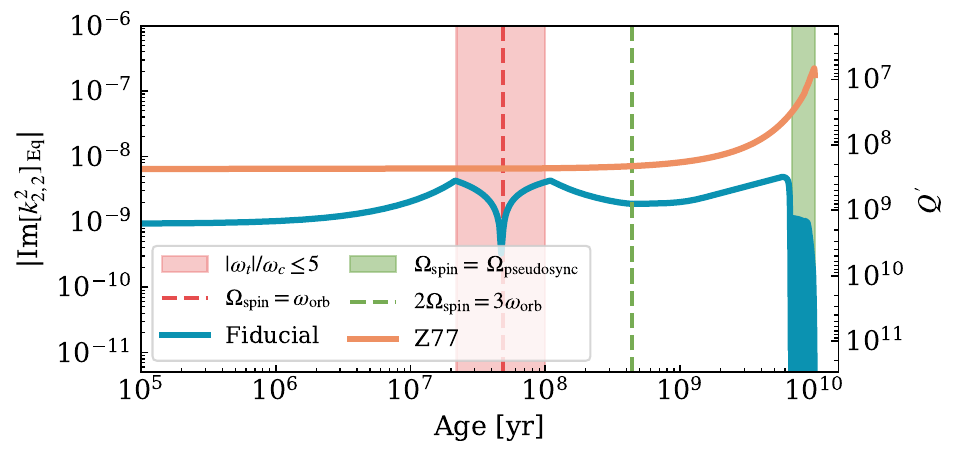}
        \caption{Absolute strength of the $\ell=2$, $m=2$, $n=2$ equilibrium tidal Love number (left y-axis) and the tidal quality factor (right y-axis) of the primary star.}
        \label{fig:int_mass_eq}
    \end{subfigure}
    \hfill
    \begin{subfigure}[t]{0.47\textwidth}
        \centering
        \includegraphics[width=0.89\columnwidth]{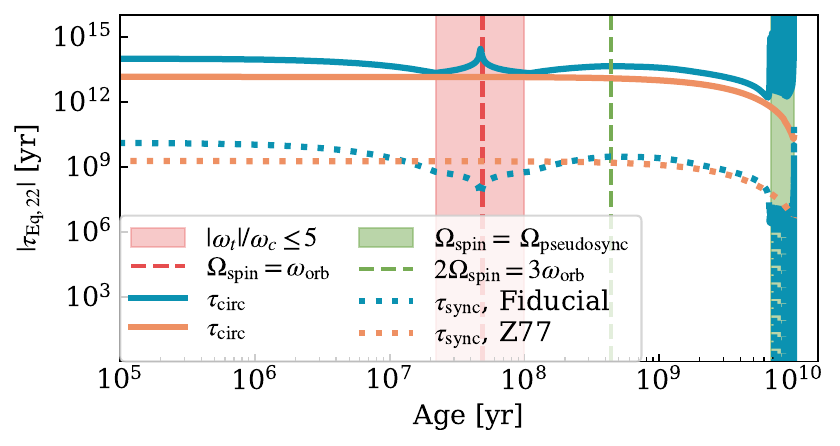}
        \caption{Circularization and synchronization timescales from equilibrium tides on the primary.}
        \label{fig:int_mass_eq_tau}
    \end{subfigure}
    \hfill
    \begin{subfigure}[t]{0.47\textwidth}
        \centering
        \includegraphics[trim={7 0 0 0}, width=0.9\textwidth]{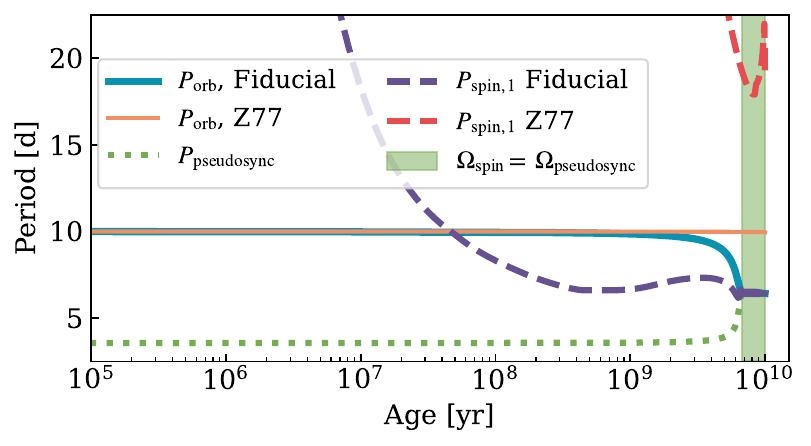}
        \caption{Orbital and primary spin period evolution under the fiducial and Z77 models.}
        \label{fig:int_mass_period}
    \end{subfigure}
    \hfill
    \begin{subfigure}[t]{0.47\textwidth}
        \centering
        \includegraphics[trim={7 0 0 0}, width=0.9\textwidth]{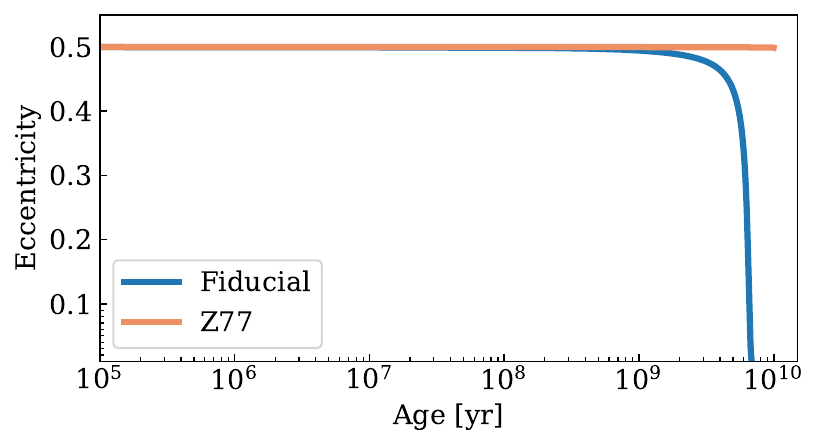}
        \caption{Eccentricity evolution under the fiducial and Z77 models.}
        \label{fig:int_mass_ecc}
    \end{subfigure}
    \caption{Stellar and tidal evolution for a 1 $M_\odot$ + 1 $M_\odot$ binary with $P_{\rm orb, ZAMS}$=10 days and $e_{\rm ZAMS}=0.5$. The binary is simulated until the end of MS, at roughly 10 Gyr.}
    \label{fig:int_mass}
\end{figure*}

For our individual binary, we simulate a system with $M_{\rm 1, ZAMS}=1 M_\odot$, $M_{\rm 2, ZAMS}=1 M_\odot$, $P_{\rm orb, ZAMS}~=~10$~days, and $e_{\rm ZAMS} = 0.5$, such that we can see circularization and synchronization within the main-sequence lifetime. The initial stellar spins are set to zero by default. The evolution is shown in Fig.~\ref{fig:int_mass}. This particular system ends as a helium white dwarf binary, although we limit our discussion to only the main sequence lifetime ($\sim 10$ Gyr) for this section.

The component stars in this binary have a radiative core and convective envelope at ZAMS, and their envelope gradually diminishes over the MS lifetime as $(1-\tau)^{1/4}$, where $\tau$ is the fractional age on MS ranging from 0 at ZAMS to 1 at TAMS \citep{Hurley:2002rf}. We show the evolution of the core and envelope in Fig.~\ref{fig:int_mass_kipp}. The primary (and secondary) experiences equilibrium tides due to viscous dissipation, dynamical tides due to gravity waves (from the envelope), and also dynamical tides due to inertial waves once the spin exceeds the critical threshold of $2\Omega_{\rm spin} \geq \omega_{\rm orb}$. The fraction of time spent under each tidal process is shown in Fig.~\ref{fig:int_mass_tides}. 

Dynamical tides (shown in Fig.~\ref{fig:int_mass_dyn} and Fig.~\ref{fig:int_mass_dyn_tau}) dominate the evolution of the binary in the fiducial model, whereas equilibrium tides (shown in Fig.~\ref{fig:int_mass_eq} and Fig.~\ref{fig:int_mass_eq_tau}) are dominant in the Z77 simulation. We must note that the Z77 dynamical tides model is meant for stars with convective cores, and is technically irrelevant for this binary. We still plot the magnitude of the (2,2) tidal Love number obtained by na\"ively applying their equations for radiative core stars. Our model predicts an imaginary Love number that is $\sim 7$ orders of magnitude higher than the ill-applied Zahn model, and  1 or 2 orders of magnitude above equilibrium tides from viscous dissipation alone.

Given the radiative-convective boundary, IGW dissipation from the envelope acts throughout binary evolution. The dynamical tidal evolution under our fiducial model is initially dominated by IGW dissipation, shown as the gray dotted curve in \revision{Fig.~\ref{fig:int_mass_dyn}}. The strength of (2,2) IGW dissipation with a radiative core scales as $\omega_{t, 22}^{8/3}$ (as per Eq.~\eqref{eq:igw_rad_core_conv_envelope}), and the magnitude of IGW dissipation decreases gradually until the point of synchronization (red dashed line). After this point, $\omega_{t ,22}$ flips sign and the absolute value of Im$[k_{2,2}^2]_{\rm Dyn, \; IGW}$ appears to `bounce'. \revision{Beyond this point, all (2,2) tides, whether dynamical or equilibrium, contribute with a flipped sign to Eqs.~\eqref{eq:tidal_dadt} - \eqref{eq:tidal_dOmegadt}, and the overall evolution of the  semi-major axis, eccentricity, and stellar rotation depends on the combined contribution from all tidal modes.}
The IW component of dynamical tidal dissipation (yellow dotted curve) becomes relevant when $2 \Omega_{\rm spin} \geq \omega_{\rm orb}$ \revision{(starting at the violet dashed vertical line in Fig.~\ref{fig:int_mass_dyn})}. Once initiated, IW dissipation remains the dominant tidal dissipation mechanism for the binary until the end of its MS evolution. Since the magnitude of IW dissipation does not explicitly depend on $\omega_{t, 22}$, its contribution to Im$[k_{2,2}^2]_{\rm Dyn}$ remains smooth through the $\Omega_{\rm spin}=\omega_{\rm orb}$ transition point in Fig.~\ref{fig:int_mass_dyn}, \revision{even as its sign flips}.

Looking to the Z77 model for comparison, the strength of dynamical tides is set almost entirely by $\left( R_*/a \right)^6$. As such, the (2,2) component of the imaginary Love number stays largely constant throughout the binary evolution, and only grows sharply when the star begins to expand towards the end of the main sequence.
Compared to Z77, our tidal model predicts dynamical tides that are stronger and also vary more substantially over the course of stellar and binary evolution.

Compared to the low mass binary in Sec.~\ref{sec:low_mass}, the primary now has lower convective density as well as lower tidal frequency, both of which prevent equilibrium tides from becoming much stronger. On the other hand, the simpler Z77 equilibrium tides model does not explicitly depend on the density of the convective shell, and the dissipation is much stronger due to the larger stellar radius. Therefore, as opposed to Sec.~\ref{sec:low_mass}, the solar mass binary initially experiences $\sim$~10 times weaker equilibrium tides in the fiducial model than in the Z77 model. As the system approaches synchronization and $|\omega_{t,22}|/\omega_c$ decreases below 5, there is a sudden change in the slope of Im[$k_{2,2}^2$]$_{\rm eq}$ due to the frequency-dependent viscosity, with the overall scaling going from $|\omega_{t, 22}|^{-1}$ to $|\omega_{t, 22}|^{1/2}$. This is visible in the red shaded regions in Fig.~\ref{fig:int_mass_eq} and Fig.~\ref{fig:int_mass_eq_tau}. Through a combination of equilibrium and dynamical tides, the primary stellar rotation of the fiducial simulation becomes synchronized with the orbit during this regime, and Im[$k_{2,2}^2$]$_{\rm eq}$ flips in sign (at the vertical dashed red line in Fig.~\ref{fig:int_mass_eq} and Fig.~\ref{fig:int_mass_eq_tau}). As with dynamical tides, we see this sign flip as a `bounce'.

As the component stars in our fiducial simulation continue spinning up past $\Omega_{\rm spin} = \omega_{\rm orb}$ due to non-zero orbital eccentricity, $|\omega_{t,22}|/\omega_c$ once again exceeds 5 at roughly $10^8$ yr. The slope of the (2,2) component of the equilibrium tide returns to its former behavior outside the red shaded region.

Once the spin exceeds $2 \Omega_{\rm spin} \geq 3\omega_{\rm orb}$ (past the dashed green line in Fig.~\ref{fig:int_mass_eq} and Fig.~\ref{fig:int_mass_eq_tau}), \revision{the sign of Im[$k_{2,3}^2$]$_{\rm Eq}$ flips and stops the spin from growing any further (see Eq.~\eqref{eq:tidal_dOmegadt}).} The change of the spin-up behavior also affects the evolution of Im[$k_{2,2}^2$]$_{\rm Eq}$, and we see an inflection point in the Love number evolution. This is an example of how the inclusion of higher-order tidal dissipation modes has a complex effect on the overall tidal evolution of a binary, and even other modes. 

We show the evolution of spin and orbital periods in Fig.~\ref{fig:int_mass_period}. Until $\sim 6\times 10^9$ yr, the pseudo-synchronization period limit for the component stars in our fiducial simulation (dotted green curve) remains lower than the component spins (dashed purple curve). In other words, tidal dissipation is not strong enough to spin up the stars to their pseudo-synchronous limit in our simulations. Due to substantially weaker dynamical tides, the Z77 binary spin (dashed red curve) never even manages to reach synchronization, much less pseudo-synchronization. 

Towards the end of MS evolution, however, the eccentricity in the fiducial simulation is dissipated entirely by dynamical tides (see Fig.~\ref{fig:int_mass_ecc}). Under angular momentum conservation, the semi-latus rectum $a (1-e^2)$ of the binary should remain constant throughout its evolution. With an initial semi-major axis of $a_{\rm ZAMS} = 0.11445$~AU and an initial eccentricity of $e_{\rm ZAMS} = 0.5$, the final semi-major axis for the circularized orbit should be $a_f = 0.08584$~AU. The final orbital separation in our simulation is slightly wider at $0.08528$~AU, given that some angular momentum is transferred to the component stars via rotation, while some angular momentum is lost due to stellar winds.

Since synchronization in a circular binary requires slower rotating component stars than for an eccentric binary, we see the pseudo-synchronization period increase in Fig.~\ref{fig:int_mass_period}. The part of our simulation where the stellar spins are clamped to the pseudo-synchronization limit is shown as the shaded green region across Fig.~\ref{fig:int_mass}. This region is accompanied by dynamical and equilibrium tides alternating in sign to preserve pseudo-synchronization. As per usual, any noisiness in the data should be attributed to the finite time stepping limitation of COMPAS rather than physical effects.

\subsection{Convective core main sequence stars}
\begin{figure}
    \centering
    \includegraphics[width=\columnwidth]{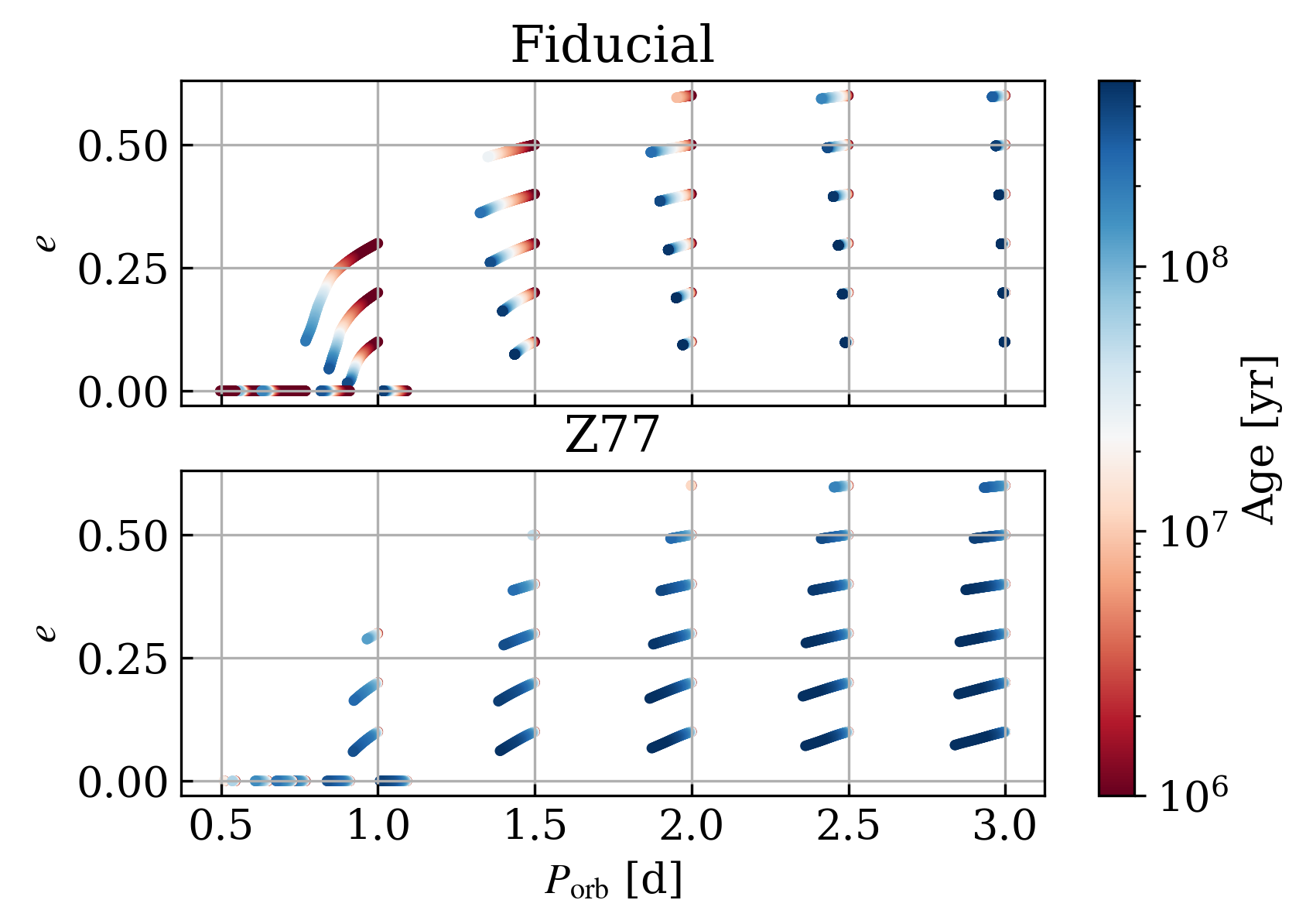}
    \caption{Time evolution of $e$ and $P_{\rm orb}$ under dynamical tides for $2.5M_\odot+2.5M_\odot$ binaries over a grid of initial orbital periods and initial eccentricities. The colors depict the age of each system in yr. The binaries are only evolved up to the end of their MS lifetimes, or until the binary merges, whichever comes first.}
    \label{fig:high_mass_grid_ecc_porb}
\end{figure}

ZAMS stars above $1.25 M_\odot$ have convective cores and radiative envelopes in COMPAS. As per our fiducial model, these stars should experience only dynamical tides due to IGW dissipation, since we ignore tidal dissipation inside convective cores. We simulate a grid over various initial orbital periods and eccentricities for a $2.5 + 2.5 M_\odot$ mass configuration in Fig.~\ref{fig:high_mass_grid_ecc_porb}. Again, these stars are initialized to be non-rotating at ZAMS, and evolved up to either the end of MS or until the stars merge. Low period and high eccentricity binaries merge promptly, so we do not see any evolution for grid locations in the upper left region of the plot. 

With our fiducial tides, binaries with orbital periods above $\sim 1$~day do not experience strong enough tides from IGW dissipation to lose appreciable eccentricity. However, shorter period binaries spin up faster, which causes the binary to circularize more efficiently. Comparing this grid to Fig.~\ref{fig:int_mass_grid_ecc_porb}, it is evident that IGW dissipation from convective core boundaries is significantly weaker than from convective envelopes in our tidal models. 

The Z77 binaries experience stronger dynamical tides from IGW dissipation than our fiducial model. As a result, they dissipate more eccentricity and orbital separation across the grid compared to the fiducial grid. The difference between the fiducial and Z77 models for convective core IGW dissipation lies in the assumed core size, which we will examine in greater detail in the rest of this section.

To observe the full tidal dissipation story, we simulate a $2.5 M_\odot + 2.5 M_\odot$ binary with $P_{\rm orb, ZAMS} = 3$~days and $e_{\rm orb, ZAMS} = 0.5$, representing typical short-period binaries in this mass range~\citep{prevot1961vitesses, Pourbaix2000, Behr_2011}. Our simulation ends as an MS+MS stellar merger after $\sim$420\,Myr. 

\begin{figure*}
    \centering
    
    \begin{subfigure}[t]{0.47\textwidth}
        \centering
        \includegraphics[trim={10 0 -30 0}, width=\textwidth]{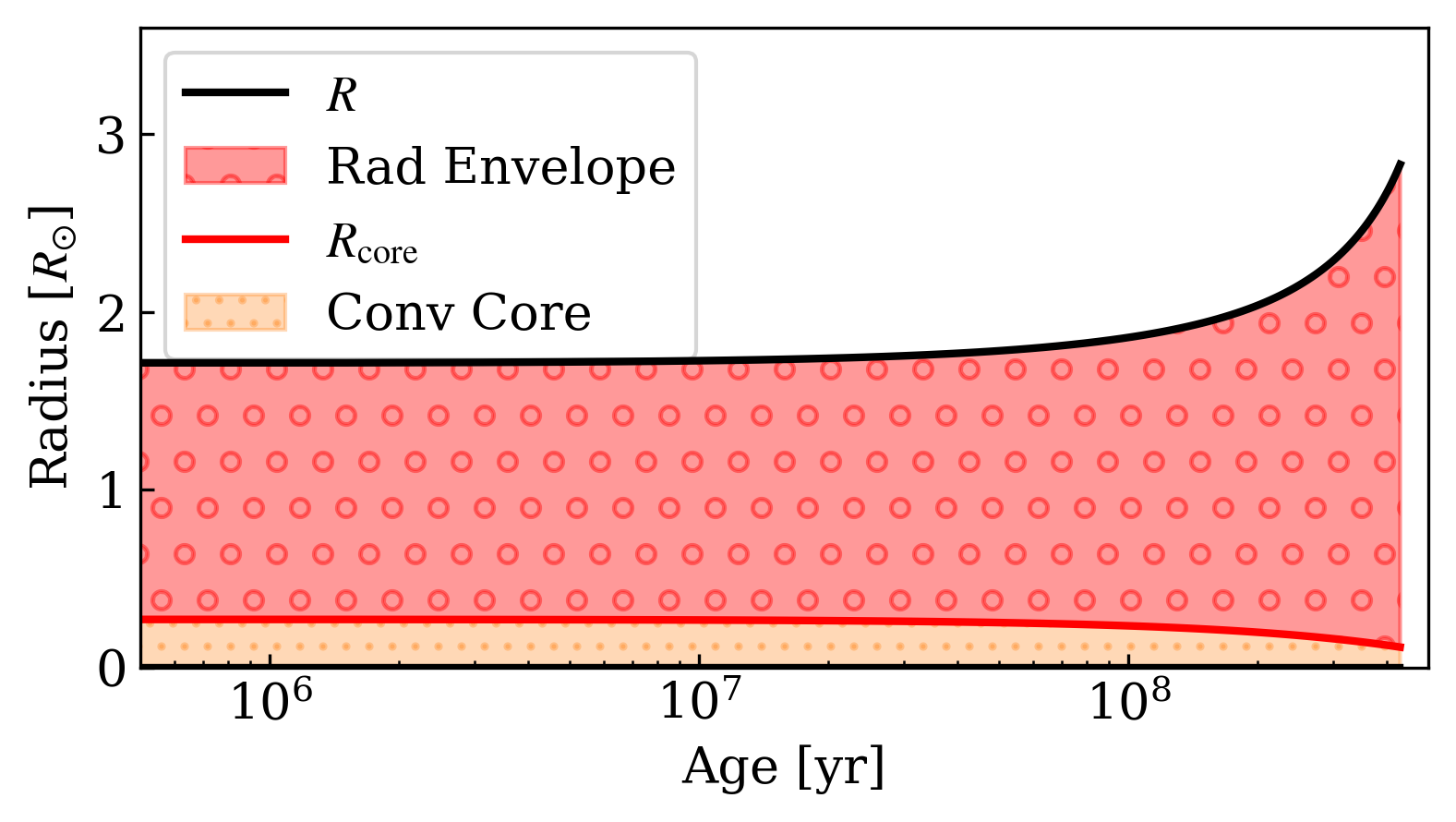}
        \caption{Kippenhahn diagram of the primary star.\\$\;$}
        \label{fig:high_mass_ms_kipp}
    \end{subfigure}
    \hfill
    \begin{subfigure}[t]{0.47\textwidth}
        \centering
        \includegraphics[trim={0 0 0 0}, width=\textwidth, height=0.20\textheight]{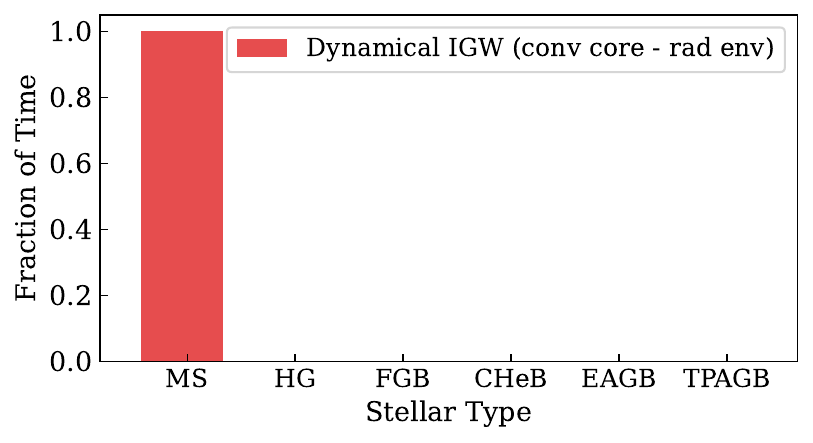}
        \caption{Fraction of time that each tidal mechanism is applicable, per stellar type encountered over the evolution.}
        \label{fig:high_mass_ms_tides}
    \end{subfigure}
        
    \begin{subfigure}[t]{0.47\textwidth}
        \centering
        \includegraphics[trim={27 -10 20 0}, width=\textwidth]{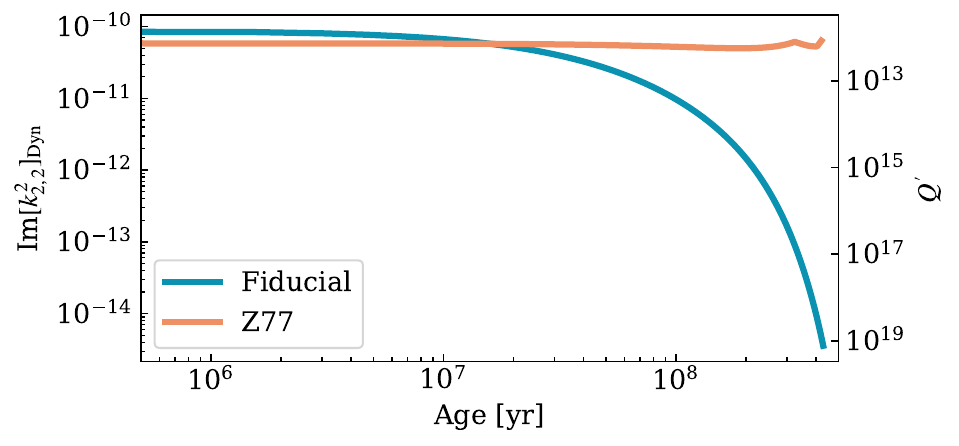}
        \caption{Absolute strength of the $\ell=2$, $m=2$, $n=2$ dynamical tidal Love number (left y-axis) and the tidal quality factor (right y-axis) of the primary star.}
        \label{fig:high_mass_ms_dyn}
    \end{subfigure}
    \hfill
    \begin{subfigure}[t]{0.47\textwidth}
        \centering
        \includegraphics[trim={14 -10 0 0}, width=\textwidth, height=0.20\textheight]{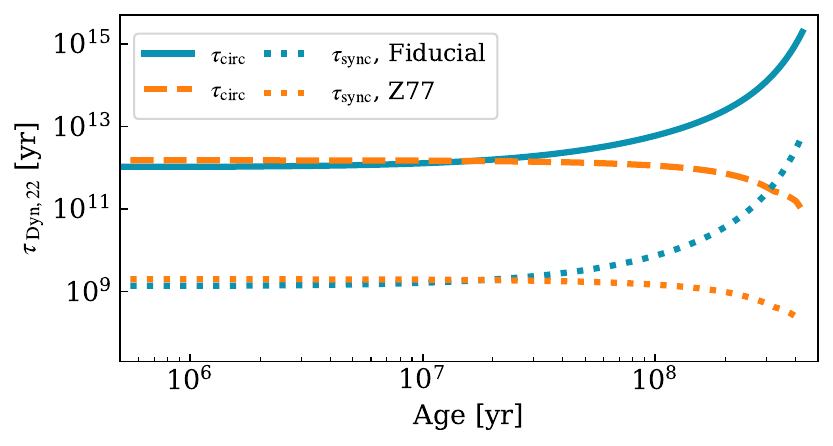}
        \caption{Dynamical tidal timescales of circularization and synchronization across the prescriptions from this work.}
        \label{fig:high_mass_ms_tau}
    \end{subfigure}
    \begin{subfigure}[t]{0.47\textwidth}
        \centering
        \includegraphics[trim={14 0 -20 10}, width=\textwidth]{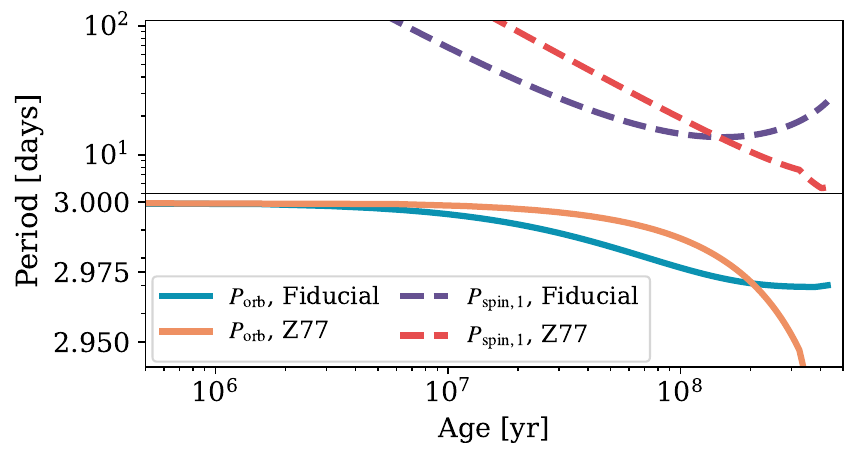}
        \caption{Orbital and primary spin period evolution under the fiducial and Z77 models, split into two vertical panels to be able to see the orbital and rotational periods together.}
        \label{fig:high_mass_period}
    \end{subfigure}
    \hfill
    \begin{subfigure}[t]{0.47\textwidth}
        \centering
        \includegraphics[trim={14 0 0 0}, width=\textwidth]{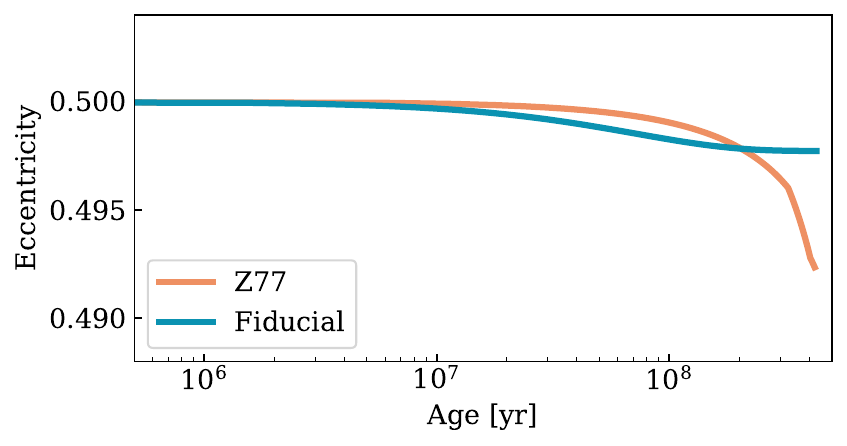}
        \caption{Eccentricity evolution under the fiducial and Z77 models.}
        \label{fig:high_mass_ecc}
    \end{subfigure}
    
    \caption{Stellar and tidal evolution for a 2.5 $M_\odot$ + 2.5 $M_\odot$ binary with $P_{\rm orb, ZAMS}$=3 days and $e_{\rm ZAMS}=0.5$. The binary is simulated until both MS stars merge at $\sim 420$\,Myr.}
    \label{fig:high_mass_ms}
\end{figure*}

On the MS, the only relevant tide in the fiducial simulation is IGW dissipation at the convective core and radiative envelope boundary (see Fig.~\ref{fig:high_mass_ms_kipp} and Fig.~\ref{fig:high_mass_ms_tides}). The mass and radius of the radiative envelope in COMPAS grow with stellar age, following the fits described in \citet{Hurley:2000pk}. At the same time, the convective core shrinks over time on the MS as per Eq.~\eqref{eq:rc_fit_shikauchi} and Eq.~\eqref{eq:rc_tams_evolution}. With these core and envelope prescriptions, the strength of dynamical tides decreases drastically as $\left({R_c}/{R_*}\right)^9$ over time in our models. Physically, this corresponds to a vanishing convective-radiative boundary from which to excite internal gravity waves. We see this as a gradually declining dynamical tidal Love number in Fig.~\ref{fig:high_mass_ms_dyn}, and consequently increasing synchronization and circularization timescales in Fig.~\ref{fig:high_mass_ms_tau}. 

Evidently from these timescales, dynamical tides in our model are not strong enough to circularize or synchronize the binary within the MS lifetime. We can see the primary spin period (dashed purple curve) steadily shorten over time in Fig.~\ref{fig:high_mass_period}, but it never reaches the orbital period. The binary also retains most of its eccentricity, with a final eccentricity of $0.4988$ after 420\,Myr, as shown in Fig.~\ref{fig:high_mass_ecc}.

The Z77 model attempts to model the effect of the convective core on dynamical tides with an $E_2$ term, which they claim should scale as $(R_c/R_*)^8$ to first order. However, the commonly used fit by \citet{Hurley:2002rf} depends only on the total mass of the star, and is based on stellar simulations presented in \citet{Zahn:1975A&A....41..329Z} (see Appendix \ref{app:z77_model} for more details). By over-simplifying the dependence of IGW dissipation on the stellar structure, the Z77 model runs the risk of mis-characterizing tidal response for high mass MS stars. Indeed, the strength of IGW dissipation in the Z77 model does not decrease with the vanishing core, and ends up becoming several orders of magnitude larger than our fiducial value as time goes on. As a result, the binary with the Z77 model reaches higher component spin rates (red dashed curve in Fig.~\ref{fig:high_mass_period}) and a lower final eccentricity of $0.4923$, as shown in Fig.~\ref{fig:high_mass_ecc}.

\subsection{Giant branch stars}
\label{sec:high_mas_agb}
The stellar structure during evolved phases can be complicated, with the possibility of thin radiative shells surrounded by extended convective envelopes. There are long-standing discrepancies between the theoretical and observed circularization periods of AGB stellar binaries.
The theoretical circularization period for stars that go through an AGB phase, such as the progenitors of Barium stars and WDs, has been estimated to be as large as $\sim 4000$~days \citep{pols2003can,Izzard:2010cu} from simulations. These results utilize fits from \citet{Hurley:2000pk} and \citet{Hurley:2002rf} for their stellar and tidal evolution models, which should be functionally equivalent to our Z77 model. These period estimates differ starkly from the observed circularization periods of binaries containing a giant type star, which range from $\sim 30$~days to $\sim 200$~days \citep{verbunt1995tidal, bluhm2016new}. One hypothesis to explain this difference suggests that eccentric WD binaries can be explained by a late-stage eccentricity boost from natal kicks or prompt mass-loss at shorter-than-orbital timescales, which would provide a mechanism for circularized GB binaries to become eccentric \citep{Izzard:2010cu}. 

\begin{figure}
    \centering
    \includegraphics[width=\columnwidth]{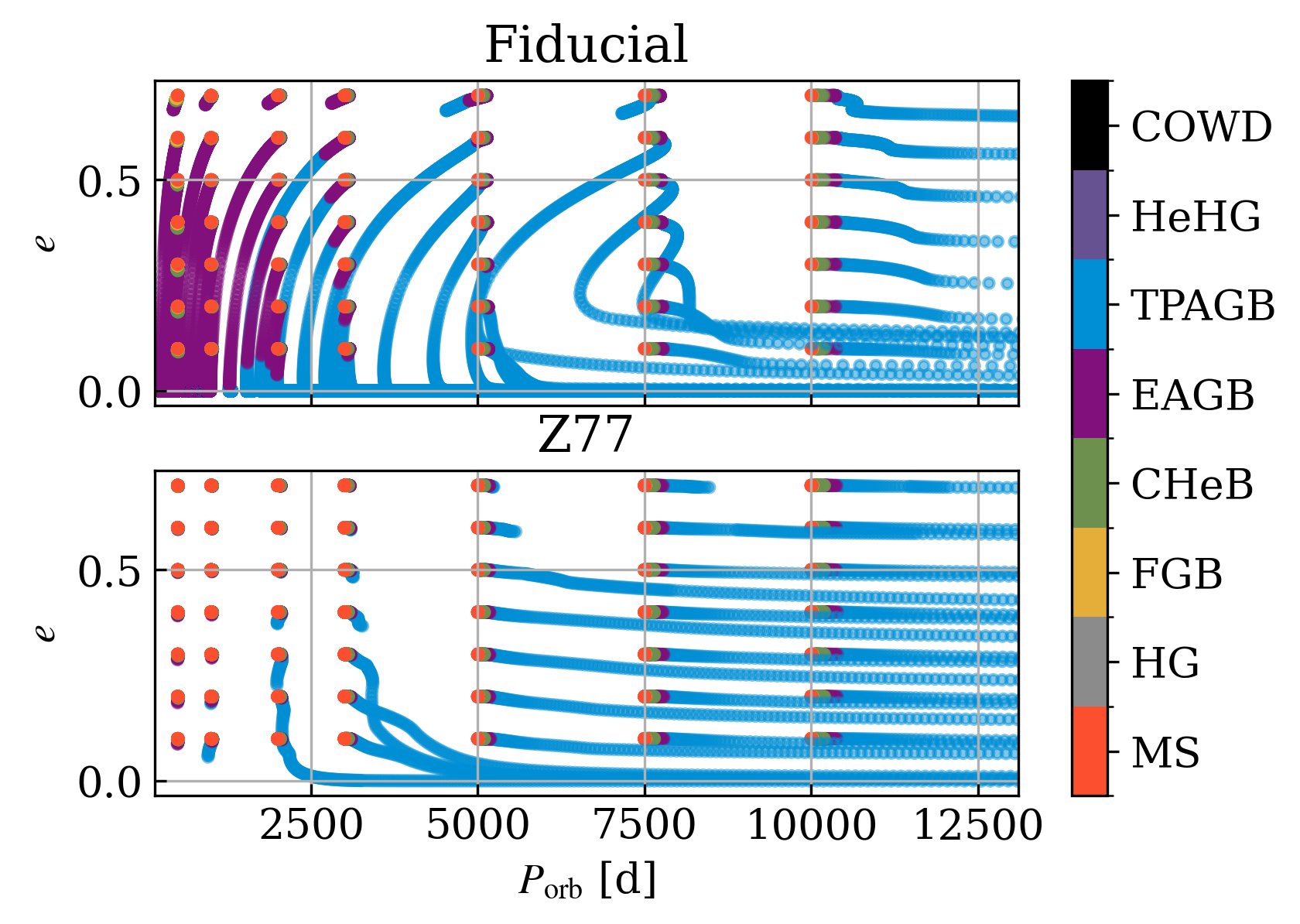}
    \caption{Time evolution of $e$ and $P_{\rm orb}$ for $3M_\odot+3M_\odot$ binaries over a grid of initial orbital periods and initial eccentricities. The colors represent the various stellar types encountered by each binary over its evolution. The binaries end in either stellar merger or as carbon-oxygen white dwarf (COWD) binaries at $\sim 450$\,Myr.}
    \label{fig:high_mas_agb_grid_ecc_porb}
\end{figure}

We simulate a set of $3M_\odot+3M_\odot$ binaries at various initial eccentricities and orbital periods, all with zero initial stellar rotation, in Fig.~\ref{fig:high_mas_agb_grid_ecc_porb}. These binaries should not be compared directly against simulations from \citet{Izzard:2010cu}, since our binaries are equal mass and will contain two giant type stars in the AGB phase rather than one. However, the circularization behavior should still be correct up to the order of magnitude.

Most binaries in our grid spend $\sim 350$\,Myr on the MS, after which they go through HG, FGB, CHeB, EAGB, and TPAGB phases, until they finally end their evolution as carbon-oxygen white dwarves (COWDs) at $\sim 450$\,Myr. The shorter evolutionary tracks in the upper left section of each panel in Fig.~\ref{fig:high_mas_agb_grid_ecc_porb} represent stellar mergers that take place before a COWD binary can form, due to a combination of small orbital separations and high eccentricities. 
Stars in COMPAS experience some of the strongest mass loss during the TPAGB phase. The orbital period of a binary that gradually loses mass in winds scales as $P_{\rm orb} \propto (M_1 + M_2)^{-2}$. This is why the orbital periods of most TPAGB binaries in Fig.~\ref{fig:high_mas_agb_grid_ecc_porb} increase dramatically over their evolution.

The strength of tides for AGB stars is higher in our models compared to Z77, as we can observe binaries losing most of their eccentricities even at orbital periods of $\sim 5000$ days in the fiducial simulations. On the other hand, the Z77 grid only seems to become circular below $\sim 3000$ day orbits (bottom left in the Z77 panel of Fig.~\ref{fig:high_mas_agb_grid_ecc_porb}).

\begin{figure*}
    \centering
    \begin{subfigure}[t]{0.46\textwidth}
        \centering
        \includegraphics[trim={-7 0 0 0}, width=0.99\columnwidth]{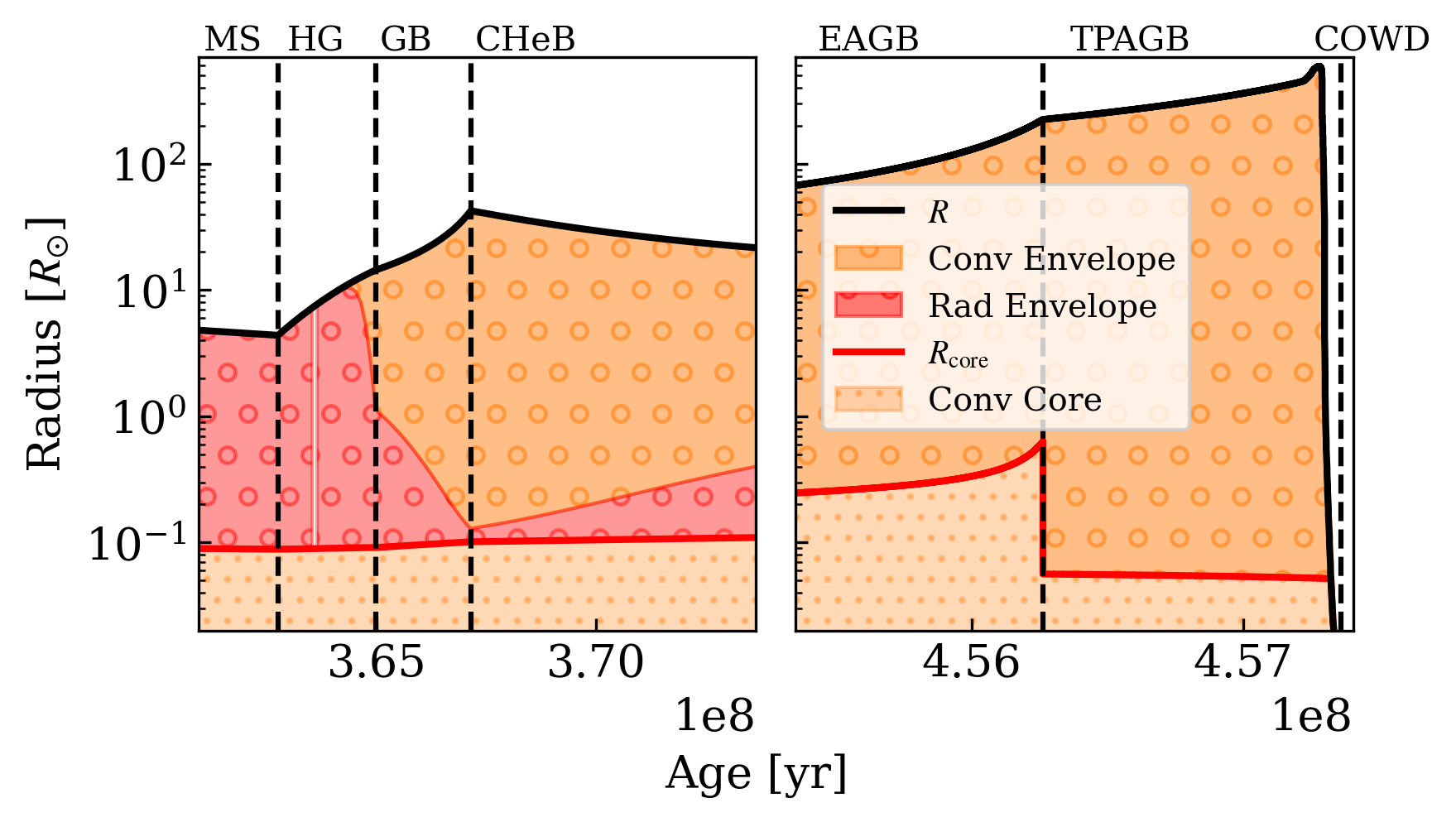}
        \caption{Kippenhahn diagram of the primary star.}
        \label{fig:high_mas_agb_kipp}
    \end{subfigure}
    \hfill
    \begin{subfigure}[t]{0.46\textwidth}
        \centering
        \includegraphics[trim={15 0 10 0}, width=0.95\columnwidth]{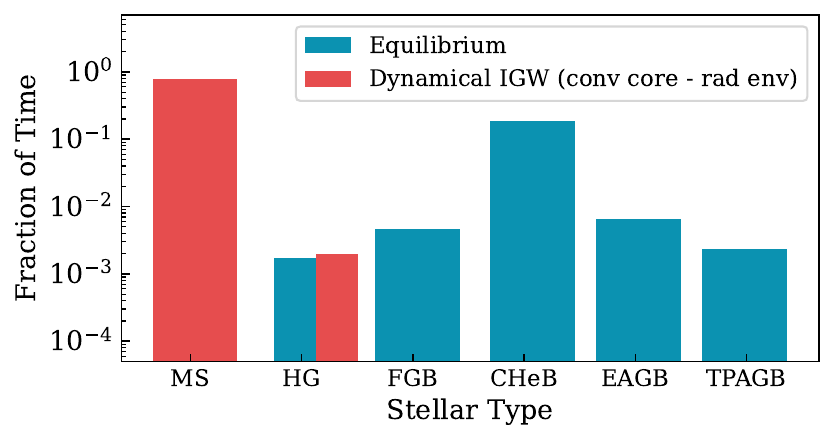}
        \caption{Fraction of time that each tidal mechanism is applicable, per stellar type encountered over the evolution. Unequal bar heights for a given stellar type mean that tidal mechanisms were relevant for different amounts of time.}
        \label{fig:high_mas_agb_tides}
    \end{subfigure}
    
     \begin{subfigure}[t]{0.46\textwidth}
        \centering
        \includegraphics[trim={-10 0 30 0}, width=0.95\columnwidth]{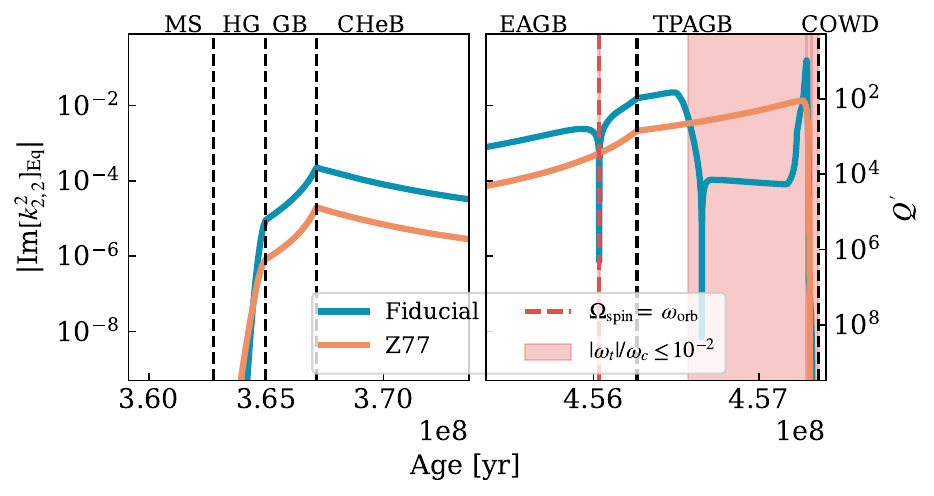}
        \caption{Strength of the $\ell=2, m=2, n=2$ component of equilibrium tides on the primary. }
        \label{fig:high_mas_agb_eq_imk22}
    \end{subfigure}
    \hfill
    \begin{subfigure}[t]{0.46\textwidth}
        \centering
        \includegraphics[trim={0 0 20 0}, width=0.95\columnwidth]{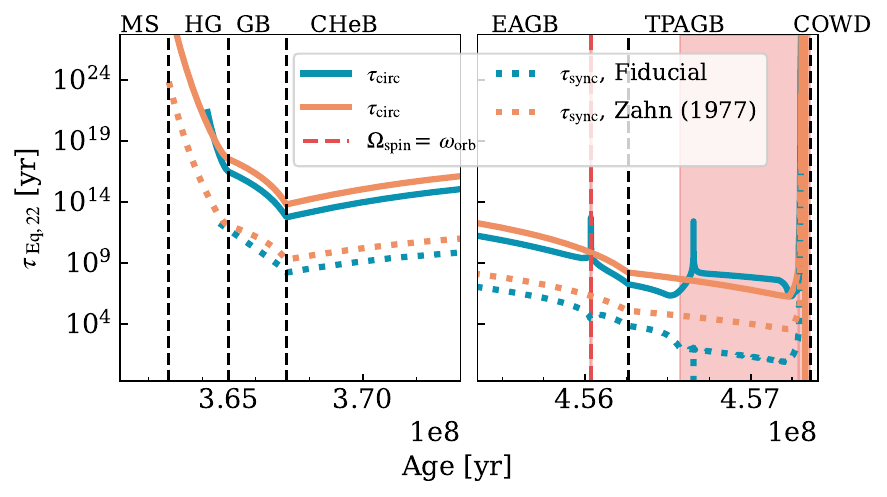}
        \caption{Circularization and synchronization timescales from equilibrium tides on the primary.}
        \label{fig:high_mas_agb_eq_tau}
    \end{subfigure}
    
    \begin{subfigure}[t]{0.46\textwidth}
        \centering
        \includegraphics[trim={0 0 30 0}, width=0.95\columnwidth]{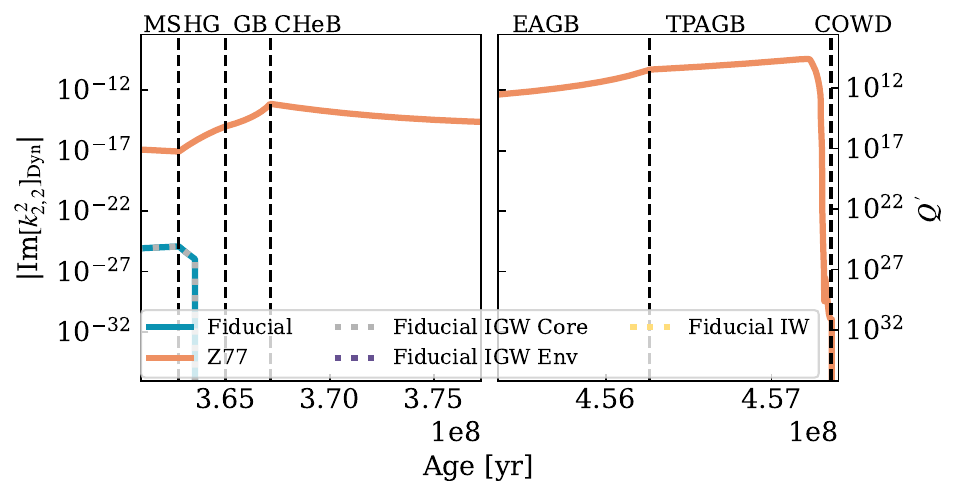}
        \caption{Strength of dynamical tides on the primary.\\ $\;$}
        \label{fig:high_mas_agb_dyn_imk22}
    \end{subfigure}
    \hfill
    \begin{subfigure}[t]{0.46\textwidth}
        \centering
        \includegraphics[width=0.95\columnwidth]{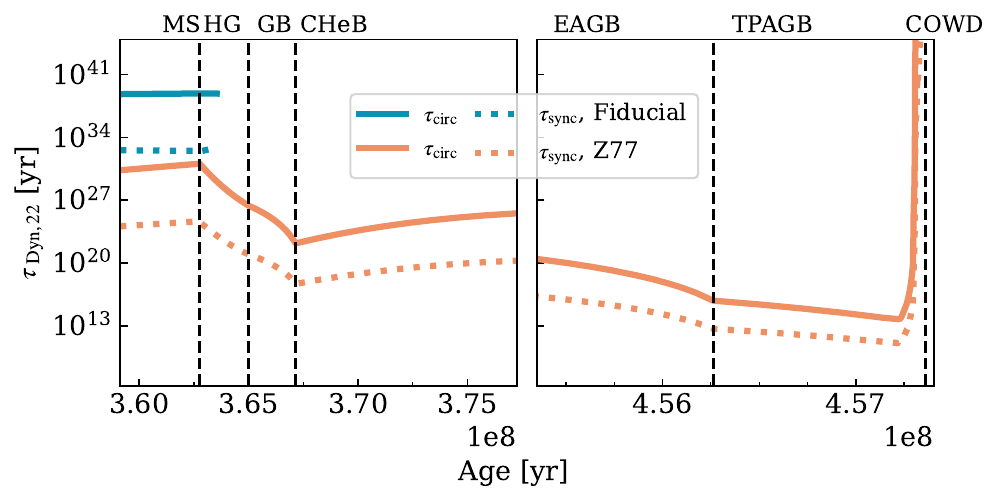}
        \caption{Circularization and synchronization timescales from dynamical tides on the primary.}
        \label{fig:high_mas_agb_dyn_tau}
    \end{subfigure}
       \begin{subfigure}[t]{0.46\textwidth}
        \centering
        \includegraphics[trim={-10 0 30 0}, width=0.95 \columnwidth]{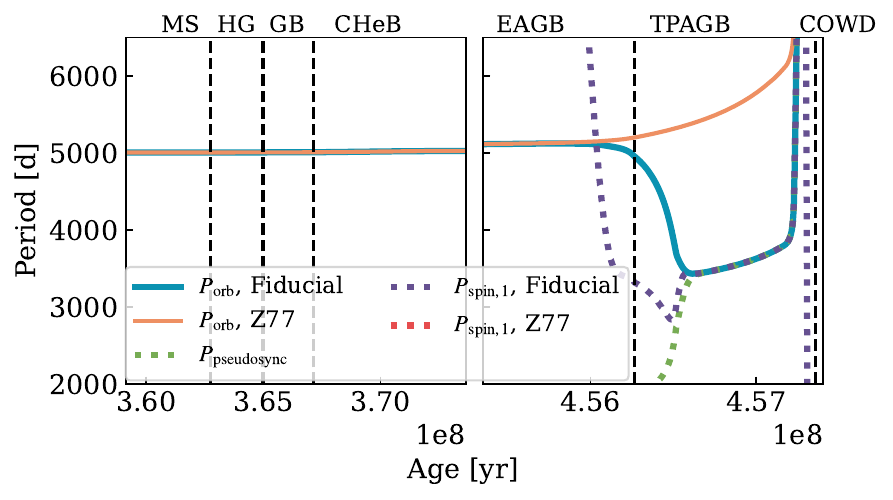}
        \caption{Orbital and primary spin period evolution under the fiducial and Z77 models.}
        \label{fig:high_mas_agb_period}
    \end{subfigure}
    \hfill
    \begin{subfigure}[t]{0.46\textwidth}
        \centering
        \includegraphics[trim={-10 0 30 0}, width=0.95\columnwidth]{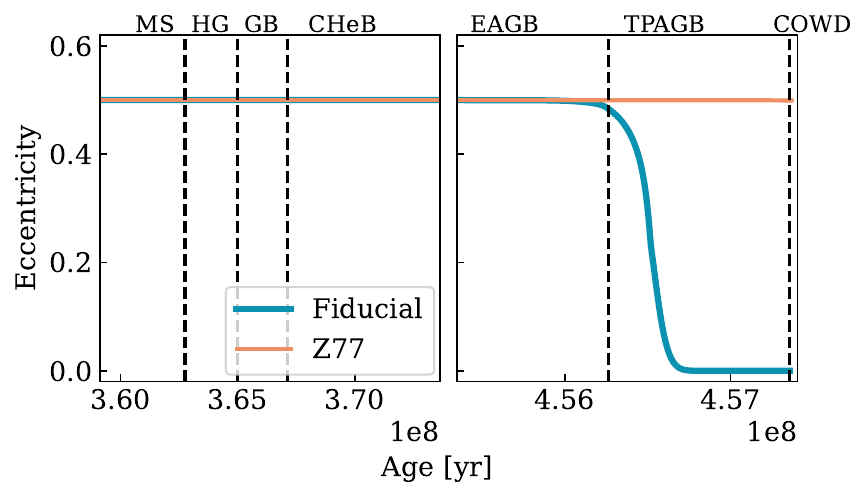}
        \caption{Eccentricity evolution under the fiducial and Z77 models.}
        \label{fig:high_mas_agb_ecc}
    \end{subfigure}
    \caption{Stellar and tidal evolution for a 3 $M_\odot$ + 3 $M_\odot$ binary with $P_{\rm orb, ZAMS}$=5000 days and $e_{\rm ZAMS}=0.5$. The binary is simulated until a WD+WD formation at $\sim 458$\,Myr. The dashed black lines show the boundaries between various stellar phases, and the phases are labeled above each plot.}
    \label{fig:high_mas_agb}
\end{figure*}

To understand the behavior of giant type binaries under our tidal models, we simulate a binary with component masses $M_1 = M_2 = 3 M_\odot$, initialized to an initial period of $P_{\rm orb, ZAMS}$ = 5000 days, with $e_{\rm ZAMS} = 0.5$. Based on Fig.~\ref{fig:high_mas_agb_grid_ecc_porb}, this binary should circularize with our fiducial tidal model but not with Z77. We show the detailed evolution in Fig.~\ref{fig:high_mas_agb}.

On the main sequence, the primary (as well as the secondary) has a convective core and a radiative envelope, as shown in Fig.~\ref{fig:high_mas_agb_kipp}. This leads to entirely dynamical tides from IGW dissipation on the MS, which we can also see in the bar chart in Fig.~\ref{fig:high_mas_agb_tides}.

The primary exits the main sequence after $\sim 362$\,Myr and moves onto the HG phase, during which a convective envelope begins to develop (see Fig.~\ref{fig:high_mas_agb_kipp}). As the convective envelope grows in size, equilibrium tides become increasingly efficient for both the fiducial and the Z77 simulations, shown in Fig.~\ref{fig:high_mas_agb_eq_imk22}. The relevant timescales are shown in Fig.~\ref{fig:high_mas_agb_eq_tau}. 

For the rest of the star's evolution, the presence of an extended convective envelope results in equilibrium tides from viscous dissipation being dominant over dynamical tides from IGW dissipation. The relative dominance of equilibrium tidal dissipation holds for both, the fiducial and the Z77 simulations, and can be observed by comparing Fig.~\ref{fig:high_mas_agb_eq_imk22} and Fig.~\ref{fig:high_mas_agb_eq_tau} to Fig.~\ref{fig:high_mas_agb_dyn_imk22} and Fig.~\ref{fig:high_mas_agb_dyn_tau}. However, our fiducial model is stronger than Z77 by a factor of $\sim10$ during the giant phases. \revision{The convection length scale in our fiducial model from Eq.~\eqref{eq:lc_convective} scales linearly with the depth of the convective envelope. As discussed in Sec.~\ref{sec:equilibrium_tides}, this causes our fiducial model to predict strong equilibrium tides for stars with extended convective envelopes. A more pessimistic estimate of the convection length would lead to weaker predicted equilibrium tides for evolved stars.}

Equilibrium tides for this binary are strong enough to synchronize the rotation of the star to the orbital period once the stars evolve onto the EAGB phase. We can see this by observing the crossing point of the primary spin period (dotted purple curve) and the binary orbital period (solid blue curve) in Fig.~\ref{fig:high_mas_agb_period}, which happens at $\sim 455$\,Myr. As usual, synchronization is associated with the flipping of signs of Im$[k_{2,2}^2]_{\rm Eq}$, which shows up as a bounce at the dashed red line in Fig.~\ref{fig:high_mas_agb_eq_imk22}. Since this binary starts with significant eccentricity, tides can spin the star up past the orbital frequency and up to the pseudo-synchronization frequency. For the next $0.8$\,Myr or so, equilibrium tides continue to become stronger due to the radial expansion of the convective envelope in both the fiducial and the Z77 simulations. Once the eccentricity dissipates as shown in Fig.~\ref{fig:high_mas_agb_ecc}, the pseudo-synchronization limit reduces the the usual synchronization limit and the binary becomes tidally locked at $\Omega_{\rm spin}=\omega_{\rm orb}$. During this phase, the $|\omega_{t,22}|/\omega_c \leq 10^{-2}$ condition is satisfied (shaded red region in Fig.~\ref{fig:high_mas_agb_eq_imk22}), and equilibrium tides in the fiducial model become significantly weaker. Z77 tides, which do not contain a frequency-dependent viscosity, instead continue to become stronger with radial expansion.

Dynamical tides due to IGW dissipation from a convective core boundary in the fiducial model completely die out as the radiative envelope vanishes, as we can observe in Fig.~\ref{fig:high_mas_agb_dyn_imk22}. For the rest of the star's life, the presence of a convective envelope keeps the radiative shell from extending to the surface, a necessary condition for IGW dissipation from convective core-radiative shell boundaries in our fiducial model. 
The $2 \Omega_{\rm spin} \geq \omega_{\rm orb}$ condition for IW dissipation is never satisfied in this binary, so there is no IW dissipation visible in Fig.~\ref{fig:high_mas_agb_dyn_imk22}.
The Z77 binary, on the other hand, experiences dynamical tides in accordance with its overall mass and radius, as outlined in Eq.~\eqref{eq:z77_dyn_imklm} and Eq.~\eqref{eq:z77_dyn_e2}.

Finally, there is a key point in the evolution of our binaries around $458$\,Myr when the mass of the convective envelope begins to become significantly stripped due to stellar winds. The primary expands as a result of mass loss, which further drives stellar winds. During this phase, the primary star loses a significant amount of angular momentum due to winds, due to which $\omega_{t,22} = 2 (\omega_{\rm orb} - \Omega_{\rm spin})$ increases, causing the (2,2) component of equilibrium tides to become stronger. At the same time, the binary gradually expands over time due to the non-conservative mass loss exactly as we observed in the evolutionary tracks in Fig.~\ref{fig:high_mas_agb_grid_ecc_porb}. This behavior is now explicitly visible in Fig.~\ref{fig:high_mas_agb_period}.

\section{Conclusions}
\label{sec:conclusion}

In this work, we have presented a systematic implementation of modern tidal dissipation theory into a rapid binary population synthesis framework. Even with our simplified implementation, we preserve the dominant tidal dissipation behaviors from detailed stellar models across a wide range of stellar and binary evolution scenarios. 

In agreement with recent computations of stars with convective envelopes, we find that equilibrium tides can vary by 1–2 orders of magnitude over binary evolution due to frequency-dependent viscosity. Equilibrium tides provide the dominant contribution for low-mass MS stars and evolved stars with extended convective envelopes (such as AGB, CHeB, etc.), but not solar-type MS stars. We also note that our fiducial implementation of equilibrium tides is typically stronger than the commonly used models from \citet{zahn_tidal_1977} and \citet{hut_tidal_1981}, although this difference is sensitive to underlying uncertainties in the convective length scale and frequency-dependent viscosity reduction.

For stars with radiative cores, such as solar-type MS stars, our fiducial model predicts $6-7$ orders of magnitude stronger dynamical tides when compared to the models from \citet{Zahn:1975A&A....41..329Z} and \citet{Hurley:2002rf}. Whereas those prescriptions do not apply to stars with radiative cores and convective envelopes, our prescription should be consistent with other models that correctly account for stellar structure, such as \citet{Goodman:1998yg, Terquem:1998ya, Ahuir:2021, Esseldeurs2024A&A...690A.266E} among others. With the updated dynamical tides model, we expect solar-type binaries to be circularized more efficiently than the \citet{Zahn:1975A&A....41..329Z} or \citet{Hurley:2002rf} models might suggest. 

The tidal models presented in this work reflect some of the most updated theoretical modeling efforts in the field of binary stellar astrophysics. However, several uncertainties remain to be addressed. As discussed briefly in Sec.~\ref{sec:high_mas_agb}, theoretically derived circularization periods for AGB binaries, including in this work, are still much higher than observations would suggest. Updates to equilibrium tidal dissipation prescriptions may help close this discrepancy. There are also several avenues for improving our models beyond the underlying tidal physics. In particular, the treatment of relative inclinations, high eccentricities (e.g. \citealp{moe2018,Grishin:2021hcp}), and differential rotation remains to be implemented. In the meantime, our model may be used as a self-consistent tidal evolution framework for population synthesis which aims to utilize as much stellar information as available to make physically informed computations regarding tidal dissipation efficiency. Our model should allow for rapid simulations of binary stars with reasonably accurate tidal evolution, allowing for a wide range of population-level predictions. In Paper II of this series, we will examine the implications of these tidal models on populations of compact object binaries and their mass transfer histories, spins, and merger rates.

Data and configuration files for our simulations are available upon reasonable request.

\section{Acknowledgments}
V.K. is grateful to Bore Gao and Morgan MacLeod for helpful discussions.
V.K.~and I.M.~acknowledge support from the Australian Research Council (ARC) Centre of Excellence for Gravitational Wave Discovery (OzGrav), through project number CE230100016. 
V.K. and E.B. are supported by NSF Grants No.~AST-2307146, No.~PHY-2513337, No.~PHY-090003, and No.~PHY-20043, by NASA Grant No.~21-ATP21-0010, by John Templeton Foundation Grant No.~62840, by the Simons Foundation [MPS-SIP-00001698, E.B.], by the Simons Foundation International [SFI-MPS-BH-00012593-02], and by Italian Ministry of Foreign Affairs and International Cooperation Grant No.~PGR01167.
E.G. acknowledges support from the ARC Discovery Early Career Research Award (DECRA) DE260101802 and the ARC Discovery Program DP240103174 (PI: Heger).
Part of the numerical work was carried out at the Advanced Research Computing at Hopkins (ARCH) core facility (\url{https://www.arch.jhu.edu/}), which is supported by the NSF Grant No.~OAC-1920103.  This research was supported in part by grant NSF PHY-2309135 to the Kavli Institute for Theoretical Physics (KITP) and grant PHY-2210452 to the Aspen Center for Physics.

\appendix

\section{Comparisons to literature}
\label{app:comparison}

\subsection{Equilibrium tides}
\label{app:equilibrium_comparison}

We would like to ensure that our expressions for the turbulent viscosity do not sacrifice accuracy for simplicity. It is commonly understood that the turbulent viscosity inside convective regions should be reduced if the convective turnover time $t_c$ is longer than the tidal period $P_{\rm tide}$. For comparison against our fiducial model, we consider two commonly used expressions for the effective turbulent viscosity as presented in \citet{zahn1989tidal}. The first expression comes from \citet{zahn1966marees}, and replaces the mean free path by the distance covered during half a tidal period
\begin{equation}
    \nu (\omega_t)_{\rm Z66} = \frac{1}{3} u_c l_c \min[1, u_c P_{\rm tide} / 2 l_c].
    \label{eq:nu_z66}
\end{equation}
The other approach comes from \citet{goldreich1977turbulent} and introduces a stronger reduction factor to the effective viscosity, which is given by
\begin{equation}
    \nu (\omega_t)_{\rm G77} = \frac{1}{3} u_c l_c \min[1, (u_c P_{\rm tide} / 2 \pi l_c)^2].
    \label{eq:nu_g77}
\end{equation}

To compare our fiducial viscosity model against the above prescriptions, we simulate a binary with mass $M_{1, \rm ZAMS} = M_{2, \rm ZAMS} = 1 M_\odot$, metallicity $Z = 0.02$, and initial orbital period $P_{\rm orb, ZAMS}=1$ day. We then plug the viscosity prescriptions of Eq.~\eqref{eq:nu_z66} and Eq.~\eqref{eq:nu_g77} into Eq.~\eqref{eq:imknm_equilibrium} to compute Im$[k^2_{2,2}]_{\rm Eq}$ under the two reference prescriptions. To compute $P_{\rm tide, 22} = 2 \pi / \omega_{t, 22}$, $u_c$, and $l_c$, we refer to the COMPAS outputs at each time step as per Sec.~\ref{sec:equilibrium_tides}. 

Among all three prescriptions, the \citet{goldreich1977turbulent} model shows the strongest reduction at high tidal frequencies, our fiducial model lies in the middle, and the \citet{zahn1966marees} model has the lowest frequency-dependent suppression.

Our plot may also be compared to Fig.~5 of \citet{Ogilvie:2014dwa}, which shows the frequency-dependent equilibrium tidal Love numbers under the same viscosity prescriptions but for different stellar models. With COMPAS, the equilibrium tidal dissipation is $\sim 1$ order of magnitude stronger than \citet{Ogilvie:2014dwa} across all viscosity prescriptions. This discrepancy is primarily caused by our approximate expression for viscous dissipation in Eq.~\eqref{eq:dnu_integral}, which depends on the turbulent effective viscosity but also other stellar quantities which we cannot access with rapid population synthesis software.

\begin{figure}
    \centering    
    \includegraphics[width=0.98\columnwidth, trim={10 0 10 0}]{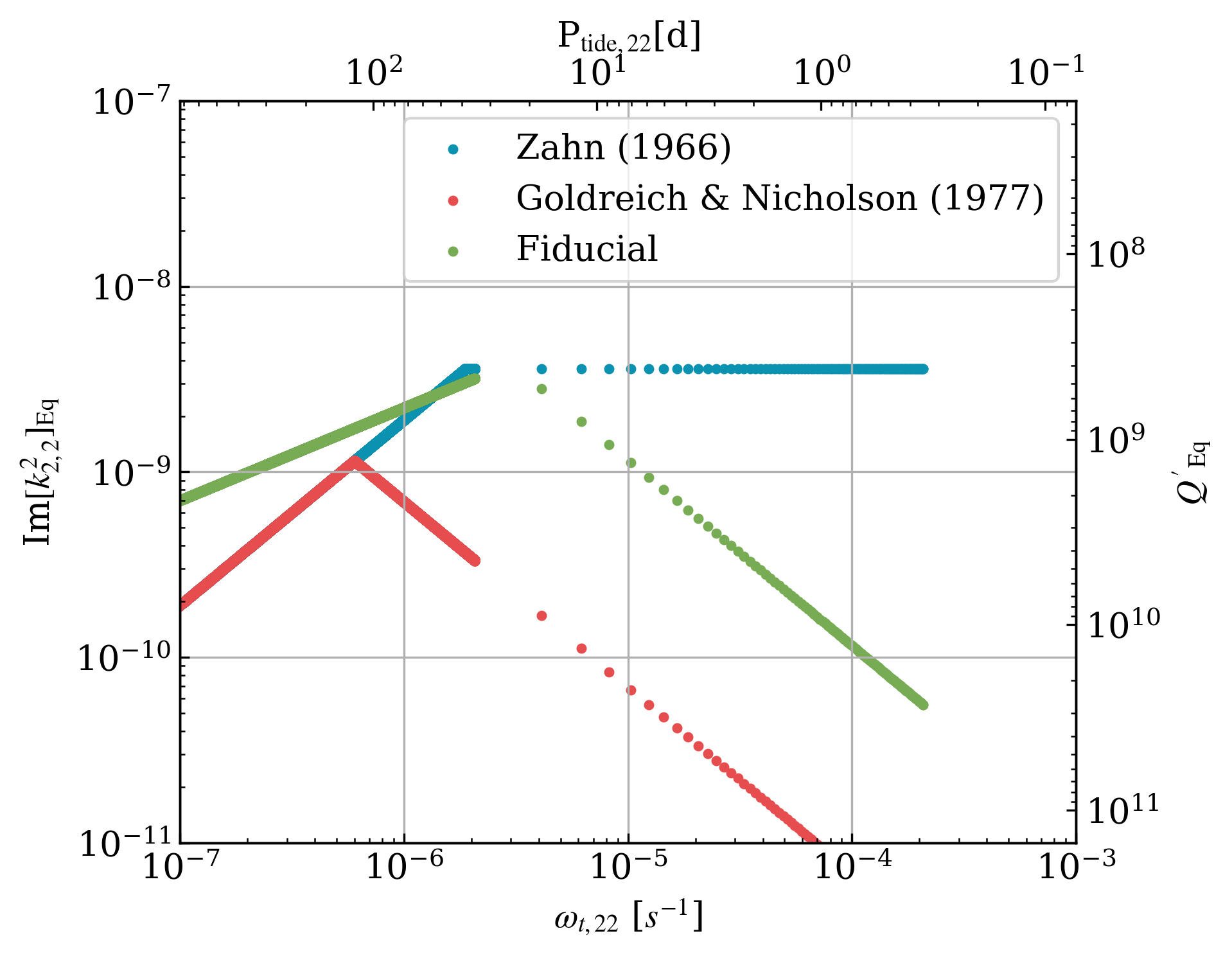}
    \caption{Strength of equilibrium tidal dissipation in our models as a function of the tidal frequency for a binary with initial mass $M_{1, \rm ZAMS} = M_{2, \rm ZAMS} = 1 M_\odot$, metallicity $Z = 0.02$, and initial orbital period $P_{\rm orb, ZAMS}=1$ day. The blue and red points are computed by plugging $\nu$ from Eq.~\eqref{eq:nu_z66} and Eq.~\eqref{eq:nu_g77} in Eq.~\eqref{eq:imknm_equilibrium}, respectively. Each scatter point represents a time step in COMPAS.}
    \label{fig:ogilvie_comparison}
\end{figure}

We can also compare our simulation against \citet{Barker:2020MNRAS.498.2270B}, who, for a $1 M_\odot + 1M_\odot$ system at a tidal period of $P_{\rm tide, 22} = 2 \pi/ \omega_{t, 22} \approx 1$ day, finds that the tidal quality factor should be roughly $10^{10} - 10^{11}$. Looking at Fig.~\ref{fig:ogilvie_comparison}, our tidal quality factor at $P_{\rm tide, 22} = 1$~day is $8.8 \times 10^9$. Our results are nearly compatible with \citet{Barker:2020MNRAS.498.2270B}, if on the lower end of their estimate. We should note that \citet{Penev:2011wj} obtain a much lower tidal quality factor at $P_{\rm tide} = 1$~day in their empirical study. Indeed, \citet{Barker:2020MNRAS.498.2270B} acknowledge that the viscosity prescription remains somewhat uncertain at high frequencies ($P_{\rm tide} \approx 1$\,day), and the tidal predictions can vary by up to an order of magnitude. 

To conclude, our simplified model for viscous dissipation agrees reasonably well with the detailed stellar simulations it tries to emulate, but may be much higher than empirical estimates.

\subsection{Dynamical tides}
\label{app:dynamical_comparison}
As a verification of our dynamical tides implementation for MS stars, we simulate a $1 M_\odot + 0.0009 M_\odot$ system at $P_{\rm orb, ZAMS} = 1$ day to compare against the Sun + Hot Jupiter system studied in \citet{Ahuir:2021}. As in their study, we do not include the effect of tides on the binary separation (and hence, tidal period), and instead only consider the effect of single stellar evolution on the tidal Love number.
To reproduce their results, we must first obtain the spin evolution shown in Fig.~7 of \citet{Ahuir:2021} so that we may correctly estimate the tidal period. As a simple fit to their plot, we construct the following piecewise linear model of the surface rotation rate
\begin{equation}
    \frac{\Omega(t)}{\Omega_\odot} =
    \begin{cases}
    10 + 90\left(\frac{t}{t_0}\right), & t < t_0, \\
    100 - 99 \left(\frac{t}{t_0}\right) & t \ge t_0,
    \end{cases}
\end{equation}
where $t_0 = 4 \times 10^{7}$ yr.
We then compute the resulting tidal dissipation from IGWs, IWs, and viscous dissipation with the stellar and binary properties from COMPAS. 
\begin{figure}
    \centering  \includegraphics[width=0.98\columnwidth, trim={10 0 10 0}]{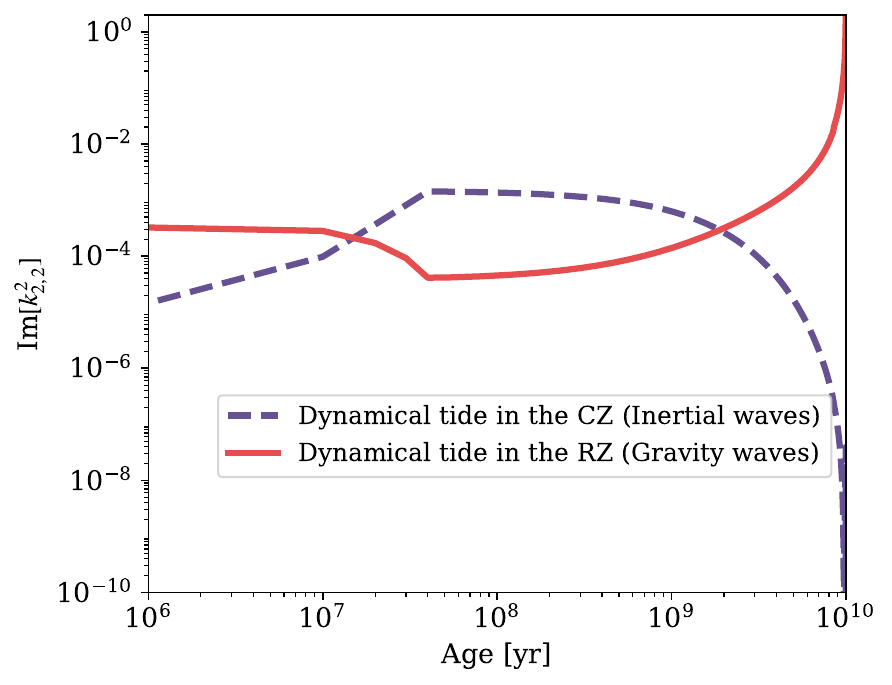}
    \caption{Tidal dissipation through various channels for a Sun + Hot Jupiter binary system. \revision{The dashed and solid lines show the dynamical tide contribution from inertial waves (convective zone) and internal gravity waves (radiative zone, RZ), respectively}.}
    \label{fig:ahuir_comparison}
\end{figure}
The results for IW and IGW tidal contributions are shown in Fig.~\ref{fig:ahuir_comparison}, and should be compared against Fig.~9 of \citet{Ahuir:2021}. On the MS, our results agree very well with their $1 M_\odot$ curve, despite our use of simplified stellar models.

\section{Z77 tidal dissipation model}
\label{app:z77_model}

Several tidal dissipation models are commonly cited in binary population synthesis studies, including \citet{zahn_tidal_1977}, \citet{hut_tidal_1981}, and \citet{Hurley:2002rf}. As pointed out in \citet{Sciarini:2023ylk}, these models may use mutually incompatible assumptions and expressions for tidal timescales. To avoid ambiguity, we will write out the exact equations we use for our Z77 reference model in this work. As far as possible, our reference model will be based on \citet{zahn_tidal_1977}, with occasional help from \citet{Hurley:2002rf} to estimate stellar constants.

\subsection{Equilibrium tides}
The imaginary component of the equilibrium tidal Love number in \citet{zahn_tidal_1977} can be obtained by re-expressing Eq.~(4.1) and Eq.~(4.2) of that work as
\begin{equation}
    {\rm Im}[k_{2,n}^m]_{\rm Eq, Z77} = 2 \left(\frac{k_2}{T} \right)_{\rm c} \frac{R_*^3}{G M_*} (n \omega_{\rm orb} - m \Omega_{\rm spin}).
\end{equation}
Here, $\left(\frac{k_2}{T} \right)_{\rm c}$ is the ratio of the apisidal motion constant $k_2$ to the convective friction time $T$, which encapsulates the strength of viscous dissipation in the star. To estimate this quantity, we turn to \citet{Hurley:2002rf}, who provide the following expression:
\begin{equation}
    \left(\frac{k_2}{T} \right)_{\rm c} = \frac{2}{21} \frac{f_{\rm conv}}{\tau_{\rm conv}} \frac{M_{\rm env}}{M_*}.
\end{equation}
In the equation above,
\begin{equation}
    \tau_{\rm conv} = \left( \frac{M_{\rm env} R_{\rm conv, ext} (R_* - \frac{1}{2} R_{\rm conv, ext})}{3 L_*} \right)^{1/3}
\end{equation}
is the eddy turnover timescale, where $R_{\rm conv, ext}$ is the radial extent of the convective envelope and $L_*$ is the luminosity of the star. The term $f_{\rm conv}$ is calculated using the following equations:
\begin{align}
    f_{\rm conv} &= {\rm min} \left[1, \left( \frac{P_{\rm tide}}{2 \tau_{\rm conv}}\right)^2 \right], \\
    \frac{1}{P_{\rm tide}} &= \left| \frac{1}{P_{\rm orb}}  - \frac{1}{P_{\rm spin}} \right|,
\end{align}
where
\begin{align}
    P_{\rm orb} &= \frac{2 \pi}{\omega_{\rm orb}},\\
    P_{\rm spin} &= \frac{2 \pi}{\Omega_{\rm spin}}.
\end{align}
\revision{By comparing against Eq.~\eqref{eq:nu_z66} and Eq.~\eqref{eq:nu_g77}, the above expression for the frequency-dependent viscosity suppression factor $f_{\rm conv}$ follows the theory from \citet{goldreich1977turbulent}, rather than \citet{zahn1966marees}, where the definition would instead scale as $f_{\rm conv} \sim\min \left[ 1,\left( \frac{P_{\rm tide}}{2 \tau_{\rm conv}}\right) \right]$.} The (2,2) components of the synchronization and circularization timescales for equilibrium tides under the \citet{zahn_tidal_1977} prescription are given by Eq.~(4.6) and Eq.~(4.7) of their work:
\begin{align}
    \frac{1}{\tau_{\rm sync, Eq, 22}} &= 6 \left(\frac{k_2}{t_{\rm F}} \right) q^2 \frac{M_* R_*^2}{I} \left(\frac{R_*}{a}\right)^6, \\
    \frac{1}{\tau_{\rm circ, Eq, 22}} &= \frac{63}{4}\left(\frac{k_2}{t_{\rm F}} \right) q (1+q) \left(\frac{R_*}{a}\right)^8. 
\end{align}
Instead of using the above timescales, we instead calculate the complete secular evolution terms by inputting ${\rm Im}[k_{\ell,n}^m]_{\rm Eq, Z77}$ into our Eqs.~\eqref{eq:tidal_dadt}, \eqref{eq:tidal_dedt}, and \eqref{eq:tidal_dOmegadt}. One can easily verify that the $(2,2)$ components of the resulting timescales under the $e \approx 0$ condition reduce to the above expressions.

As an aside, we must note that \citet{Hurley:2002rf} also provide their own equations of tidal timescales, which should be based on \citet{zahn_tidal_1977}. Although the \citet{Hurley:2002rf} and \citet{zahn_tidal_1977} models agree on their respective (2,2) equilibrium tidal Love number terms (${\rm Im}[k_{2,2}^2]_{\rm Eq}$) by construction, they use different equations to define the synchronization ($\tau_{\rm sync, Eq}$) and circularization ($\tau_{\rm circ, Eq}$) timescales. In this work, we only refer to the Z77 model to avoid confusion.

\subsection{Dynamical tides}
\citet{zahn_tidal_1977} considers the damping of gravity waves excited from the boundary between a convective core and a radiative envelope. The imaginary Love number for the dynamical tides raised by this mechanism follows from Eq.~(5.5) of his paper, and is given by
\begin{equation}
    {\rm Im}[k_{2,n}^m]_{\rm Dyn, Z77} = E_2 s_{nm}^{8/3}
    \label{eq:z77_dyn_imklm}
\end{equation}
where $s_{nm} = \frac{R_*^3}{G M_*}  (n \omega_{\rm orb} - m \Omega_{\rm spin})$, and $E_2$ is a coefficient that is sensitive to the internal structure of the star. Although \citet{zahn_tidal_1977} does not provide an expression for this term, \citet{Hurley:2002rf} fit the following power law equation to Table 1 of \citet{Zahn:1975A&A....41..329Z}
\begin{equation}
    E_2 = 1.592 \times 10^{-9} \left( \frac{M_*}{M_\odot} \right)^{2.84}.
    \label{eq:z77_dyn_e2}
\end{equation}
In the circular limit, the timescales are given by Eq.~(5.6) and Eq.~(5.9) of \citet{zahn_tidal_1977},
\begin{align}
    \frac{1}{\tau_{\rm sync, Dyn, 22}} &= 3 \left(\frac{G M_*}{R_*^3}\right)^{1/2} \frac{M_* R_*^2}{I} E_2 \; q^2 \left(\frac{R_*}{a}\right)^6 s_{22}^{5/3}\\
    \frac{1}{\tau_{\rm circ, Dyn, 22}} &= \frac{21}{2} \left( \frac{G M_*}{R_*^3} \right)^{1/2} q (1+q)^{11/6} E_2 \left(\frac{R_*}{a}\right)^{21/2}. 
\end{align}
Again, we rely instead on Eqs.~\eqref{eq:tidal_dadt}, \eqref{eq:tidal_dedt}, and \eqref{eq:tidal_dOmegadt} to obtain the full equations of secular binary evolution.

We should again note that the tidal prescription provided in \citet{Hurley:2002rf} uses the same definition of the dynamical circularization timescale as \citet{zahn_tidal_1977}. However, their synchronization timescale equation contains an inconsistency \citep{Sciarini:2023ylk}.

\subsection{Summary of differences between the fiducial and Z77 models}
\label{app:fiducial_vs_z77}
\revision{\textit{Equilibrium tides:} Both models treat equilibrium tidal dissipation via turbulent convective viscosity, but they differ in two important respects. First, the suppression of viscosity in the fast-tides regime: the Z77 model, as implemented here following \citet{Hurley:2002rf}, adopts the \citet{goldreich1977turbulent} quadratic suppression $f_{\rm conv} \sim 1$ for slow tides and $f_{\rm conv} \sim (P_{\rm tide}/\tau_{\rm conv})^2$ for fast tides. The fiducial model adopts the frequency-dependent effective viscosity from \citet{Duguid:10.1093/mnras/staa2216} and \citet{vidal2020efficiency}, which should have similar scalings for slow and fast tides, but an additional intermediate scaling of $f_{\rm conv} \sim (P_{\rm tide}/\tau_{\rm conv})^{1/2}$ for a range of tidal frequencies in between. 
Second, the fiducial model computes the dissipation by approximating the dissipation integral over the stellar interior, rather than relying on a global $k_2/T$ factor estimated from envelope-averaged quantities as in \citet{Hurley:2002rf}.}

\revision{\textit{Dynamical tides -- radiative cores and convective envelopes:} The Z77 model does not include dynamical tides from inertial waves (IW) in the convective envelope, as the formalism of \citet{zahn_tidal_1977} was developed for stars with radiative envelopes and convective cores. The fiducial model includes IW dissipation in the convective zone, following \citet{Barker:2010ru} and \citet{Ogilvie:2012gm}, which can exceed equilibrium tide dissipation by orders of magnitude when the tidal frequency lies within the inertial wave range $\omega_{\rm orb} \leq 2\Omega_{\rm spin}$. The fiducial model also uses the prescription from \citet{Ahuir:2021} to model IGW dissipation, which depends on the radial extent, mass, and density of the radiative zone.}

\revision{\textit{Dynamical tides -- convective cores:} For stars with convective cores and radiative envelopes, both models include dynamical tides from internal gravity waves (IGW) excited at the convective-radiative boundary. However, the Z77 model relies on the structural coefficient $E_2$ fit from \citet{Hurley:2002rf} to account for the core mass, which can be inaccurate far from the mass range over which the fit was calibrated. The fiducial model instead follows \citet{Kushnir:2017} which depends explicitly on the convective core radius. The resulting IGW dissipation should be more sensitive to stellar evolution than the Z77 model.}

\bibliography{refs}{}

@ARTICLE{am2025,
       author = {{Alvarado-Montes}, Jaime A. and {Sucerquia}, Mario and {Zuluaga}, Jorge I. and {Schwab}, Christian},
        title = "{Orbital Decay of the Ultra-hot Jupiter TOI-2109b: Tidal Constraints and Transit-timing Analysis}",
      journal = {Astrophys. J.},
         year = 2025,
        month = jul,
       volume = {988},
       number = {1},
          eid = {66},
        pages = {66},
          doi = {10.3847/1538-4357/ade057},
archivePrefix = {arXiv},
       eprint = {2505.18941},
 primaryClass = {astro-ph.EP},
       adsurl = {https://ui.adsabs.harvard.edu/abs/2025ApJ...988...66A}
}

@ARTICLE{moe2018,
       author = {{Moe}, Maxwell and {Kratter}, Kaitlin M.},
        title = "{Dynamical Formation of Close Binaries during the Pre-main-sequence Phase}",
      journal = {Astrophys. J.},
         year = 2018,
        month = feb,
       volume = {854},
       number = {1},
          eid = {44},
        pages = {44},
          doi = {10.3847/1538-4357/aaa6d2},
archivePrefix = {arXiv},
       eprint = {1706.09894},
 primaryClass = {astro-ph.SR},
       adsurl = {https://ui.adsabs.harvard.edu/abs/2018ApJ...854...44M}
}

@article{Hurley:2002rf,
    author = "Hurley, Jarrod R. and Tout, Christopher A. and Pols, Onno R.",
    title = "{Evolution of binary stars and the effect of tides on binary populations}",
    eprint = "astro-ph/0201220",
    archivePrefix = "arXiv",
    doi = "10.1046/j.1365-8711.2002.05038.x",
    journal = "Mon. Not. Roy. Astron. Soc.",
    volume = "329",
    pages = "897",
    year = "2002"
}

@article{hut_tidal_1981,
	title = {Tidal evolution in close binary systems.},
	volume = {99},
	issn = {0004-6361},
	url = {https://ui.adsabs.harvard.edu/abs/1981A&A....99..126H/abstract},
	language = {en},
	urldate = {2023-02-08},
	journal = {Astronomy and Astrophysics},
	author = {Hut, P.},
	month = jun,
	year = {1981},
	pages = {126--140},
}

@article{zahn_tidal_1977,
	title = {Tidal friction in close binary systems.},
	volume = {57},
	issn = {0004-6361},
	language = {en},
	urldate = {2023-02-08},
	journal = {Astronomy and Astrophysics},
	author = {Zahn, J.-P.},
	month = may,
	year = {1977},
	pages = {383--394},
}

@article{Qin:2018vaa,
    author = "Qin, Y. and Fragos, T. and Meynet, G. and Andrews, J. and S\o{}rensen, M. and Song, H. F.",
    title = "{The spin of the second-born black hole in coalescing binary black holes}",
    eprint = "1802.05738",
    archivePrefix = "arXiv",
    primaryClass = "astro-ph.SR",
    doi = "10.1051/0004-6361/201832839",
    journal = "Astron. Astrophys.",
    volume = "616",
    pages = "A28",
    year = "2018"
}

@ARTICLE{Press:1977,
       author = {{Press}, W.~H. and {Teukolsky}, S.~A.},
        title = "{On formation of close binaries by two-body tidal capture.}",
      journal ="Astrophys. J.",
         year = 1977,
        month = apr,
       volume = {213},
        pages = {183-192},
          doi = {10.1086/155143},
       adsurl = {https://ui.adsabs.harvard.edu/abs/1977ApJ...213..183P}
}

@article{Sciarini:2023ylk,
    author = {Sciarini, Luca and Ekstr\"om, Sylvia and Eggenberger, Patrick and Meynet, Georges and Fragos, Tassos and Song, Han Feng},
    title = "{Dynamical tides in binaries: Inconsistencies in the implementation of Zahn\textquoteright{}s prescription}",
    eprint = "2312.08437",
    archivePrefix = "arXiv",
    primaryClass = "astro-ph.SR",
    reportNumber = "aa48424-23",
    doi = "10.1051/0004-6361/202348424",
    journal = "Astron. Astrophys.",
    volume = "681",
    pages = "L1",
    year = "2024"
}

@article{Grishin:2021hcp,
    author = "Grishin, Evgeni and Perets, Hagai B.",
    title = "{Chaotic dynamics of wide triples induced by galactic tides: a novel channel for producing compact binaries, mergers, and collisions}",
    eprint = "2112.11475",
    archivePrefix = "arXiv",
    primaryClass = "astro-ph.SR",
    doi = "10.1093/mnras/stac706",
    journal = "Mon. Not. Roy. Astron. Soc.",
    volume = "512",
    number = "4",
    pages = "4993--5009",
    year = "2022"
}

@article{Kushnir:2017,
    author = {Kushnir, Doron and Zaldarriaga, Matias and Kollmeier, Juna A. and Waldman, Roni},
    title = "{Dynamical tides reexpressed}",
    journal = {Monthly Notices of the Royal Astronomical Society},
    volume = {467},
    number = {2},
    pages = {2146-2149},
    year = {2017},
    month = {02},
    issn = {0035-8711},
    doi = {10.1093/mnras/stx255},
    url = {https://doi.org/10.1093/mnras/stx255},
    eprint = {https://academic.oup.com/mnras/article-pdf/467/2/2146/10874796/stx255.pdf},
}

@article{Ahuir:2021,
	author = {Ahuir, J. and Mathis, S. and Amard, L.},
	doi = {10.1051/0004-6361/202040174},
	journal = {Astronomy and Astrophysics},
	pages = {A3},
	title = {Dynamical tide in stellar radiative zones - General formalism and evolution for low-mass stars},
	url = {https://doi.org/10.1051/0004-6361/202040174},
	volume = 651,
	year = 2021}

@article{Ogilvie:2014dwa,
    author = "Ogilvie, Gordon I.",
    title = "{Tidal dissipation in stars and giant planets}",
    eprint = "1406.2207",
    archivePrefix = "arXiv",
    primaryClass = "astro-ph.SR",
    doi = "10.1146/annurev-astro-081913-035941",
    journal = "Ann. Rev. Astron. Astrophys.",
    volume = "52",
    pages = "171--210",
    year = "2014"
}

@ARTICLE{Barker:2020MNRAS.498.2270B,
       author = {{Barker}, A.~J.},
        title = "{Tidal dissipation in evolving low-mass and solar-type stars with predictions for planetary orbital decay}",
      journal = {Monthly Notices of the Royal Astronomical Society},
         year = 2020,
        month = oct,
       volume = {498},
       number = {2},
        pages = {2270-2294},
          doi = {10.1093/mnras/staa2405},
archivePrefix = {arXiv},
       eprint = {2008.03262},
 primaryClass = {astro-ph.EP},
       adsurl = {https://ui.adsabs.harvard.edu/abs/2020MNRAS.498.2270B}
}

@article{Duguid:10.1093/mnras/staa2216,
    author = {Duguid, Craig D and Barker, Adrian J and Jones, C A},
    title = "{Convective turbulent viscosity acting on equilibrium tidal flows: new frequency scaling of the effective viscosity}",
    journal = {Monthly Notices of the Royal Astronomical Society},
    volume = {497},
    number = {3},
    pages = {3400-3417},
    year = {2020},
    month = {07},
    issn = {0035-8711},
    doi = {10.1093/mnras/staa2216},
    url = {https://doi.org/10.1093/mnras/staa2216},
    eprint = {https://academic.oup.com/mnras/article-pdf/497/3/3400/33648559/staa2216.pdf},
}

@article{Hurley:2000pk,
    author = "Hurley, Jarrod R. and Pols, Onno R. and Tout, Christopher A.",
    title = "{Comprehensive analytic formulae for stellar evolution as a function of mass and metallicity}",
    eprint = "astro-ph/0001295",
    archivePrefix = "arXiv",
    doi = "10.1046/j.1365-8711.2000.03426.x",
    journal = "Mon. Not. Roy. Astron. Soc.",
    volume = "315",
    pages = "543",
    year = "2000"
}

@article{Picker:2024vqb,
    author = "Picker, Lewis and Hirai, Ryosuke and Mandel, Ilya",
    title = "{Fits for the Convective Envelope Mass in Massive Stars}",
    eprint = "2402.13180",
    archivePrefix = "arXiv",
    primaryClass = "astro-ph.SR",
    doi = "10.3847/1538-4357/ad4a5d",
    journal = "Astrophys. J.",
    volume = "969",
    number = "1",
    pages = "1",
    year = "2024"
}

@article{Ogilvie:2012gm,
    author = "Ogilvie, Gordon I.",
    title = "{Tides in rotating barotropic fluid bodies: the contribution of inertial waves and the role of internal structure}",
    eprint = "1211.0837",
    archivePrefix = "arXiv",
    primaryClass = "astro-ph.EP",
    doi = "10.1093/mnras/sts362",
    journal = "Mon. Not. Roy. Astron. Soc.",
    volume = "429",
    pages = "613",
    year = "2013"
}

@ARTICLE{Zahn:1975A&A....41..329Z,
    author = {{Zahn}, J. -P.},
    title = "{The dynamical tide in close binaries.}",
    journal = {Astronomy and Astrophysics},
    year = 1975,
    month = jul,
    volume = {41},
    pages = {329-344}
}

@article{Hansen:2010nq,
    author = "Hansen, Brad",
    title = "{Calibration of Equilibrium Tide Theory for Extrasolar Planet Systems}",
    eprint = "1009.3027",
    archivePrefix = "arXiv",
    primaryClass = "astro-ph.SR",
    doi = "10.1088/0004-637X/723/1/285",
    journal = "Astrophys. J.",
    volume = "723",
    pages = "285--299",
    year = "2010"
}

@article{Love1909,
    author = {Love, A. E. H.},
    title = "{The yielding of the Earth to disturbing forces}",
    journal = {Monthly Notices of the Royal Astronomical Society},
    volume = {69},
    number = {6},
    pages = {476-480},
    year = {1909},
    month = {04},
    issn = {0035-8711},
    doi = {10.1093/mnras/69.6.476},
    url = {https://doi.org/10.1093/mnras/69.6.476},
    eprint = {https://academic.oup.com/mnras/article-pdf/69/6/476/3236445/mnras69-0476.pdf},
}

@article{Gaia:2018ydn,
    author = "Brown, A. G. A. and others",
    collaboration = "Gaia",
    title = "{Gaia Data Release 2}: {Summary of the contents and survey properties}",
    eprint = "1804.09365",
    archivePrefix = "arXiv",
    primaryClass = "astro-ph.GA",
    doi = "10.1051/0004-6361/201833051",
    journal = "Astron. Astrophys.",
    volume = "616",
    pages = "A1",
    year = "2018"
}

@article{Meibom_2005,
doi = {10.1086/427082},
url = {https://dx.doi.org/10.1086/427082},
year = {2005},
month = {feb},
publisher = {},
volume = {620},
number = {2},
pages = {970},
author = {Søren Meibom and Robert D. Mathieu},
title = {A Robust Measure of Tidal Circularization in Coeval Binary Populations: The Solar-Type Spectroscopic Binary Population in the Open Cluster M35*},
journal = {The Astrophysical Journal}
}

@article{Behr_2011,
doi = {10.1088/0004-6256/142/1/6},
url = {https://dx.doi.org/10.1088/0004-6256/142/1/6},
year = {2011},
month = {may},
publisher = {The American Astronomical Society},
volume = {142},
number = {1},
pages = {6},
author = {Bradford B. Behr and Andrew T. Cenko and Arsen R. Hajian and Robert S. McMillan and Marc Murison and Jeff Meade and Robert Hindsley},
title = {STELLAR ASTROPHYSICS WITH A DISPERSED FOURIER TRANSFORM SPECTROGRAPH. II. ORBITS OF DOUBLE-LINED SPECTROSCOPIC BINARIES},
journal = {The Astronomical Journal}
}

@article{prevot1961vitesses,
  title={Vitesses radiales et {\'e}l{\'e}ments orbitaux de zeta1 Ursae Majoris},
  author={Prevot, L},
  journal={Journal des Observateurs, Vol. 44, p. 83},
  volume={44},
  pages={83},
  year={1961}
}

@article{Pourbaix2000,
	author = {{Pourbaix, D.}},
	title = {Resolved double-lined spectroscopic binaries: A neglected source of hypothesis-free
parallaxes and stellar masses},
	DOI= "10.1051/aas:2000237",
	url= "https://doi.org/10.1051/aas:2000237",
	journal = {Astron. Astrophys. Sup.},
	year = 2000,
	volume = 145,
	number = 2,
	pages = "215-222",
}

@article{Shikauchi:2024yqd,
    author = "Shikauchi, Minori and Hirai, Ryosuke and Mandel, Ilya",
    title = "{Evolution of the Convective Core Mass during the Main Sequence}",
    eprint = "2409.00460",
    archivePrefix = "arXiv",
    primaryClass = "astro-ph.SR",
    doi = "10.3847/1538-4357/adc5fa",
    journal = "Astrophys. J.",
    volume = "984",
    number = "2",
    pages = "149",
    year = "2025"
}

@article{Fuller:2017bzf,
    author = "Fuller, Jim",
    title = "{Heartbeat Stars, Tidally Excited Oscillations, and Resonance Locking}",
    eprint = "1706.05054",
    archivePrefix = "arXiv",
    primaryClass = "astro-ph.SR",
    doi = "10.1093/mnras/stx2135",
    journal = "Mon. Not. Roy. Astron. Soc.",
    volume = "472",
    number = "2",
    pages = "1538--1564",
    year = "2017"
}

@article{Penev_2018,
doi = {10.3847/1538-3881/aaaf71},
url = {https://dx.doi.org/10.3847/1538-3881/aaaf71},
year = {2018},
month = {mar},
publisher = {The American Astronomical Society},
volume = {155},
number = {4},
pages = {165},
author = {Penev, Kaloyan and Bouma, L. G. and Winn, Joshua N. and Hartman, Joel D.},
title = {Empirical Tidal Dissipation in Exoplanet Hosts From Tidal Spin-up},
journal = {The Astronomical Journal}
}

@article{El-Badry:2024vjt,
    author = "El-Badry, Kareem",
    title = "{Gaia\textquoteright{}s binary star renaissance}",
    eprint = "2403.12146",
    archivePrefix = "arXiv",
    primaryClass = "astro-ph.SR",
    doi = "10.1016/j.newar.2024.101694",
    journal = "New Astron. Rev.",
    volume = "98",
    pages = "101694",
    year = "2024"
}

@article{Fragos:2022jik,
    author = "Fragos, Tassos and others",
    title = "{POSYDON: A General-purpose Population Synthesis Code with Detailed Binary-evolution Simulations}",
    eprint = "2202.05892",
    archivePrefix = "arXiv",
    primaryClass = "astro-ph.SR",
    doi = "10.3847/1538-4365/ac90c1",
    journal = "Astrophys. J. Suppl.",
    volume = "264",
    number = "2",
    pages = "45",
    year = "2023"
}

@article{COMPASTeam:2021tbl,
    author = "{Team COMPAS: J. Riley} and others",
    title = "{Rapid Stellar and Binary Population Synthesis with COMPAS}",
    eprint = "2109.10352",
    archivePrefix = "arXiv",
    primaryClass = "astro-ph.IM",
    doi = "10.3847/1538-4365/ac416c",
    journal = "Astrophys. J. Supp.",
    volume = "258",
    number = "2",
    pages = "34",
    year = "2022"
}

@ARTICLE{COMPAS:2025,
       author = {{Team COMPAS: I. Mandel} and {Riley}, Jeff and {Boesky}, Adam and {Brcek}, Adam and {Hirai}, Ryosuke and {Kapil}, Veome and {Lau}, Mike Y.~M. and {Merritt}, JD and {Rodr{\'\i}guez-Segovia}, Nicol{\'a}s and {Romero-Shaw}, Isobel and {Song}, Yuzhe and {Stevenson}, Simon and {Vajpeyi}, Avi and {van Son}, L.~A.~C. and {Vigna-G{\'o}mez}, Alejandro and {Willcox}, Reinhold},
        title = "{Rapid Stellar and Binary Population Synthesis with COMPAS: Methods Paper II}",
      journal = {Astrophys. J. Supp. S.},
         year = 2025,
        month = sep,
       volume = {280},
       number = {1},
          eid = {43},
        pages = {43},
          doi = {10.3847/1538-4365/adf8d0},
archivePrefix = {arXiv},
       eprint = {2506.02316},
 primaryClass = {astro-ph.SR},
       adsurl = {https://ui.adsabs.harvard.edu/abs/2025ApJS..280...43T}
}

@article{Mirouh2023:10.1093/mnras/stad2048,
    author = {Mirouh, Giovanni M and Hendriks, David D and Dykes, Sophie and Moe, Maxwell and Izzard, Robert G},
    title = {Detailed equilibrium and dynamical tides: impact on circularization and synchronization in open clusters},
    journal = {Monthly Notices of the Royal Astronomical Society},
    volume = {524},
    number = {3},
    pages = {3978-3999},
    year = {2023},
    month = {07},
    issn = {0035-8711},
    doi = {10.1093/mnras/stad2048},
    url = {https://doi.org/10.1093/mnras/stad2048},
    eprint = {https://academic.oup.com/mnras/article-pdf/524/3/3978/50968227/stad2048.pdf},
}

@ARTICLE{Esseldeurs2024A&A...690A.266E,
       author = {{Esseldeurs}, M. and {Mathis}, S. and {Decin}, L.},
        title = "{Tidal dissipation in evolved low- and intermediate-mass stars}",
      journal = {Astron. Astrophys.},
         year = 2024,
        month = oct,
       volume = {690},
          eid = {A266},
        pages = {A266},
          doi = {10.1051/0004-6361/202449648},
archivePrefix = {arXiv},
       eprint = {2407.10573},
 primaryClass = {astro-ph.SR},
       adsurl = {https://ui.adsabs.harvard.edu/abs/2024A&A...690A.266E}
}

@article{LIGOScientific:2018mvr,
    author = "Abbott, B. P. and others",
    collaboration = "LIGO Scientific, Virgo",
    title = "{GWTC-1: A Gravitational-Wave Transient Catalog of Compact Binary Mergers Observed by LIGO and Virgo during the First and Second Observing Runs}",
    eprint = "1811.12907",
    archivePrefix = "arXiv",
    primaryClass = "astro-ph.HE",
    reportNumber = "LIGO-P1800307",
    doi = "10.1103/PhysRevX.9.031040",
    journal = "Phys. Rev. X",
    volume = "9",
    number = "3",
    pages = "031040",
    year = "2019"
}

@article{LIGOScientific:2020ibl,
    author = "Abbott, R. and others",
    collaboration = "LIGO Scientific, Virgo",
    title = "{GWTC-2: Compact Binary Coalescences Observed by LIGO and Virgo During the First Half of the Third Observing Run}",
    eprint = "2010.14527",
    archivePrefix = "arXiv",
    primaryClass = "gr-qc",
    reportNumber = "P2000061",
    doi = "10.1103/PhysRevX.11.021053",
    journal = "Phys. Rev. X",
    volume = "11",
    pages = "021053",
    year = "2021"
}

@article{KAGRA:2021vkt,
    author = "Abbott, R. and others",
    collaboration = "KAGRA, VIRGO, LIGO Scientific",
    title = "{GWTC-3: Compact Binary Coalescences Observed by LIGO and Virgo during the Second Part of the Third Observing Run}",
    eprint = "2111.03606",
    archivePrefix = "arXiv",
    primaryClass = "gr-qc",
    reportNumber = "LIGO-P2000318",
    doi = "10.1103/PhysRevX.13.041039",
    journal = "Phys. Rev. X",
    volume = "13",
    number = "4",
    pages = "041039",
    year = "2023"
}

@article{Idini_2021,
doi = {10.3847/PSJ/abe715},
url = {https://dx.doi.org/10.3847/PSJ/abe715},
year = {2021},
month = {apr},
publisher = {The American Astronomical Society},
volume = {2},
number = {2},
pages = {69},
author = {Idini, Benjamin and Stevenson, David J.},
title = {Dynamical Tides in Jupiter as Revealed by Juno},
journal = {The Planetary Science Journal}
}

@article{Dewberry_2022,
doi = {10.3847/1538-4357/ac3ede},
url = {https://dx.doi.org/10.3847/1538-4357/ac3ede},
year = {2022},
month = {feb},
publisher = {The American Astronomical Society},
volume = {925},
number = {2},
pages = {124},
author = {Dewberry, Janosz W. and Lai, Dong},
title = {Dynamical Tidal Love Numbers of Rapidly Rotating Planets and Stars},
journal = {The Astrophysical Journal}
}

@article{Barker:2010ru,
    author = "Barker, A. J. and Ogilvie, G. I.",
    title = "{On internal wave breaking and tidal dissipation near the centre of a solar-type star}",
    eprint = "1001.4009",
    archivePrefix = "arXiv",
    primaryClass = "astro-ph.EP",
    doi = "10.1111/j.1365-2966.2010.16400.x",
    journal = "Mon. Not. Roy. Astron. Soc.",
    volume = "404",
    pages = "1849",
    year = "2010"
}

@ARTICLE{Maxted2023Univ....9..498M,
       author = {{Maxted}, Pierre F.~L. and {Triaud}, Amaury H.~M.~J. and {Martin}, David V.},
        title = "{The EBLM Project{\textemdash}From False Positives to Benchmark Stars and Circumbinary Exoplanets}",
      journal = {Universe},
         year = 2023,
        month = nov,
       volume = {9},
       number = {12},
          eid = {498},
        pages = {498},
          doi = {10.3390/universe9120498},
archivePrefix = {arXiv},
       eprint = {2311.16677},
 primaryClass = {astro-ph.SR},
       adsurl = {https://ui.adsabs.harvard.edu/abs/2023Univ....9..498M}
}

@article{Chabrier:1997vx,
    author = "Chabrier, Gilles and Baraffe, Isabelle",
    title = "{Structure and evolution of low-mass stars}",
    eprint = "astro-ph/9704118",
    archivePrefix = "arXiv",
    journal = "Astron. Astrophys.",
    volume = "327",
    pages = "1039--1053",
    year = "1997"
}

@article{Raghavan2010survey,
  title={A survey of stellar families: multiplicity of solar-type stars},
  author={Raghavan, Deepak and McAlister, Harold A and Henry, Todd J and Latham, David W and Marcy, Geoffrey W and Mason, Brian D and Gies, Douglas R and White, Russel J and Ten Brummelaar, Theo A},
  journal={The Astrophysical Journal Supplement Series},
  volume={190},
  number={1},
  pages={1},
  year={2010},
  publisher={IOP Publishing}
}

@article{bluhm2016new,
  title={New spectroscopic binary companions of giant stars and updated metallicity distribution for binary systems},
  author={Bluhm, P and Jones, MI and Vanzi, Leonardo and Soto, MG and Vos, J and Wittenmyer, RA and Drass, Holger and Jenkins, James Stewart and Olivares, F and Mennickent, RE and others},
  journal={Astronomy \& Astrophysics},
  volume={593},
  pages={A133},
  year={2016},
  publisher={EDP Sciences}
}

@article{verbunt1995tidal,
  title={Tidal circularization and the eccentricity of binaries containing giant stars.},
  author={Verbunt, F and Phinney, ES},
  journal={Astronomy and Astrophysics, v. 296, p. 709},
  volume={296},
  pages={709},
  year={1995}
}

@article{Izzard:2010cu,
    author = "Izzard, Robert G. and Dermine, Tyl and Church, Ross P.",
    title = "{White-Dwarf Kicks and Implications for Barium Stars}",
    eprint = "1008.3818",
    archivePrefix = "arXiv",
    primaryClass = "astro-ph.SR",
    doi = "10.1051/0004-6361/201015254",
    journal = "Astron. Astrophys.",
    volume = "523",
    pages = "A10",
    year = "2010"
}

@inproceedings{pols2003can,
  title={Can Standard Evolution Models Explain the Properties of Barium Stars?},
  author={Pols, OR and Karakas, AI and Lattanzio, JC and Tout, CA},
  booktitle={Symbiotic Stars Probing Stellar Evolution},
  volume={303},
  pages={290},
  year={2003}
}

@article{Bavera:2020inc,
    author = "Bavera, Simone S. and Fragos, Tassos and Qin, Ying and Zapartas, Emmanouil and Neijssel, Coenraad J. and Mandel, Ilya and Batta, Aldo and Gaebel, Sebastian M. and Kimball, Chase and Stevenson, Simon",
    title = "{The origin of spin in binary black holes: Predicting the distributions of the main observables of Advanced LIGO}",
    eprint = "1906.12257",
    archivePrefix = "arXiv",
    primaryClass = "astro-ph.HE",
    doi = "10.1051/0004-6361/201936204",
    journal = "Astron. Astrophys.",
    volume = "635",
    pages = "A97",
    year = "2020"
}

@article{Steinle:2022rhj,
    author = "Steinle, Nathan and Kesden, Michael",
    title = "{Signatures of spin precession and nutation in isolated black-hole binaries}",
    eprint = "2206.00391",
    archivePrefix = "arXiv",
    primaryClass = "astro-ph.HE",
    doi = "10.1103/PhysRevD.106.063028",
    journal = "Phys. Rev. D",
    volume = "106",
    number = "6",
    pages = "063028",
    year = "2022"
}

@article{Levine:1999cy,
    author = "Levine, Alan M. and Rappaport, Saul A. and Zojcheski, Goce",
    title = "{Orbital decay in lmc x-4}",
    eprint = "astro-ph/9911173",
    archivePrefix = "arXiv",
    reportNumber = "CSR-99-18",
    doi = "10.1086/309398",
    journal = "Astrophys. J.",
    volume = "541",
    pages = "194",
    year = "2000"
}

@article{mathis2015variation,
  title={Variation of tidal dissipation in the convective envelope of low-mass stars along their evolution},
  author={Mathis, St{\'e}phane},
  journal={Astronomy \& Astrophysics},
  volume={580},
  pages={L3},
  year={2015},
  publisher={EDP Sciences}
}

@article{Goodman:1998yg,
    author = "Goodman, Jeremy and Dickson, Eric S.",
    title = "{Dynamical tide in solar-type binaries}",
    eprint = "astro-ph/9801289",
    archivePrefix = "arXiv",
    reportNumber = "POPE-744",
    doi = "10.1086/306348",
    journal = "Astrophys. J.",
    volume = "507",
    pages = "938",
    year = "1998"
}

@article{Belczynski:2005mr,
    author = "Belczynski, Krzystof and Kalogera, Vassiliki and Rasio, Frederic A. and Taam, Ronald E. and Zezas, Andreas and Bulik, Tomasz and Maccarone, Thomas J. and Ivanova, Natalia",
    title = "{Compact object modeling with the startrack population synthesis code}",
    eprint = "astro-ph/0511811",
    archivePrefix = "arXiv",
    doi = "10.1086/521026",
    journal = "Astrophys. J. Suppl.",
    volume = "174",
    pages = "223",
    year = "2008"
}

@article{Paxton:2010ji,
    author = "Paxton, Bill and Bildsten, Lars and Dotter, Aaron and Herwig, Falk and Lesaffre, Pierre and Timmes, Frank",
    collaboration = "MESA",
    title = "{Modules for Experiments in Stellar Astrophysics (MESA)}",
    eprint = "1009.1622",
    archivePrefix = "arXiv",
    primaryClass = "astro-ph.SR",
    doi = "10.1088/0067-0049/192/1/3",
    journal = "Astrophys. J. Suppl.",
    volume = "192",
    pages = "3",
    year = "2011"
}

@article{Paxton:2013pj,
    author = "Paxton, Bill and others",
    title = "{Modules for Experiments in Stellar Astrophysics (MESA): Planets, Oscillations, Rotation, and Massive Stars}",
    eprint = "1301.0319",
    archivePrefix = "arXiv",
    primaryClass = "astro-ph.SR",
    doi = "10.1088/0067-0049/208/1/4",
    journal = "Astrophys. J. Suppl.",
    volume = "208",
    pages = "4",
    year = "2013"
}

@article{Siess:2000hq,
    author = "Siess, L. and Dufour, E. and Forestini, M.",
    title = "{An internet server for update pre-main sequence tracks of low- and intermediate-mass stars}",
    eprint = "astro-ph/0003477",
    archivePrefix = "arXiv",
    journal = "Astron. Astrophys.",
    volume = "358",
    pages = "593--599",
    year = "2000"
}

@article{amard2019first,
  title={First grids of low-mass stellar models and isochrones with self-consistent treatment of rotation-From 0.2 to 1.5 M⊙ at seven metallicities from PMS to TAMS},
  author={Amard, Louis and Palacios, Ana and Charbonnel, Corinne and Gallet, F and Georgy, Cyril and Lagarde, Nadege and Siess, Lionel},
  journal={Astronomy \& Astrophysics},
  volume={631},
  pages={A77},
  year={2019},
  publisher={EDP Sciences}
}

@article{claret1997circularization,
  title={Circularization and synchronization times in Main-Sequence of detached eclipsing binaries II. Using the formalisms by Zahn.},
  author={Claret, A and Cunha, NCS},
  journal={Astronomy and Astrophysics, v. 318, p. 187-197},
  volume={318},
  pages={187--197},
  year={1997}
}

@article{barker2022tidal,
  title={Tidal dissipation due to inertial waves can explain the circularization periods of solar-type binaries},
  author={Barker, Adrian J},
  journal={The Astrophysical Journal Letters},
  volume={927},
  number={2},
  pages={L36},
  year={2022},
  publisher={IOP Publishing}
}

@article{zahn1989tidal,
  title={Tidal evolution of close binary stars. II-Orbital circularization of late-type binaries},
  author={Zahn, J-P and Bouchet, L},
  journal={Astronomy and Astrophysics (ISSN 0004-6361), vol. 223, no. 1-2, Oct. 1989, p. 112-118.},
  volume={223},
  pages={112--118},
  year={1989}
}

@article{Terquem:1998ya,
    author = "Terquem, C. and Papaloizou, J. C. B. and Nelson, R. P. and Lin, D. N. C.",
    title = "{On the tidal interaction of a solar-type star with an orbiting companion: excitation of g mode oscillation and orbital evolution}",
    eprint = "astro-ph/9801280",
    archivePrefix = "arXiv",
    doi = "10.1086/305927",
    journal = "Astrophys. J.",
    volume = "502",
    pages = "788--801",
    year = "1998"
}

@article{Ogilvie:2007tja,
    author = "Ogilvie, Gordon I. and Lin, D. N. C.",
    title = "{Tidal dissipation in rotating solar-type stars}",
    eprint = "astro-ph/0702492",
    archivePrefix = "arXiv",
    doi = "10.1086/515435",
    journal = "Astrophys. J.",
    volume = "661",
    pages = "1180--1191",
    year = "2007"
}

@article{geller2021stellar,
  title={Stellar Radial Velocities in the Old Open Cluster M67 (NGC 2682). II. The Spectroscopic Binary Population},
  author={Geller, Aaron M and Mathieu, Robert D and Latham, David W and Pollack, Maxwell and Torres, Guillermo and Leiner, Emily M},
  journal={The Astronomical Journal},
  volume={161},
  number={4},
  pages={190},
  year={2021},
  publisher={IOP Publishing}
}

@article{hambleton2018kic,
  title={KIC 8164262: a heartbeat star showing tidally induced pulsations with resonant locking},
  author={Hambleton, K and Fuller, J and Thompson, S and Pr{\v{s}}a, A and Kurtz, Donald Wayne and Shporer, A and Isaacson, H and Howard, AW and Endl, M and Cochran, W and others},
  journal={Monthly Notices of the Royal Astronomical Society},
  volume={473},
  number={4},
  pages={5165--5176},
  year={2018},
  publisher={Oxford University Press}
}

@article{dawson2018origins,
  title={Origins of hot Jupiters},
  author={Dawson, Rebekah I and Johnson, John Asher},
  journal={Annual Review of Astronomy and Astrophysics},
  volume={56},
  number={1},
  pages={175--221},
  year={2018},
  publisher={Annual Reviews}
}

@article{tokovinin2020orbits,
  title={Orbits of Five Triple Stars},
  author={Tokovinin, Andrei and Latham, David W},
  journal={The Astronomical Journal},
  volume={160},
  number={6},
  pages={251},
  year={2020},
  publisher={IOP Publishing}
}

@article{vallenari2023gaia,
  title={Gaia data release 3-summary of the content and survey properties},
  author={Vallenari, Antonella and Brown, Anthony GA and Prusti, Timo and De Bruijne, Jos HJ and Arenou, F and Babusiaux, Carine and Biermann, Michael and Creevey, Orlagh L and Ducourant, Christine and Evans, Dafydd Wyn and others},
  journal={Astronomy \& Astrophysics},
  volume={674},
  pages={A1},
  year={2023},
  publisher={EDP sciences}
}

@article{Moe:2017icj,
    author = "Moe, Maxwell and Di Stefano, Rosanne",
    title = "{Mind Your Ps and Qs: The Interrelation between Period (P) and Mass-ratio (Q) Distributions of Binary Stars}",
    eprint = "1606.05347",
    archivePrefix = "arXiv",
    primaryClass = "astro-ph.SR",
    doi = "10.3847/1538-4365/aa6fb6",
    journal = "Astrophys. J. Suppl.",
    volume = "230",
    number = "2",
    pages = "15",
    year = "2017"
}

@article{Gerosa:2013laa,
    author = "Gerosa, Davide and Kesden, Michael and Berti, Emanuele and O'Shaughnessy, Richard and Sperhake, Ulrich",
    title = "{Resonant-plane locking and spin alignment in stellar-mass black-hole binaries: a diagnostic of compact-binary formation}",
    eprint = "1302.4442",
    archivePrefix = "arXiv",
    primaryClass = "gr-qc",
    doi = "10.1103/PhysRevD.87.104028",
    journal = "Phys. Rev. D",
    volume = "87",
    pages = "104028",
    year = "2013"
}

@article{Gerosa:2018wbw,
    author = "Gerosa, Davide and Berti, Emanuele and O'Shaughnessy, Richard and Belczynski, Krzysztof and Kesden, Michael and Wysocki, Daniel and Gladysz, Wojciech",
    title = "{Spin orientations of merging black holes formed from the evolution of stellar binaries}",
    eprint = "1808.02491",
    archivePrefix = "arXiv",
    primaryClass = "astro-ph.HE",
    doi = "10.1103/PhysRevD.98.084036",
    journal = "Phys. Rev. D",
    volume = "98",
    number = "8",
    pages = "084036",
    year = "2018"
}

@article{Kumar:1996dw,
    author = "Kumar, Pawan and Quataert, Eliot J.",
    title = "{Differential rotation enhanced dissipation of tides in the psr j0045-7319 binary}",
    eprint = "astro-ph/9611005",
    archivePrefix = "arXiv",
    doi = "10.1086/310573",
    journal = "Astrophys. J. Lett.",
    volume = "479",
    pages = "L51",
    year = "1997"
}

@article{Penev:2011wj,
    author = "Penev, Kaloyan and Sasselov, Dimitar",
    title = "{Tidal Evolution of Close-in Extrasolar Planets: High Stellar Q from New Theoretical Models}",
    eprint = "1102.3187",
    archivePrefix = "arXiv",
    primaryClass = "astro-ph.SR",
    doi = "10.1088/0004-637X/731/1/67",
    journal = "Astrophys. J.",
    volume = "731",
    pages = "67",
    year = "2011"
}

@article{vidal2020efficiency,
  title={Efficiency of tidal dissipation in slowly rotating fully convective stars or planets},
  author={Vidal, J{\'e}r{\'e}mie and Barker, Adrian J},
  journal={Monthly Notices of the Royal Astronomical Society},
  volume={497},
  number={4},
  pages={4472--4485},
  year={2020},
  publisher={Oxford University Press}
}

@article{goldreich1977turbulent,
  title={Turbulent viscosity and Jupiter's tidal Q},
  author={Goldreich, Peter and Nicholson, Philip D},
  journal={Icarus},
  volume={30},
  number={2},
  pages={301--304},
  year={1977},
  publisher={Elsevier}
}

@article{zahn1966marees,
  title={Les mar{\'e}es dans une {\'e}toile double serr{\'e}e},
  author={Zahn, Jean-Paul},
  journal={Annales d'Astrophysique, Vol. 29, p. 313},
  volume={29},
  pages={313},
  year={1966}
}

@article{Peale:1999ay,
    author = "Peale, S. J.",
    title = "{Origin and evolution of the natural satellites}",
    doi = "10.1146/annurev.astro.37.1.533",
    journal = "Ann. Rev. Astron. Astrophys.",
    volume = "37",
    pages = "533--602",
    year = "1999"
}

@article{lecoanet2017conversion,
  title={Conversion of internal gravity waves into magnetic waves},
  author={Lecoanet, Daniel and Vasil, Geoffrey M and Fuller, Jim and Cantiello, Matteo and Burns, Keaton J},
  journal={Monthly Notices of the Royal Astronomical Society},
  volume={466},
  number={2},
  pages={2181--2193},
  year={2017},
  publisher={Oxford University Press}
}

@article{fuller2015asteroseismology,
  title={Asteroseismology can reveal strong internal magnetic fields in red giant stars},
  author={Fuller, Jim and Cantiello, Matteo and Stello, Dennis and Garcia, Rafael A and Bildsten, Lars},
  journal={Science},
  volume={350},
  number={6259},
  pages={423--426},
  year={2015},
  publisher={American Association for the Advancement of Science}
}

@article{duguid2024efficient,
  title={An efficient tidal dissipation mechanism via stellar magnetic fields},
  author={Duguid, Craig D and de Vries, Nils B and Lecoanet, Daniel and Barker, Adrian J},
  journal={The Astrophysical Journal Letters},
  volume={966},
  number={1},
  pages={L14},
  year={2024},
  publisher={The American Astronomical Society}
}

@article{terquem2021new,
  title={On a new formulation for energy transfer between convection and fast tides with application to giant planets and solar type stars},
  author={Terquem, Caroline},
  journal={Monthly Notices of the Royal Astronomical Society},
  volume={503},
  number={4},
  pages={5789--5806},
  year={2021},
  publisher={Oxford University Press}
}

@article{terquem2021circularization,
  title={The circularization time-scales of late--type binary stars},
  author={Terquem, Caroline and Martin, Scott},
  journal={Monthly Notices of the Royal Astronomical Society},
  volume={507},
  number={3},
  pages={4165--4177},
  year={2021},
  publisher={Oxford University Press}
}

@article{barker2021interaction,
  title={On the interaction between fast tides and convection},
  author={Barker, Adrian J and Astoul, Aur{\'e}lie AV},
  journal={Monthly Notices of the Royal Astronomical Society: Letters},
  volume={506},
  number={1},
  pages={L69--L73},
  year={2021},
  publisher={Oxford University Press}
}
\bibliographystyle{aasjournalv7}

\end{document}